\documentclass[fleqn,preprint,aip,pof]{revtex4-1}
\usepackage[utf8]{inputenc}
\usepackage{amsmath}
\usepackage{amsfonts}
\usepackage{amssymb}
\usepackage{verbatim}
\usepackage{graphicx}
\usepackage{float}
\usepackage{subfigure}
\usepackage{hyperref}
\usepackage{threeparttable}
\renewcommand{\vec}[1]{\boldsymbol{#1}}

\begin{document}


\title{Unified Gas-kinetic Wave-Particle Methods II: Multiscale Simulation on Unstructured Mesh}

\author{Yajun Zhu}
\email{zhuyajun@mail.nwpu.edu.cn}
\affiliation
{National Key Laboratory of Science and Technology on Aerodynamic Design and Research, Northwestern Polytechnical University, Xi'an, Shaanxi 710072, China}

\author{Chang Liu}
\email{maliu@ust.hk}
\affiliation
{Department of Mathematics, Hong Kong University of Science and Technology, Hong Kong, China}

\author{Chengwen Zhong}
\email{zhongcw@nwpu.edu.cn}
\affiliation
{National Key Laboratory of Science and Technology on Aerodynamic Design and Research, Northwestern Polytechnical University, Xi'an, Shaanxi 710072, China}

\author{Kun Xu}
\email{makxu@ust.hk}
\affiliation
{Department of Mathematics, Hong Kong University of Science and Technology, Hong Kong, China}
\affiliation
{HKUST Shenzhen Research Institute, Shenzhen 518057, China}


\begin{abstract}
In this paper, we present a unified gas-kinetic wave-particle (UGKWP) method on unstructured mesh for multiscale
simulation of continuum and rarefied flow.
Inheriting from the multicale transport in the unified gas-kinetic scheme (UGKS),
the integral solution of kinetic model equation is employed in the construction of UGKWP method
to model the flow physics in the cell size and time step scales.
A novel wave-particle adaptive formulation is introduced in the UGKWP method to describe the flow dynamics in each control volume.
The local gas evolution is constructed through the dynamical interaction of the deterministic hydrodynamic wave and the stochastic kinetic particle.
Within the resolution of cell size and time step, the decomposition, interaction, and evolution of the hydrodynamic wave and the kinetic particle
depend on the ratio of the time step to the local particle collision time.
In the rarefied flow regime, the flow physics is mainly recovered  by the discrete particles and the UGKWP method performs as a stochastic particle method. In the continuum flow regime, the flow behavior is solely followed  by macroscopic variable evolution and the UGKWP method becomes a gas-kinetic hydrodynamic flow solver for the viscous and heat-conducting Navier--Stokes solutions.
In the transition regime, both kinetic particle and hydrodynamic wave contribute adaptively in UGKWP to capture 
the peculiar non-equilibrium flow physics in a
most efficient way through optimized weighting functions between wave and particle according to the ratio of time step to the local particle collision time.
In different flow regimes, the Sod shock tube, lid-driven cavity flow, laminar boundary layer,
and high-speed cylinder flow, are computed  to validate the UGKWP method on unstructured mesh.
The UGKWP method can get the same UGKS solutions in all Knudsen regimes without the requirement of the time step and mesh size being less than
than the particle collision time and mean free path.
With an automatic wave-particle decomposition, the UGKWP method becomes very efficient.
For example, at Mach number $30$ and Knudsen number $0.1$, in comparison with UGKS several-order-of-magnitude reductions in computational cost and memory requirement have been achieved by UGKWP.
\end{abstract}


\maketitle

\section{Introduction}\label{sec:introduction}
There are mainly two kinds of numerical methods for non-equilibrium gas flow simulations,
i.e., the stochastic particle method and the deterministic method.
The stochastic method employs discrete particles to simulate the statistical behavior of molecular gas dynamics.
Since the Lagrangian-type stochastic particle method can be regarded as the best adaptive strategy in velocity space discretization, it requires low computational memory and gains high efficiency in rarefied flow computations, especially for high-speed flows in multidimensional cases.
Due to the particle implementation, the stochastic method is very robust and shows great advantages on numerical modeling for complex flow physics, such as gas mixture and chemical reaction.
However, by using a finite number of simulation particles to recover the gas distribution function,
it suffers from the statistical noises especially for low-speed flow with small temperature variation.
Moreover, in the continuum flow regime at small Knudsen number,
the intensive particle collisions make the computational cost very high by explicitly following
particle's interaction physics.
On the other hand, the deterministic method employs discrete distribution function to solve the kinetic equations.
It can give accurate solutions without the statistical noises and is able to achieve high efficiency with equation-based numerical
acceleration techniques.
In order to capture non-equilibrium distribution, the whole velocity space has to be discretized with a high resolution,
which leads to huge memory consumption and computational cost for high speed rarefied flow in three dimensional cases.
In addition, for both the stochastic particle and the deterministic methods, once the gas evolution process is splitted into
the collisionless trasport  and instant collision, a numerical dissipation being proportional to the time step
will be introduced implicitly. Therefore, the mesh size and the time step in these schemes
have to be less than the mean free path and the particle collision time in order to properly control the numerical error and
reduce its contamination  to the physical dissipative effect.
Otherwise, only the Euler limit can be recovered in these schemes once the physical dissipation is overwhelmingly taken  by the numerical one
in the continuum regime, such as the simulation of boundary layer or the cavity flow at high Reynolds numbers.
This would severely constrain the applications of these kinetic methods to the  continuum flow simulations.

The direct simulation Monte Carlo (DSMC) method \cite{bird1994book} is the most successful particle simulation method
in the engineering applications of rarefied flow \cite{oran1998direct}.
In the past decades, great efforts have been made to further improve the DSMC method on the aspects of accuracy and efficiency.
In order to reduce the statistical error, the information preservation (IP) method \cite{fan2001statistical,shen2006rarefied,sun2002direct} and the low-variance deviational simulation Monte Carlo (LVDSMC) method \cite{baker2005variance,homolle2007low} have been developed for low-speed microflows.
By matching the moments of macroscopic equations, the moment-guided Monte Carlo method \cite{degond2011moment} is able to reduce the fluctuations as well.
In order to address the stiffness problem of the collision term in the continuum regimes, several asymptotic preserving (AP) Monte Carlo methods  \cite{pareschi2001time,pareschi2000asymptotic,ren2014asymptotic,dimarco2011exponential} have been developed so that the Euler solution can be obtained in the hydrodynamic limit without the requirement on the time step and mesh size as that in the traditional DSMC method.
In order to reduce the numerical diffusion error resulting from free molecular transport process, a low diffusion particle method \cite{burt2008low} has been constructed for inviscid flow simulations.
For further improving the computational efficiency, the stochastic particle methods based on kinetic model equations, such as the Bhatnagar--Gross--Krook (BGK), the ellipsoidal statistical BGK (ES-BGK) models \cite{nanbu1981simulation,gallis2000application,macrossan2001particle,tumuklu2016particle,fei2018particle}, and the Fokker-Planck (FP) model \cite{jenny2010solution,gorji2011fokker,gorji2015fokker}, have been constructed to simulate monatomic and diatomic gas flows.
By taking account of the collision into the transport process, the particle methods of Jenny et al. \cite{jenny2010solution,gorji2011fokker,gorji2015fokker} and Fei et al. \cite{fei2018particle} are capable for flow
simulation without the restriction on the time step and the mesh size to the kinetic level.

On the other hand,  many studies have been carried out on the discrete velocity methods (DVM)  to solve the Boltzmann equation \cite{chu1965kinetic,aristov2012direct,tcheremissine2005direct,li2004gkua,li2019,xu2010unified,guo2013discrete}.
In order to reduce the computational cost and increase the numerical efficiency,
many acceleration methods have been applied, including implicit algorithms \cite{yang1995rarefied,mieussens2000discrete,mieussens2000implicit,zhu2016implicit,zhu2017unified,zhu2019implicit}, high-order/low-order (HOLO) methods \cite{taitano2014moment,chacon2017multiscale}, memory reduction techniques \cite{chen2017memory,yang2018memory}, fast evaluation of the full Boltzmann collision term \cite{mouhot2006fast,wu2013deterministic}, and adaptive refinement method \cite{chen2012amr}.
Asymptotic preserving schemes \cite{filbet2010class,dimarco2013imex} are also developed to release the stiffness of the collision term in the small Knudsen number cases.
However, for most AP schemes only the Euler solution in the hydrodynamic limit is recovered.
By employing the integral solution of kinetic model equation, the unified gas-kinetic scheme (UGKS) \cite{xu2010unified,xu2015direct} and discrete UGKS (DUGKS) \cite{guo2013discrete} are constructed with a true multiscale transport process which couples particles' free streaming and collision.
Similar to the gas-kinetic scheme (GKS) in the hydrodynamic regime \cite{xu2001gks},
the NS solutions can be preserved by the UGKS without the cell resolution to the kinetic scale.
The multiscale transport modeling in UGKS is general and can be applied in many transport problems, such as aerospace application \cite{jiang}, radiative transfer \cite{sun2015gray,sun2015asymptotic}, phonon transport \cite{guo2016phonon,luo2017discrete}, plasma \cite{liu2017plasma}, and granular flow \cite{liu2019granular}.
Since the time step and the mesh size in UGKS are not limited to be smaller than the particle collision time and the mean free path, the UGKS can achieve higher efficiency in the near continuum flow regime.
However, as a deterministic method based on the discrete velocity distribution function,
the computational cost and memory requirement of the UGKS are very high for multi-dimensional hypersonic rarefied flow
due to the massive number of discrete points in the particle velocity space.

Since the UGKS and the particle method have different, but complementary advantages and disadvantages,
a particle implementation of the UGKS is preferred to design a more powerful tool for simulating multiscale transport.
In this paper, we will present in details the construction of the unified gas-kinetic wave-particle (UGKWP) method on unstructured mesh.
The direct particle implementation of the multiscale UGKS is the unified gas-kinetic particle (UGKP) method \cite{li2018ugkp},
and the UGKWP method  is a further improvement by incorporating waves into UGKP 
in order to reduce the computational cost and memory requirement \cite{liu2018ugkwp}. 
In the UGKP method, the multiscale transport process in UGKS is recovered in the particles' evolution process, 
where the collision effect is taken into account so that this particle method can present accurate solutions in different flow regimes without the DSMC requirement for the mesh size and time step.
With the particle implementation, the capability of the UGKS to solve high-speed rarefied flow problems has been much enhanced.
Since the particles' interaction makes the distribution function to approach to the equilibrium state,
some of the simulation particles can be replaced by the evolution of analytical distribution function after some collisions.
Therefore, a wave-particle formulation is introduced in the construction of the UGKWP method, where these equilibrium particles will be
expressed and computed in a deterministic way instead of by discrete simulation particles.
As a result, for the near continuum flow, only a few particles are required and most computation can be handled analytically, so the computational efficiency could be greatly increased and the statistical noises from discrete particles could be efficiently reduced.
The UGKWP method can adaptively become the particle simulation method in highly rarefied flow regimes and the hydrodynamic flow solver, same as the
gas-kinetic scheme \cite{xu2001gks}, in the continuum regimes.
It should be pointed out that different from the hybrid methods \cite{wadsworth1990one,sun2004hybrid,degond2010multiscale} which are based on the domain decomposition and solver hybridization,
the UGKWP describes the physical state by an adaptive wave-particle decomposition in  each cell with a unified treatment
in the whole computational domain.
Specifically, in the UGKWP method the physical state in a finite volume cell is separated into the hydrodynamic waves and discrete particles, which are expressed and transported through an analytic distribution for waves and the particles for the non-equilibrium part, respectively.
According to the numerical resolution and local flow physics, the evolutions of the hydrodynamic waves and discrete particles
are coupled dynamically.
Therefore, with the multiscale transport process and the wave-particle adaptive property, the UGKWP method not only recovers  the multiscale nature of UGKS, but also achieves multi-efficiency in all Knudsen number regimes. More specifically, the UGKWP method performs
as a stochastic particle method in highly rarefied flows and becomes a hydrodynamic fluid solver for viscous flow simulations in the continuum limit.
In the current paper, the UGKWP method is constructed and validated on the unstructured mesh.
Numerical test cases, including the Sod shock tube, cavity flow, laminar boundary layer, and high-speed cylinder flow, will be computed across different flow regimes. For the hypersonic rarefied flow, such as Mach number $30$, the UGKWP has several-order-of-magnitude computational cost and memory reductions in comparison with UGKS.

The remainder of the paper is organized as follows.
Section \ref{sec:numerical_method} presents the construction of the UGKWP method on the unstructured mesh.
In Section \ref{sec:cases}, numerical test cases are carried out to validate the present UGKWP method.
Discussion and conclusion will be given in the last section.

\section{Numerical Method}\label{sec:numerical_method}
In this section, we will introduce the unified gas-kinetic wave-particle (UGKWP) methods in details.
Since the unified gas-kinetic particle (UGKP) method is a direct particle implementation of the unified gas-kinetic scheme (UGKS), and the UGKWP method is an enhanced UGKP method by employing the adaptive wave-particle decomposition, the UGKP method and UGKS will be introduced first before presenting the UGKWP method.

\subsection{Unified gas-kinetic scheme (UGKS) }\label{sec:UGKS}
In the framework of finite volume method, the UGKS considers the conservations in the discretized space and time for both the macroscopic flow variables and microscopic gas distribution function.
Specifically, for a discrete finite volume cell $i$ and discretized time step $\Delta t = t^{n+1} - t^n$, the governing equations are
\begin{equation}\label{eq:macro_conservation}
\vec{w}_i^{n+1} = \vec{w}_i^{n} - \dfrac{\Delta t}{\Omega_i}\sum_{j \in N(i)} {\vec{F}_{ij} S_{ij}},
\end{equation}
and
\begin{equation}\label{eq:micro_conservation}
f_i^{n+1} = f_i^{n} - \dfrac{1}{\Omega_i} \sum_{j \in N(i)}{\int_{0}^{\Delta t}{\vec{u} \cdot \vec{n}_{ij} f_{ij}(t) S_{ij} dt}} + \int_{0}^{\Delta t} {\mathcal{J}(f,f) dt},
\end{equation}
where $\vec{w}$ is the conservative flow variables, i.e., the densities of mass, momentum and energy $(\rho, \rho \vec{V}, \rho E)$, and $f$ is the gas distribution function.
$N(i)$ denotes the set of the interface-adjacent neighboring cells of cell $i$, and cell $j$ is one of the neighbors.
The interface between cells $i$ and $j$ is represented by the subscript $ij$.
Hence, $S_{ij}$ and $\vec{n}_{ij}$ are referred to as the area and normal vector of the interface $ij$, $\vec{F}_{ij}$ denotes the macroscopic fluxes across the interface, and $f_{ij}(t)$ is the time-dependent distribution function on the interface.
In addition, $\Omega_i$ is the volume of cell $i$,  $\vec{u}$ denotes the microscopic velocity, and $\mathcal{J}(f,f)$ is the collision term.
The connection between the macroscopic and microscopic governing equations (\ref{eq:macro_conservation}) and (\ref{eq:micro_conservation}) is that all the macroscopic variables can be obtained from the moments of distribution function, such as conservative variables and macroscopic fluxes
\begin{equation}\label{eq:var_macro_to_micro}
\vec{w}_i = \int{f_i \vec{\psi}(\vec{u}) d\vec{u}},
\end{equation}
and
\begin{equation}\label{eq:flux_macro_to_micro}
\vec{F}_{ij} = \dfrac{1}{\Delta t}\int_{0}^{\Delta t}{\int {\vec{u} \cdot \vec{n}_{ij} f_{ij}(t) \vec{\psi}(\vec{u}) d\vec{u}} dt},
\end{equation}
where $\vec{\psi}(\vec{u}) = (1, \vec{u}, \frac{1}{2} |\vec{u}|^2)$.
The collision term satisfies the compatibility condition
\begin{equation}\label{eq:compatibility_condition}
\int {\mathcal{J}(f,f) \vec{\psi}(\vec{u}) d\vec{u}} = \vec{0} .
\end{equation}
In this paper, the BGK relaxation model \cite{bhatnagar1954model} is considered for the collision term, i.e.,
\begin{equation}\label{eq:bgk_model}
\mathcal{J}(f,f) = \dfrac{g- f}{\tau},
\end{equation}
where the relaxation time or the mean collision time $\tau$ is computed by the ratio of dynamic viscosity to pressure, i.e., $\tau = \mu / p$.
The equilibrium state $g$ is the Maxwellian distribution
\begin{equation}\label{eq:equilibrium_vdf}
g = \rho \left(\dfrac{\lambda}{\pi}\right)^{\frac{d}{2}} {\exp}\left[{-\lambda (\vec{u}-\vec{V})^2}\right],
\end{equation}
where $d$ is degree of freedoms, and $\lambda$ is related to the temperature $T$ by $\lambda = m_0 /2 k_B T$.
Here $m_0$ and $k_B$ are the molecular mass and the Boltzmann constant.

It should be noted that Eq.~(\ref{eq:macro_conservation}) and Eq.~(\ref{eq:micro_conservation}) are the fundamental physical laws on the scale of mesh size and time step, which describe the general conservations of the macroscopic flow variables and microscopic gas distribution function.
For finite volume method, the evolution of flow physics relies mainly on the construction of the flux function at the cell interfaces.
In UGKS, the time-dependent distribution function $f_{ij}(t)$ for the flux function is constructed from the integral solution of kinetic model equation,
\begin{equation}\label{eq:integral_solution}
f(\vec{x}_0, t) = \dfrac{1}{\tau} \int_{t_0}^{t} {g(\vec{x}^{\prime}, t^\prime) e^{-(t-t^\prime)/\tau} dt^\prime} + e^{-(t-t_0)/\tau} f_0(\vec{x}_0 - \vec{u} (t-t_0)),
\end{equation}
where $\vec{x}_0$ is the point for the evaluation of local gas distribution function evolution,
such as the center of a cell interface for flux evaluation.
$f_0(\vec{x})$ is the initial distribution function around $\vec{x}_0$ at the beginning of each step $t_0 = 0$, and $g(\vec{x}, t)$ is the equilibrium state distributed around $\vec{x}_0$ and $t_0$.

Specifically, for second order accuracy we have
\begin{equation}\label{eq:g_expansion}
g(\vec{x},t) = g_0 + g_t t + g_{\vec{x}} \cdot \vec{x},
\end{equation}
and
\begin{equation}\label{eq:f_expansion}
f_0(\vec{x}) = f_0 + f_{\vec{x}} \cdot \vec{x}.
\end{equation}
The time dependent distribution function at the cell interface can be constructed
\begin{equation}\label{eq:fij_flux}
\begin{aligned}
 f_{ij}(t) &= c_1 g_0 + c_2 g_{\vec{x}} \cdot \vec{u} + c_3 g_t + c_4 f_0 + c_5 f_{\vec{x}} \cdot\vec{u}\\
 &= f_{ij}^{eq}(t) + f_{ij}^{fr}(t)
\end{aligned}
\end{equation}
with the coefficients
\begin{equation}\label{eq:fij_coefficients}
\begin{aligned}
c_1 &=  1 - e^{-t/\tau}, \\
c_2 &= t e^{-t/\tau} - \tau (1-e^{-t/\tau}), \\
c_3 &= t -  \tau (1-e^{-t/\tau}),\\
c_4 &=  e^{-t/\tau}, \\
c_5 &= -t e^{-t/\tau},\\
\end{aligned}
\end{equation}
where $f_{ij}^{eq}(t)$ and $f_{ij}^{fr}(t)$ are the terms related to the equilibrium state $g(\vec{x}, t)$ and initial distribution function $f_0(\vec{x})$, respectively.
The initial gas distribution function $f_0(\vec{x})$ in Eq.~(\ref{eq:f_expansion}) is obtained from the spatial reconstruction of the distribution function at $t^n$.
The equilibrium state $g_0$ is computed from the compatibility condition
\begin{equation}\label{eq:collided_g0}
\vec{w}_0 = \int {g_0 \vec{\psi}(\vec{u}) d\vec{u}} = \int {f_0 \vec{\psi}(\vec{u}) d\vec{u}},
\end{equation}
and the spatial and temporal derivatives of the equilibriums state can be obtained by
\begin{equation}\label{eq:equilibrium_derivative}
\begin{aligned}
\int {g_{\vec{x}} \vec{\psi}(\vec{u}) d\vec{u}} &= \vec{w}_{\vec{x}}, \\
\int {g_t \vec{\psi}(\vec{u}) d\vec{u}} &= -\int {\vec{u} g_{\vec{x}} \vec{\psi}(\vec{u}) d\vec{u}},\\
\end{aligned}
\end{equation}
where $\vec{w}_{\vec{x}}$ is the spatial derivatives of the conservative variables obtained from reconstruction.

Eq.~(\ref{eq:integral_solution}) and Eq.~(\ref{eq:fij_flux}) give a transition from the initial distribution function to the equilibrium state with the increment of time, which couples the particles' free transport and collision processes.
With an accumulating effect of the particle collision, the modeling scale changes from the kinetic scale to hydrodynamic scale.
For a discretized space and time, the integral solution will adapt the physical solution on the scale of mesh size and time step according to the relation between the numerical resolution and flow physics, such as the ratio of the time step to the mean collision time $\Delta t / \tau$.
Specifically, the integrated flux over a time step gives
\begin{equation}\label{eq:fij_integrated_flux}
\begin{aligned}
\int_{0}^{\Delta t}{\vec{u} \cdot \vec{n}_{ij} f_{ij}(t) dt}
&= \vec{u} \cdot \vec{n}_{ij} \left( q_1 g_0 + q_2 g_{\vec{x}} \cdot \vec{u} + q_3 g_t \right) \\
& + \vec{u} \cdot \vec{n}_{ij} \left( q_4 f_0 + q_5 f_{\vec{x}} \cdot\vec{u} \right) \\
&= \mathcal{F}_{ij}^{eq} + \mathcal{F}_{ij}^{fr}
\end{aligned}
\end{equation}
with the coefficients
\begin{equation}\label{eq:flux_coefficients}
\begin{aligned}
q_1 &=  \Delta t - \tau (1 - e^{-\Delta t/\tau}), \\
q_2 &= 2\tau^2 (1 - e^{-\Delta t/\tau}) - \tau \Delta t - \tau \Delta t e^{-\Delta t/\tau}, \\
q_3 &= \dfrac{\Delta t^2}{2} - \tau \Delta t  + \tau^2 (1 - e^{-\Delta t/\tau}),\\
q_4 &=  \tau (1 - e^{-\Delta t/\tau}), \\
q_5 &=  \tau \Delta t  e^{-\Delta t/\tau} - \tau^2 (1 - e^{-\Delta t/\tau}).\\
\end{aligned}
\end{equation}
Here $\mathcal{F}_{ij}^{eq}$ and $\mathcal{F}_{ij}^{fr}$ are referred to as the equilibrium flux and the free transport flux, respectively.
We can see that when $\Delta t \gg \tau$, only the terms $\mathcal{F}_{ij}^{eq}$  with $q_1 \approx \Delta t$ and $q_3 \approx \Delta t^2 /2$ are remained for equilibrium wave interaction; when $\Delta t \ll \tau$, only  $\mathcal{F}_{ij}^{fr}$ with $q_4 \approx \Delta t$ and $q_5 \approx -\Delta t^2 / 2$ are left for non-equilibrium particle free transport.
With adaptive variation of $\Delta t / \tau$ in different regions, UGKS is able to provide multiscale flow solutions on the numerical scales.
In comparison with the kinetic method based on particles' free transport mechanics, the mesh size and time step of the UGKS are not constrained by the particles' mean free path and mean collision time.
With the scale-adaptive flux function, the UGKS is an efficient deterministic method for multiscale flow simulations in all regimes.

The algorithms for one time step evolution of the UGKS from $t^n$ to $t^{n+1}$ can be summarized as follows.
\begin{description}
\item[Step 1] Reconstruct the microscopic gas distribution function $f^n$, and obtain the initial gas distribution function $f_0(\vec{x})$ in Eq.~(\ref{eq:f_expansion}).
\item[Step 2] Obtain the equilibrium state $g_0$ at cell interface from the initial distribution function $f_0$ by the compatibility condition (\ref{eq:collided_g0}).
\item[Step 3] Reconstruct the macroscopic flow variables $\vec{w}^n$ and obtain the spatial and temporal derivatives of the equilibrium state $g_{\vec{x}}$, $g_t$ from Eq.~(\ref{eq:equilibrium_derivative}) with the reconstructed $\vec{w}_{\vec{x}}$.
\item[Step 4] Compute the microscopic and macroscopic fluxes across cell interfaces by Eq.~(\ref{eq:fij_integrated_flux}) and Eq.~(\ref{eq:flux_macro_to_micro}).
\item[Step 5] Update the conservative flow variables $\vec{w}^{n+1}$ and the microscopic gas distribution function $f^{n+1}$ by Eq.~(\ref{eq:macro_conservation}) and Eq.~(\ref{eq:micro_conservation}).
\end{description}
Detailed implementation and analysis of the UGKS can be found in Refs.~\onlinecite{xu2010unified,huang2012unified,xu2015direct}.

\subsection{Unified gas-kinetic particle (UGKP) method}\label{sec:UGKP}
Since the particles' tracking and interaction can be regarded as an optimal strategy for the grid point adaption in the particle velocity space,
the stochastic particle methods obtain very high efficiency for simulation of high-speed rarefied flows in three dimensional case.
Therefore, in this section the particle implementation of the UGKS with multiscale transport process will be carried out to construct the UGKP method.
\begin{figure}[H]
	\centering
	\includegraphics[width=0.5\textwidth]{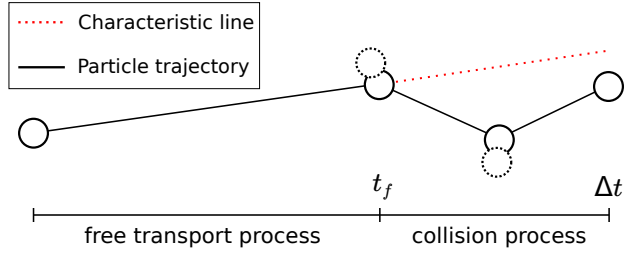}
	\caption{\label{fig:molecule}Particles' evolution on the numerical scale of a time step  $\Delta t$. $t_f$ is the free transport time before the particle encounters the first collision.}
\end{figure}

The physical picture for particles' evolution in a time step $\Delta t$ is illustrated in Fig.~\ref{fig:molecule}.
It shows that one particle will keep free transport until it encounters other particle and gets collided, and then it will continue its
moving and colliding process.
Before its first collision, the particle's trajectory and the characteristic line are the same.
The particle retains its initial discrete velocity.
Once collision happens, the particle velocity changes and we cannot get the exact location and velocity of the particle unless it is traced step by step for each individual collision.
It should be emphasized  that Fig.~\ref{fig:molecule} describes the physics on the numerical scale instead of the kinetic scales of mean free path and collision time.
The free transport time $t_f$ changes with the local physics, and multiple or a huge number of collisions are allowed within the time interval $(t_f, \Delta t)$, which specifies different flow regimes.
Similar to the UGKS, direct modeling of the flow physics on the time step  $\Delta t$ scale is the key to construct a multiscale scheme.

Here we re-write the integral solution (\ref{eq:fij_flux}) along the characteristic line for the end point $(\vec{x}_e, t_e)$ as
\begin{equation}\label{eq:particle_evolution}
\begin{aligned}
f(\vec{x}_e, t_e)
&= (1-e^{-t_e/\tau}) g(\vec{x}^{\prime},t^{\prime}) + e^{-t_e/\tau} f_0(\vec{x}_e -\vec{u} t_e)\\
&= (1-e^{-t_e/\tau}) g_p + e^{-t_e/\tau} f_p\\
\end{aligned}
\end{equation}
where
\begin{equation}
\vec{x}^{\prime} = \vec{u} \left(\dfrac{t_e e^{-t_e / \tau}}{1-e^{-t_e / \tau}} -\tau\right), \quad
t^{\prime} = \left(\dfrac{t_e}{1-e^{-t_e / \tau}} -\tau\right),
\end{equation}
and $t_e$ is related to the time step for a numerical scheme.
The point $(\vec{x}^{\prime}, t^\prime)$ locates on the characteristic line moving from
the mid point to the end point as the increasing of the ratio $t_e / \tau$.
Eq.~(\ref{eq:particle_evolution}) describes that the discrete distribution function at time $t_e$ is a combination of the initial distribution function $f_p$ and the Taylor expansion of the equilibrium state $g_p$.
In analogy to the discrete distribution function in the deterministic methods, it is straightforward to evolve the particle in one time step through Eq.~(\ref{eq:particle_evolution}) by changing the mass-weight of particle instead of its microscopic velocity.
However, this treatment will lose one of the most important advantages of particle methods, i.e., the adaptive property in the velocity space.
In Eq.~(\ref{eq:particle_evolution}), the probability of maintaining the initial distribution function through particles' free transport is given by $e^{-t/\tau}$.
Statistically, the free transport time for each particle can be determined.
Therefore, the motion of all the particles before their first collision can be accurately tracked.
Although the subsequent collision and motion are not exactly followed, in a systematic point of view,
all the particles encountering collision in a local region will approach to an equilibrium distribution $g_p$ according to the kinetic model.

In the framework of finite volume method, we will construct the UGKP method by directly modeling  the above physical processes on the scale of mesh size and time step \cite{xu2015direct}.
Specifically, the free transport process of each particle before its first collision within a time step will be accurately tracked
and the effect of collision is to annihilate the particles and be recovered by re-sampling them from a specific Maxwellian equilibrium state.
In order to give a clear description, we will use the terms of free transport process and collision process as shown in Fig.~\ref{fig:molecule} to denote the stages before and after the particles' first collision in one time step, while the whole evolution process in one time step is denoted as the transport process or the multiscale transport process.
As a finite volume method, the computation of the UGKP method for a time step evolution would contain
\begin{description}
	\item[Macro level] Compute the numerical fluxes to update the conservative flow variables, which includes
	\begin{description}
		\item[(Pa)] computing the fluxes contributed from the particles' motion in the free transport process;
		\item[(Pb)] computing the fluxes contributed from the particles' motion in the collision process.
	\end{description}
	\item[Micro level] Evolve the gas distribution function, i.e., update the simulation particles, which includes
	\begin{description}
		\item[(Pc)] tracking all the particles' motion in the free transport process;
		\item[(Pd)] re-sampling collisional particles in the collision process.
	\end{description}
\end{description}

\subsubsection{Free transport process}\label{sec:transport}

Eq.~(\ref{eq:particle_evolution}) gives the cumulative distribution for particles' free transport time
\begin{equation}\label{eq:cumulative_distribution}
\mathcal{G}(t) = e^{-t / \tau},
\end{equation}
so the free transport time of a particle within a time step $\Delta t$ can be determined by
\begin{equation}\label{eq:free_transport_time}
t_f ={\min}(- \tau {\rm ln}(r_0), \Delta t)
\end{equation}
where $r_0$ is a random number generated from a uniform distribution between $(0,1)$.
Given with the free transport time $t_f$, the particle can be accurately tracked by
\begin{equation}\label{eq:free_transport_displacement}
\vec{x}_p = \vec{x}_p^n + \vec{u} t_f,
\end{equation}
where the micro velocity $\vec{u}$ of the particle remains unchanged during the free transport process.
Different from the DSMC method with $t_f = \Delta t$, where all the particles free stream with a whole time step, the free transport time in the UGKP method is constrained due to particles' collision.

During the free transport process, the contribution to the numerical fluxes of cell $i$ can be obtained by counting the particles across the cell interfaces
\begin{equation}\label{eq:particle_flux}
\vec{W}_i^{fr} = \sum_{k \in P(i)} \vec{\phi}_k
\end{equation}
where $P(i)$ is the set of the particles moving across the interfaces of the cell $i$ during the free transport process.
The vector $\vec{\phi}_k = (m_p, m_p \vec{u}_k, \frac{1}{2} m_p |\vec{u}_k^2|)$ denotes the mass, momentum,and energy carried by the particle $k$.
In comparison with the multiscale transport process given in the UGKS in Eq.~(\ref{eq:fij_integrated_flux}), the free transport
process (\ref{eq:free_transport_displacement}) only recovers the fluxes contributed by the initial distribution function $f_0(\vec{x})$, i.e., $\mathcal{F}_{ij}^{fr}$ with the terms $q_4$ and $q_5$.
For comparison, the counterpart of $\vec{W}_i^{fr}$ in the deterministic UGKS would be
\begin{equation}\label{eq:free_transport_flux}
\vec{W}_i^{fr}
= -\sum_{j \in N(i)} S_{ij} \int_{0}^{\Delta t}{\int {\vec{u} \cdot \vec{n}_{ij} f_{ij}^{fr}(t)  \vec{\psi}(\vec{u}) d\vec{u}} dt} = -\sum_{j \in N(i)} S_{ij} \int { \mathcal{F}_{ij}^{fr}  \vec{\psi}(\vec{u}) d\vec{u}}.
\end{equation}

So far, we have carried out the processes {\bf (Pa)}  and {\bf (Pc)} by Eq.~(\ref{eq:particle_flux}) and Eq.~(\ref{eq:free_transport_displacement}).

\subsubsection{Collision process: macroscopic fluxes}\label{sec:macro}
In the collision process, the particles get collided and keep on moving and colliding process.
During this process, once the particles move across the cell interfaces, they will contribute to the macroscopic fluxes as well.
However, since we are not developing a full particle tracking method, the motion of simulation particles in the collision process will not
be followed explicitly.
So the macroscopic fluxes cannot be directly obtained from the discrete particles as in the free transport process.
Fortunately, these fluxes have been already given in the UGKS in Eq.~(\ref{eq:fij_integrated_flux}), i.e.,  $\mathcal{F}_{ij}^{eq}$ with the terms $q_1$, $q_2$ and $q_3$.

Hence, the macroscopic fluxes of the collision process across the cell interface $ij$ can be computed from the reconstructed macroscopic flow variables by
\begin{equation}\label{eq:equilibrium_flux}
\vec{F}_{ij}^{eq} = \int_{0}^{\Delta t}{\int {\vec{u} \cdot \vec{n}_{ij} f_{ij}^{eq}(t) \vec{\psi}(\vec{u}) d\vec{u}} dt} = \int { \mathcal{F}_{ij}^{eq} \vec{\psi}(\vec{u}) d\vec{u}}.
\end{equation}
In UGKS, $g_0$ is obtained from the reconstructed initial distribution function $f_0$ by the compatibility condition (\ref{eq:collided_g0}). However, in the particle method there is no explicit gas distribution function, so the equilibrium state $g_0$ at cell interface is computed from the reconstructed macroscopic flow variables, i.e.,
\begin{equation}\label{eq:collided_glr}
\int {g_0 \vec{\psi}(\vec{u}) d\vec{u}} = \int_{\vec{u} \cdot \vec{n} > 0} {g_l \vec{\psi}(\vec{u}) d\vec{u}} + \int_{\vec{u} \cdot \vec{n} < 0} {g_r \vec{\psi}(\vec{u}) d\vec{u}} ,
\end{equation}
where $g_l$ and $g_r$ are the equilibrium state on the left and right sides of cell interface, which are determined by the interpolated macroscopic flow variables $\vec{w}_l$ and $\vec{w}_r$.
Same as that in the GKS and UGKS,  $g_t$ and $g_{\vec{x}}$ can be obtained by Eq.~(\ref{eq:equilibrium_derivative}), and then the equilibrium fluxes $\vec{F}_{ij}^{eq}$ can be analytically computed.

At this moment, the multiscale fluxes (\ref{eq:fij_integrated_flux}) in the UGKS have been fully recovered by free transport fluxes (\ref{eq:particle_flux}) and collisional fluxes (\ref{eq:equilibrium_flux}).
The macroscopic flow variables can be updated by the conservation laws
\begin{equation}\label{eq:macro_update}
\vec{w}_i^{n+1} = \vec{w}_i^{n} - \dfrac{1}{\Omega_i}\sum_{j \in N(i)} {\vec{F}_{ij}^{eq} S_{ij}} + \dfrac{\vec{W}_i^{fr}}{\Omega_i}.
\end{equation}
So far, we have dealt with the process {\bf (Pb)} and finished the update on the macroscopic level.

\subsubsection{Collision process: microscopic particles}\label{sec:sample}
In the free transport process, the detailed motion of all the particles in the time interval $(0, t_f)$ have been tracked.
For the collisionless particles with $t_f = \Delta t$, the update of the microscopic state including particles' velocity and location has been finished for the current step.
While for the collisional particles with $t_f < \Delta t$, each of them will suffer at least one collision in the time interval $(t_f, \Delta t)$, and the collective effect of the collisions is to force all the collisional particles in the local region to follow a specific equilibrium distribution $g_p$.
According to the conservation, from the updated macroscopic flow variables and the streamed collisionless particles, we can easily obtain the the collisional particles at the end of the time step
\begin{equation}\label{eq:updated_wave}
\vec{w}_i^{h} = \vec{w}_i^{n+1} - \vec{w}_i^{p}
\end{equation}
where $\vec{w}_i^p$ is the conservative flow variables carried by the collisionless particles in the currently investigated cell $i$ after their solely free transport process.
Therefore, in the collision process $(t_f, \Delta t)$, these collisional particles will be deleted first due to particles' collision, and then re-sampled from their corresponding
macroscopic flow variable $\vec{w}_i^{h}$ at each end of time step.
A unique Maxwellian distribution (\ref{eq:equilibrium_vdf}) can be determined from $\vec{w}_i^{h}$.
According to the macroscopic velocity and temperature, the collisional particles can be re-sampled in the cell $i$ to recover the distribution function.
So far, we have carried out the process {\bf (Pd)}, and both the macroscopic flow variables and the microscopic particles have been updated.

\subsubsection{Summary and discussions}

\begin{figure}[H]
\centering
\subfigure[\label{fig:ugkp_a}]
{\includegraphics[width=0.24\textwidth]{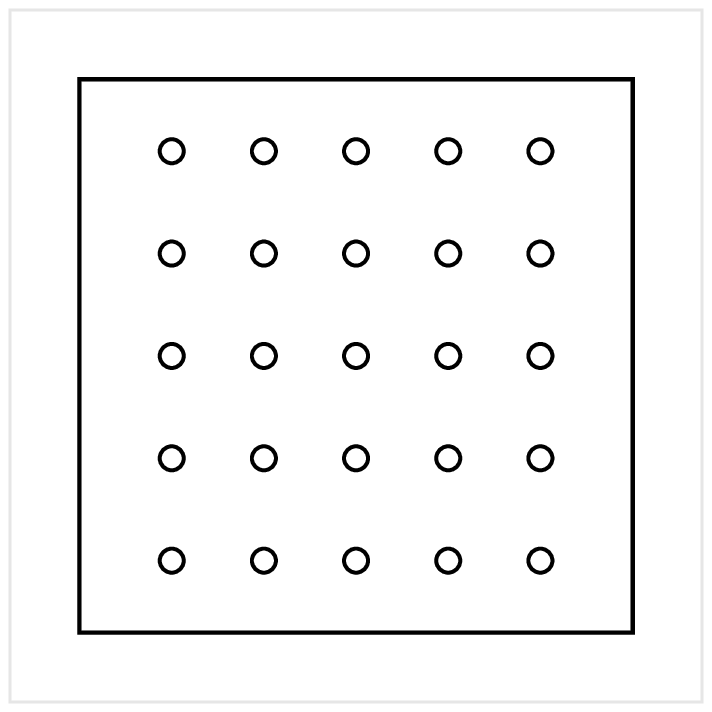}}
\subfigure[\label{fig:ugkp_b}]
{\includegraphics[width=0.24\textwidth]{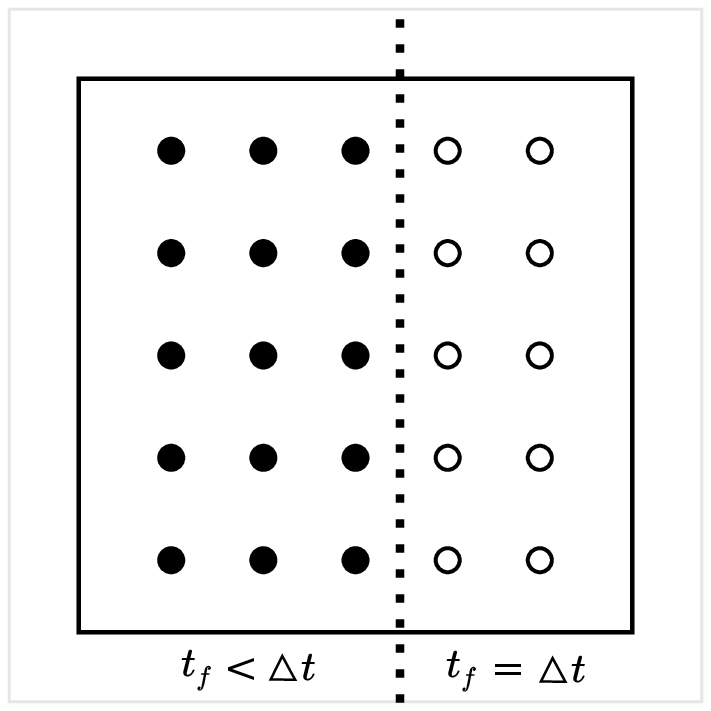}}
\subfigure[\label{fig:ugkp_c}]
{\includegraphics[width=0.24\textwidth]{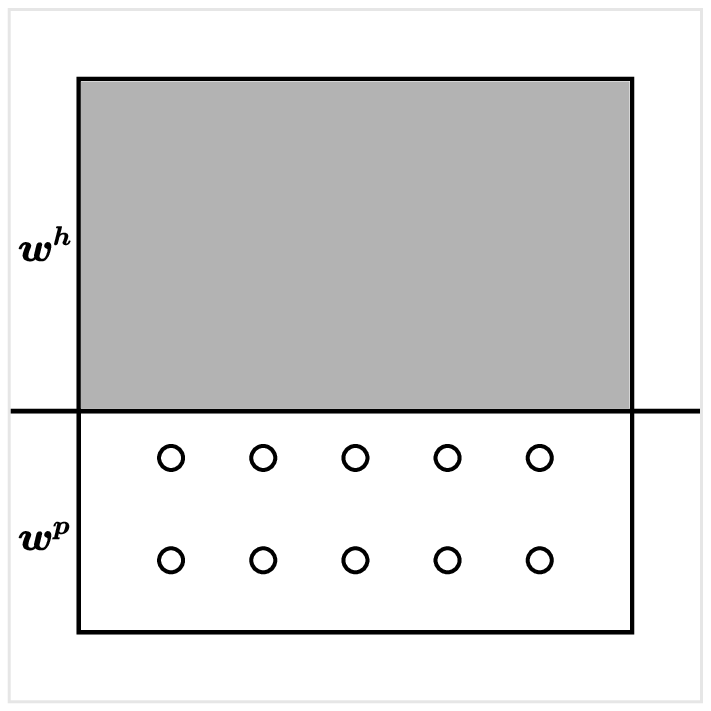}}
\subfigure[\label{fig:ugkp_d}]
{\includegraphics[width=0.24\textwidth]{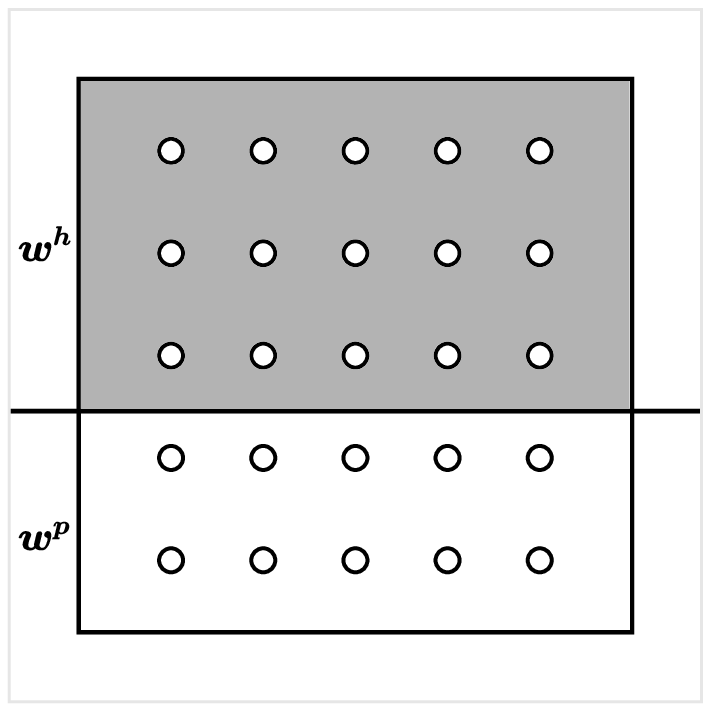}}
\caption{\label{fig:ugkp_diagram}Diagram to illustrate the composition of the particles during time evolution in the UGKP method. (a) Initial state; (b) classification of the collisionless particles (white circle) and collisional particles (solid circle) according to the free transport time $t_f$; (c) update solution at macroscopic level;(d) update solution at microscopic level.}
\end{figure}

In order to give a visual demonstration for the evolution of simulation particles, a series of diagrams are drawn in Fig.~\ref{fig:ugkp_diagram} to illustrate the composition of particles on different evolution stages within one time step.
The explanation of Fig.~\ref{fig:ugkp_diagram} and the summary of the UGKP method will be given as follows.
\begin{description}
\item[Step 1] Give an initial state with macroscopic flow variables and microscopic particles as shown in Fig.~\ref{fig:ugkp_a}, where the simulation particles could be an initial  equilibrium distribution for the start of flow simulations, or a non-equilibrium distribution evolved from the previous step in the time evolution process.
\item[Step 2] Free transport process, which includes
\begin{itemize}
\item Obtain the free transport time $t_f$ for each particle by Eq.~(\ref{eq:free_transport_time}) and classify the particles into collisionless particles (white circles in Fig.~\ref{fig:ugkp_b}) and collisional particles (solid circles in Fig.~\ref{fig:ugkp_b});
\item Move all the particles for a free transport time $t_f$ in Eq.~(\ref{eq:free_transport_displacement});
\item Cumulate the free transport fluxes $\vec{W}_i^{fr}$ in Eq.~(\ref{eq:particle_flux}) by counting the particles which move across the cell interfaces;
\item Tally the collisionless particles with $t_f = \Delta t$ after streaming all the particles and calculate $\vec{w}_i^p$ of these freely transported particles in each cell $i$
(denoted by the particles on the bottom part of Fig.~\ref{fig:ugkp_c}), and delete the collisional particles.
\end{itemize}
\item[Step 3] Collision process, which includes
\begin{itemize}
\item Reconstruct macroscopic flow variables $\vec{w}^n$ to obtain the conservative flow variables $\vec{w}_l$ and $\vec{w}_r$ on the left and right sides of cell interface;
\item Obtain the equilibrium state $g_0$ at cell interface from Eq.~(\ref{eq:collided_glr}), and compute the derivatives $g_{\vec{x}}$ and $g_t$ from the reconstructed spatial derivative of macroscopic flow variables $\vec{w}_{\vec{x}}$ by Eq.~(\ref{eq:equilibrium_derivative});
\item Compute the collisional fluxes $\vec{F}_{ij}^{eq}$ in Eq.~(\ref{eq:equilibrium_flux}), i.e., the terms with $q_1$, $q_2$ and $q_3$ in Eq.~(\ref{eq:fij_integrated_flux}).
\item Update the conservative variables $\vec{w}_i^{n+1}$ by Eq.~(\ref{eq:macro_update}), and obtain the macroscopic variables $\vec{w}_i^h$ by Eq.~(\ref{eq:updated_wave}) for the updated collisional particles (the grey area shown in Fig.~\ref{fig:ugkp_c}).
\item Re-sample the collisional particles from $\vec{w}_i^h$ to finish the update process of the microscopic particles as illustrated in Fig.~\ref{fig:ugkp_d}.
\end{itemize}
\item[Step 4] Determine the computation to the next time step.
\begin{itemize}
\item If the solution is convergent for steady flows or the pre-described evolution time is achieved for unsteady flows, stop the program;
\item Otherwise, go to Step 1 and continue the computation, where the updated state in Fig.~\ref{fig:ugkp_d} could be an initial state in Fig.~\ref{fig:ugkp_a} for next step evolution.
\end{itemize}
\end{description}
The UGKP method is a conservative finite volume method, where the simulation particles are employed to recover the underlying non-equilibrium distribution function.
On the macroscopic level, the conservative variables are updated with the fluxes by conservation laws.
On the microscopic level, all the particles are accurately tracked in the free transport process and the collisional particles are re-sampled from the updated equilibrium state in the collisional process.
The maintenance of conservation laws is the key to the success of the current particle method.
In addition, it should be noted that in the free transport process each particle moves over a free transport time $t_f$ instead of a whole time step $\Delta t$, and together with the subsequent collisional process a mult-scale transport process is constructed with recovering the multiscale nature of UGKS.

\subsection{Unified gas-kinetic wave-particle (UGKWP) method}\label{sec:ugkwp}
In this section, the concept of the wave-particle will be introduced into the UGKP method for the further development of an efficient UGKWP method.
In the UGKP method, the gas distribution function is fully represented by the simulation particles, and the collisional particles are deleted and re-sampled from a Maxwellian distribution in the collisional process.
Theoretically, this portion of gas distribution function can be expressed in an analytic way.
As shown in Fig.~\ref{fig:ugkp_c}, the gas distribution function can be recovered by hydrodynamic waves and discrete kinetic particles, which correspond to the macroscopic variables $\vec{w}^h$ and $\vec{w}^p$.
For the next time step evolution in the UGKP method, the re-sampled equilibrium particles will be re-classified into collisionless and the collisional particles according to the free transport time $t_f$ again.
In the free transport process, both types of the particles will contribute to the free transport fluxes, but only the collisionless part particles
will be remained as particles at the end of the time step to recover the non-equilibrium gas distribution function.
Therefore, only the collisionless particles in the hydrodynamic waves should be re-sampled at the end of each time step
and the contribution from these collisional particles, which are generated from the hydrodynamic wave previously,
to the free transport fluxes in the next time step can be computed analytically.

From the cumulative distribution (\ref{eq:cumulative_distribution}), we can easily obtain the expectation of the proportion of the collisionless particles in each cell, and the particles required to be sampled in the hydrodynamic waves $\vec{w}^h$ at the end of time step are
\begin{equation}\label{eq:sampled_collisionless_particles}
	\vec{w}^{hp} = \vec{w}^{h} e^{-\Delta t / \tau}.
\end{equation}
The free transport fluxes contributed from the collisional particles of $(\vec{w}^h-\vec{w}^{hp})$ can be computed analytically by
\begin{equation}\label{eq:wave_flux}
\begin{aligned}
\vec{F}^{fr,wave}
&= \vec{F}_{UGKS}^{fr}(\vec{w}^h) - \vec{F}_{DVM}^{fr}(\vec{w}^{hp})\\
&= \int {\vec{u}\cdot \vec{n}\left[(q_4 - \Delta t e^{-\Delta t / \tau} ) g_0^h + (q_5 +  \dfrac{\Delta t^2}{2} e^{-\Delta t / \tau}) \vec{u} \cdot g_{\vec{x}}^h\right] \vec{\psi}(\vec{u})d\vec{u} }
\end{aligned}
\end{equation}
where $g_0^h$ is the Maxwellian distribution determined by $\vec{w}^h$ and $g_{\vec{x}}^h$ is the spatial derivative of the Maxwellian distribution, which can be obtained from the reconstruction of $\vec{w}^h$.

Therefore, in the UGKWP method the update process for the conservative variables will be
\begin{equation}\label{eq:macro_update_wp}
\vec{w}_i^{n+1} = \vec{w}_i^{n} - \dfrac{1}{\Omega_i}\sum_{j \in N(i)} {\vec{F}_{ij}^{eq} S_{ij}}
- \dfrac{1}{\Omega_i}\sum_{j \in N(i)} {\vec{F}_{ij}^{fr,wave} S_{ij}} + \dfrac{\vec{W}_i^{fr}}{\Omega_i}.
\end{equation}
In comparison with the update formula (\ref{eq:macro_update}) in the UGKP method, the term of $\vec{F}_{ij}^{fr,wave}$ is the analytic part extracted from the particles' free transport fluxes $\vec{W}_i^{fr}$.
The combination of the last two terms in Eq.(\ref{eq:macro_update_wp}) for UGKWP is the same as the last term in Eq.(\ref{eq:macro_update}) for UGKP.

\begin{figure}[H]
\centering
\subfigure[\label{fig:ugkwp_a}]
{\includegraphics[width=0.24\textwidth]{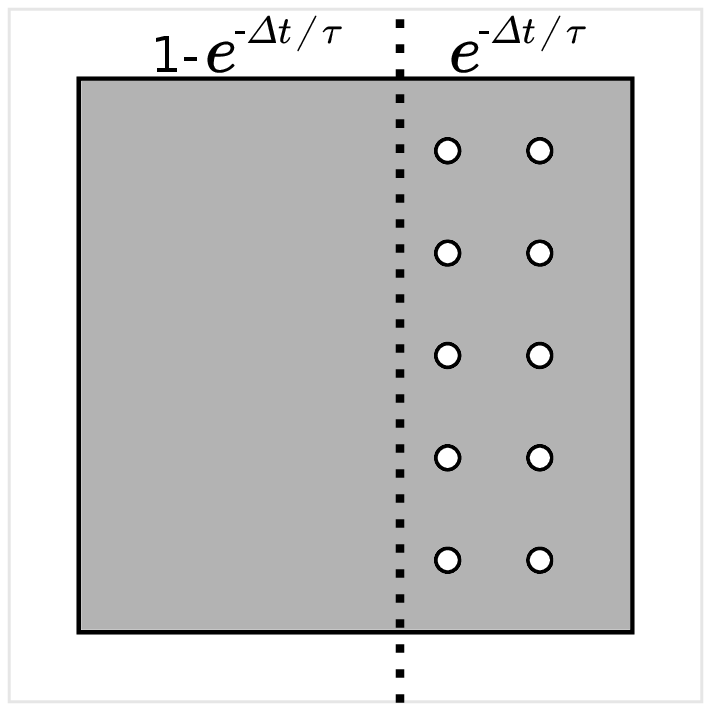}}
\subfigure[\label{fig:ugkwp_b}]
{\includegraphics[width=0.24\textwidth]{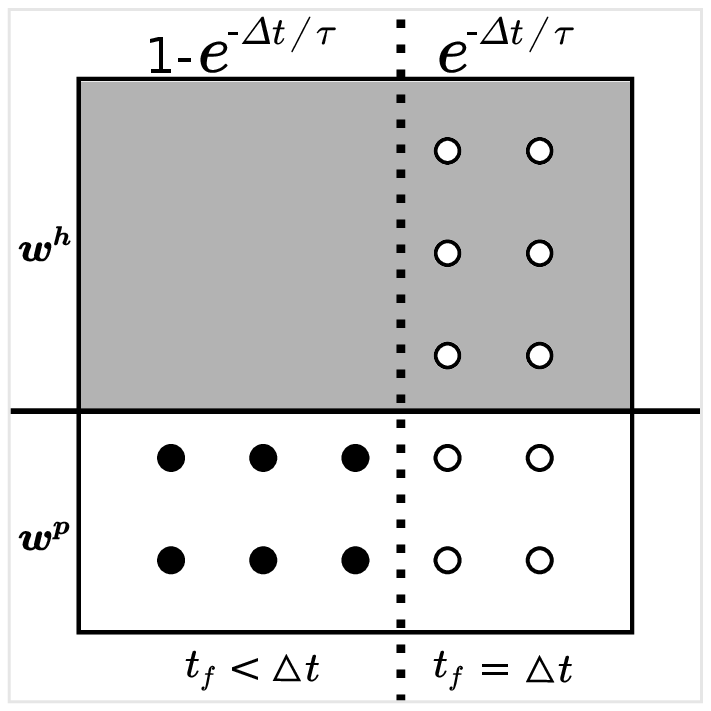}}
\subfigure[\label{fig:ugkwp_c}]
{\includegraphics[width=0.24\textwidth]{ugkp_c.eps}}
\subfigure[\label{fig:ugkwp_d}]
{\includegraphics[width=0.24\textwidth]{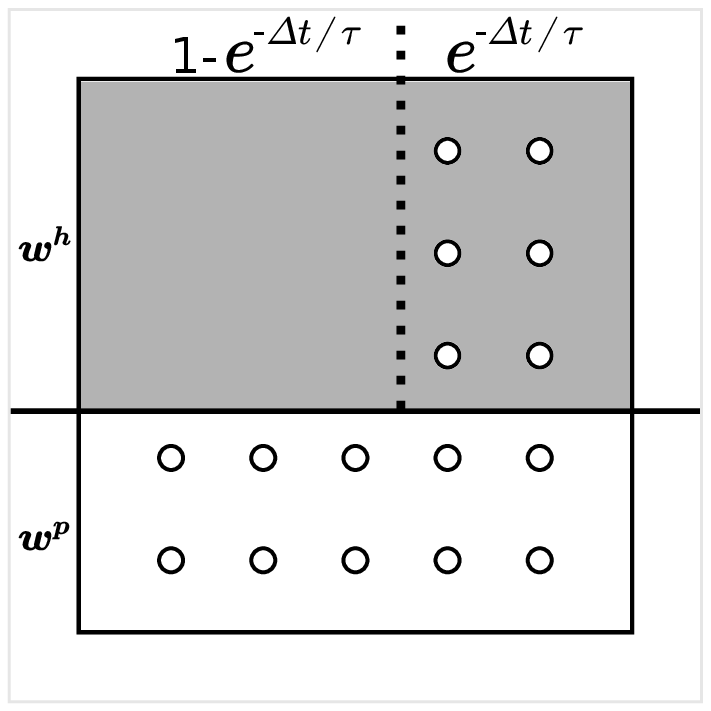}}
\caption{\label{fig:ugkwp_diagram}Diagram to illustrate the composition of the particles during time evolution in the UGKWP method. (a) Initial state; (b) classification of the collisionless and collisional particles for the part of $\vec{w}^p$; (c) update on macroscopic level; (d) update on the microscopic level.}
\end{figure}

A series of figures are drawn in Fig.~\ref{fig:ugkwp_diagram} to illustrate the evolution of the UGKWP method.
The algorithm of the UGKWP method can be summarized as follows.
\begin{description}
\item[Step 1] Get an initial state with the macroscopic flow variables $\vec{w}^n$ and the microscopic particles. The particles include the particles $\vec{w}^p$ evolved from the previous step and the collisionless particles sampled from the updated hydrodynamics waves $\vec{w}^{hp}$ as show in Fig.~\ref{fig:ugkwp_d}. For the first step, $\vec{w}^p = 0$ and $\vec{w}^h = \vec{w}^{n=0}$ as shown in Fig.~\ref{fig:ugkwp_a}.
\item[Step 2] Free transport process, which includes
\begin{itemize}
\item Classify the particles $\vec{w}^p$ into collisionless particles (white circles in Fig.~\ref{fig:ugkwp_b}) and collisional particles (solid circles in Fig.~\ref{fig:ugkwp_b}) according to the free transport time $t_f$ determined by Eq.~(\ref{eq:free_transport_time}).The free transport time of the re-sampled collisionless particles $\vec{w}^{hp}$ is always $t_f = \Delta t$ (white circles on the right top of Fig.~\ref{fig:ugkwp_b});
\item Stream all the particles over the free transport time $t_f$;
\item Cumulate the free transport fluxes $\vec{W}_i^{fr}$ in Eq.~(\ref{eq:particle_flux}) by counting the particles which move across cell interfaces;
\item Tally the updated collisionless particles $\vec{w}^p$ and delete the streamed collisional particles;
\item Compute the free transport fluxes $\vec{F}^{fr,wave}$ contributed from the un-sampled particles $(\vec{w}^h - \vec{w}^{hp})$ by Eq.~(\ref{eq:wave_flux}).
\end{itemize}
\item[Step 3] Collision process, which includes
\begin{itemize}
	\item Compute the collisional fluxes $\vec{F}_{ij}^{eq}$ as same as in the UGKP method;
	\item Update the conservative variables $\vec{w}_i^{n+1}$ by the conservation laws (\ref{eq:macro_update_wp}), and obtain the updated hydrodynamic waves $\vec{w}_i^h$ by Eq.~(\ref{eq:updated_wave}), as shown in Fig.~\ref{fig:ugkwp_c};
	\item Sample the collisionless particles $ \vec{w}_i^{hp}$ in Eq.~(\ref{eq:sampled_collisionless_particles}) for next step evolution and finish the update process of the microscopic particles, see in Fig.~\ref{fig:ugkwp_d}.
\end{itemize}
\item[Step 4] Determine the computation to the next time step.
\begin{itemize}
\item If the solution is convergent for steady flows or the pre-described evolution time is achieved for unsteady flows, stop the program;
\item Otherwise, go to Step 1 and continue the computations.
\end{itemize}
\end{description}

In the UGKWP method, the wave-particle formulation is introduced to represent the non-equilibrium gas distribution function.
The difference between the UGKP and UGKWP methods is that only the collisionless particles in the hydrodynamic waves are re-sampled at the end of time step; and for the next step evolution the free transport fluxes contributed from these un-sampled particles in the hydrodynamic waves are computed in a deterministic way.
For near continuum flows where intense collisions are involved, i.e., $t_f \ll \Delta t$,
the hydrodynamic waves will be dominant and only a few collionless particles are required to be sampled, which makes the current method very efficient.
Therefore, in different flow regimes, the wave-particle decomposition will give an optimal formulation for the non-equilibrium gas distribution function, and achieve higher efficiency both in the continuum and rarefied flow regimes.

\subsection{Unstructured mesh technique}

\subsubsection{Particle tracking}
\begin{figure}[H]
	\centering
	\includegraphics[width=0.4\textwidth]{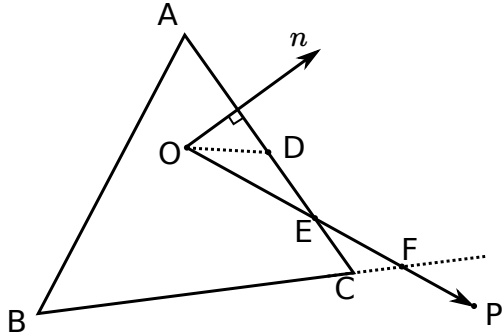}
	\caption{\label{fig:track}Illustration of particle tracking on triangular unstructured meshes.}
\end{figure}
During the free transport process, trajectories of the simulation particles are fully tracked.
For an arbitrary particle locating at point $O$ with microscopic velocity $\vec{u}$ in the triangular cell $\triangle {\rm ABC}$, its displacement in the free transport process would be $\vec{x}_{\rm OP} = \vec{u} t_f$.
The displacement $\vec{x}_{\rm OP}$ may intersect with the faces satisfying $\vec{x}_{\rm OP} \cdot \vec{n} > 0$, where $\vec{n}$ is the normal vector of cell interface.
The intersection point, e.g., point $E$ on face ${\rm AC}$ satisfies
\begin{equation}
\dfrac{\rm OE}{\rm OP} = \dfrac{\vec{x}_{\rm OE} \cdot \vec{n}}{\vec{x}_{\rm OP} \cdot \vec{n}} = \dfrac{\vec{x}_{\rm OD} \cdot \vec{n}}{\vec{x}_{\rm OP} \cdot \vec{n}} ,
\end{equation}
where point $D$ is the centroid of face $\rm AC$.
Similarly, the intersection point $F$ on face $\rm BC$ can be obtained as well.
A minimum value
\begin{equation}
\alpha = \min\left(\dfrac{\rm OE}{\rm OP},\dfrac{\rm OF}{\rm OP}\right)
\end{equation}
can be used to determine the first intersection point of the trajectory and the cell interfaces \cite{haselbacher2007efficient}.
If $\alpha > 1$, the particle is still inside the current cell, and the updated location will be $\vec{x}_P = \vec{x}_O + \vec{u} t_f$.
If $\alpha \le 1$, the particle will move out of the current cell.
For this case, we will first move the particle to the intersection point
$\vec{x}_E = \vec{x}_O + \vec{u} \alpha t_f$, and then track the particle in its neighboring cell using the same method for the remaining free transport time $(1-\alpha)t_f$.

\subsubsection{Particle sampling}
In the collision process, simulation particles will be re-sampled from a given Maxwellian distribution function to recover the gas distribution function on microscopic level.
Given with a Maxwellian distribution determined by $(\rho_s, \vec{V}_s, \lambda_s)$, the microscopic velocity for each particle to sample can be obtained from
\begin{equation}
\begin{aligned}
u &= U_s + \sqrt{-{\rm ln}(r_1) / \lambda_s} \cos(2\pi r_2),\\
v &= V_s + \sqrt{-{\rm ln}(r_1) / \lambda_s} \sin(2\pi r_2),\\
w &= W_s + \sqrt{-{\rm ln}(r_3) / \lambda_s} \cos(2\pi r_4),\\
\end{aligned}
\end{equation}
where $U_s$, $V_s$ and $W_s$ are the components of $\vec{V}_s$, $r_1$, $r_2$, $r_3$ and $r_4$ are independent random numbers generated from the uniform distribution between the interval $(0,1)$.
In the current study, a symmetric sampling process is employed.
Specifically, from a group of $r_1$, $r_2$, $r_3$ and $r_4$, a pair of simulation particles with microscopic velocity $(u,v,w)$ and $(u^\prime, v^\prime, w^\prime)$ are sampled, where the symmetric microscopic velocity is
\begin{equation}
\begin{aligned}
u^\prime &= U_s - \sqrt{-{\rm ln}(r_1) / \lambda_s} \cos(2\pi r_2),\\
v^\prime &= V_s - \sqrt{-{\rm ln}(r_1) / \lambda_s} \sin(2\pi r_2),\\
w^\prime &= W_s - \sqrt{-{\rm ln}(r_3) / \lambda_s} \cos(2\pi r_4).\\
\end{aligned}
\end{equation}

Given with a pre-set reference mass $m_r$ for the simulation particle, the number of particles to be sampled is determined by
\begin{equation}\label{eq:sampling_number}
N_s =
\begin{cases}
0, & {\rm if} \  \Omega_s \rho_s \le m_{min},\\
2 \left\lceil \dfrac{\Omega_s \rho_s}{2 m_r} \right\rceil, & {\rm if} \  \Omega_s \rho_s > m_{min},
\end{cases}
\end{equation}
where $\Omega_s$ is the cell volume, and $m_{min}$ is the minimum mass to sample.
In the sampling process, for the cases $N_s > 0$ the mass weight actually sampled for each simulation particle is
\begin{equation}
m_p = \dfrac{\Omega_s \rho_s}{N_s}
\end{equation}
which guarantees the mass density $\rho_s$ in the volume $\Omega_s$ after sampling process.

Another way is to give a preference number $N_r$ for each cell, then the reference mass $m_r$ can be approximated by
\begin{equation}\label{eq:reference_mass}
m_r = \dfrac{(\rho - \rho^h) + \rho^h  e^{-\Delta t / \tau}}{N_r} \Omega_s,
\end{equation}
and then the number of particles to sample $N_s$ can be obtained by Eq.~(\ref{eq:sampling_number}).
By this way, the number of particles in each cell can be basically controlled around the given reference number $N_r$.

\begin{figure}[H]
\centering
\includegraphics[width=0.4\textwidth]{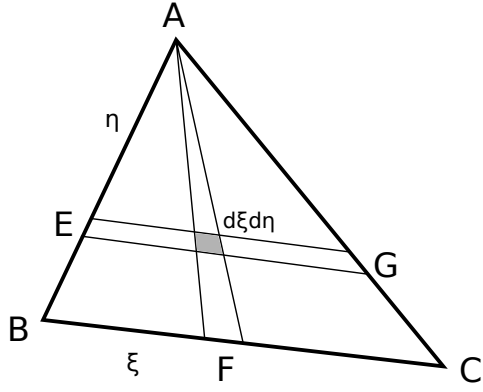}
\caption{\label{fig:triangle}Illustration of the computation of cumulative distribution function on triangular unstructured meshes.}
\end{figure}

Besides the mass weight and the microscopic velocity, the location of each simulation particle is required as well in the sampling process.
For an arbitrary triangular cell $\triangle {\rm ABC}$ as shown in Fig.\ref{fig:triangle}, a point inside can be denoted by $(\xi, \eta)$, which has ${\rm BF} = \xi {\rm BC}$, ${\rm AE} = \eta {\rm AB}$ and ${\rm EG} \parallel {\rm BC}$. The coordinates of the point $(\xi, \eta)$ in the global system are
\begin{equation}\label{eq:tri_location}
\vec{x} = \vec{x}_A (1-\eta) + \vec{x}_B (1-\xi) \eta + \vec{x}_C \xi \eta.
\end{equation}
Assuming that the density is constant inside the cell, i.e., $\rho(\vec{x}) = \rho_s$, the normalized cumulative distribution function up to the line $\xi = \xi_0$ is
\begin{equation}\label{eq:tri_cdf_a}
\mathcal{G}(\xi_0) = \left. \int_{0}^{1}{\int_{0}^{\xi} \rho(\xi, \eta)  d\xi d\eta} \middle/  \int_{0}^{1}{\int_{0}^{1} \rho(\xi, \eta) d\xi d\eta} \right. = \xi_0,
\end{equation}
and along the line $\xi = \xi_0$ the cumulative distribution function up to the point $(\xi_0, \eta_0)$ is
\begin{equation}\label{eq:tri_cdf_b}
\mathcal{H}(\xi_0, \eta_0) = \left. \int_{0}^{\eta_0} {\rho(\xi_0, \eta) } d\eta \middle/  \int_{0}^{1}{\rho(\xi_0, \eta) d\eta} \right. = \eta_0^2.
\end{equation}
Therefore, generating two random numbers $r_1$ and $r_2$ from a standard uniform distribution, the particle location can be determined by Eq.(\ref{eq:tri_location}) with $\xi = r_1$ and $\eta = \sqrt{r_2}$.

In the current study, although piecewise constant of density is assumed in a finite volume cell during the particle sampling process, we find the spatial accuracy is almost not reduced.
This is due to the fact that only the portion $e^{-\Delta t /\tau}$ of the hydrodynamic waves are sampled, and the remaining part of hydrodynamic waves is computed analytically with second order accuracy.
In the continuum regimes, although the hydrodynamic waves are dominant, the collisionless particles in the hydrodynamic waves are very few due to small value of $e^{-\Delta t /\tau}$; while in the rarefied flows, the hydrodynamic waves only take a small portion of the physical state due to mild collisions, so the particles in the hydrodynamic waves required to be re-sampled are not many as well.
If the spatial derivative of density is considered in $\rho(\vec{x})$, the cumulative distribution can be derived as well, which would be more complicated than Eq.~(\ref{eq:tri_cdf_a}) and Eq.~(\ref{eq:tri_cdf_b}).
Acceptance-rejection strategy \cite{bird1994book,shen2006rarefied} can be applied to handle the location sampling.

\section{Numerical examples}\label{sec:cases}

In this section, the UGKWP method will be tested in a wide range of multiscale flow problems. The performance of the method will be evaluated quantitatively.

\subsection{Sod test case}
On a two dimensional triangular mesh, the Sod shock tube problem has been computed at different Knudsen numbers to valid the current UGKWP method in the continuum and rarefied flows.
\begin{figure}[H]
	\centering
	\includegraphics[width=0.8\textwidth]{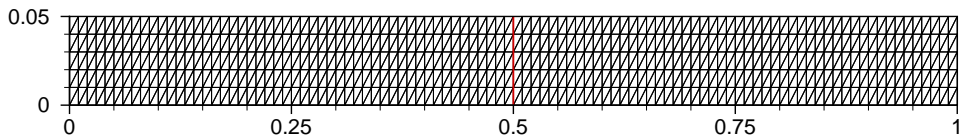}
	\caption{\label{fig:sod_mesh} Unstructured mesh for numerical computations of the Sod shock tube problem.}
\end{figure}
The initial condition is
\begin{equation}
(\rho, U, V, p) =
\begin{cases}
(1,0, 0, 1), & 0 < x < 0.5,\\
(0.125, 0, 0, 0.1) & 0.5 < x < 1.
\end{cases}
\end{equation}
As shown in Fig.~\ref{fig:sod_mesh}, the spatial discretization is carried out by an unstructured mesh with $100\times5\times2$ triangular cells.
For the UGKWP computation, the pre-set reference mass $m_r$ for a simulation particle is $10^{-7}$; while for the UGKS simulation, $100\times100$ velocity points are used to discretize the velocity space.
The top and bottom boundaries are treated as symmetric planes.
The results at the time  $t = 0.12$ for the cases at ${\rm Kn} = 10^{-4}$, $10^{-3}$, $10^{-2}$, $0.1$, $1$ and $10$ in all flow regimes are presented.

\begin{figure}[H]
\subfigure[Density]{\includegraphics[width=0.32\textwidth]{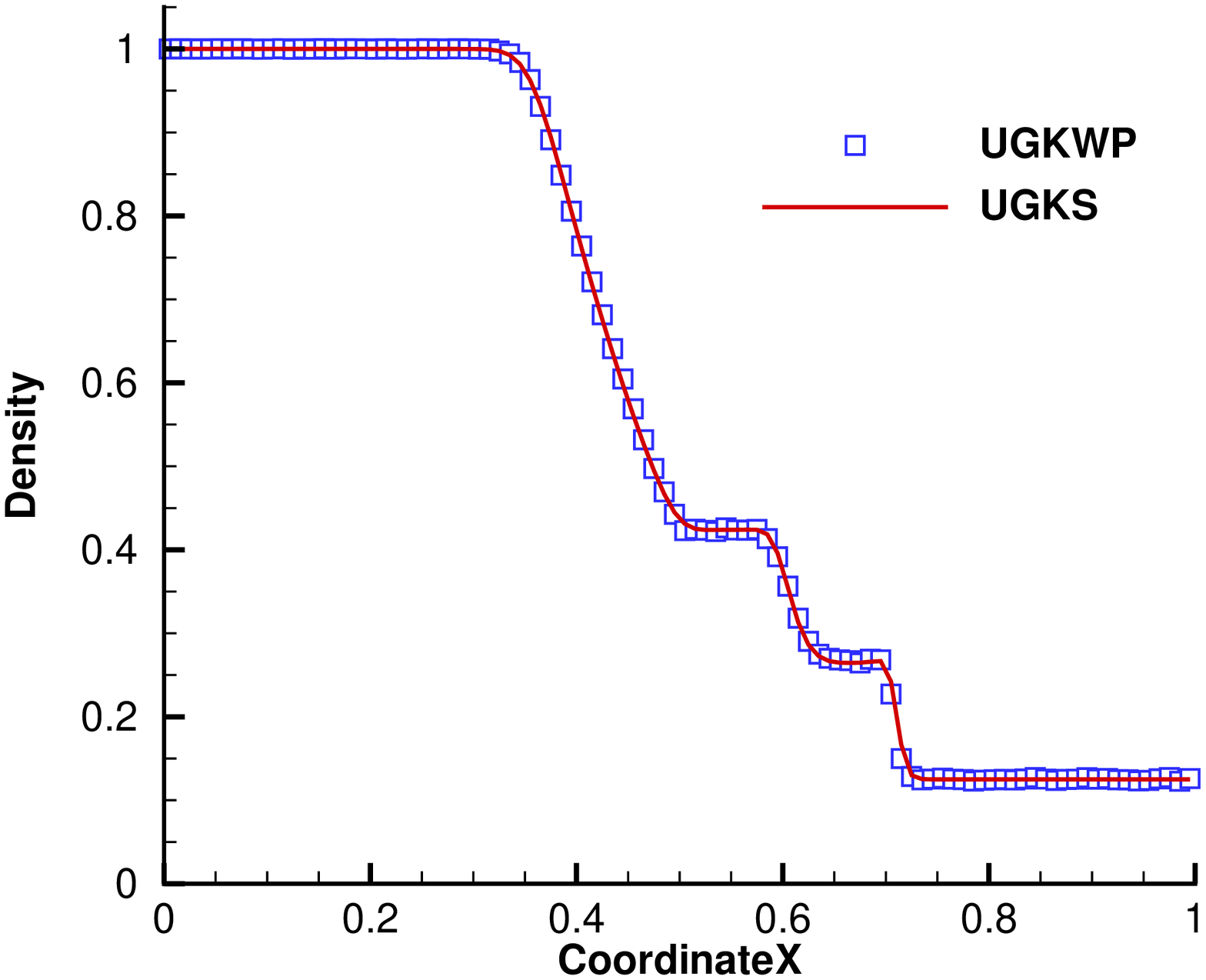}}
\subfigure[Velocity]{\includegraphics[width=0.32\textwidth]{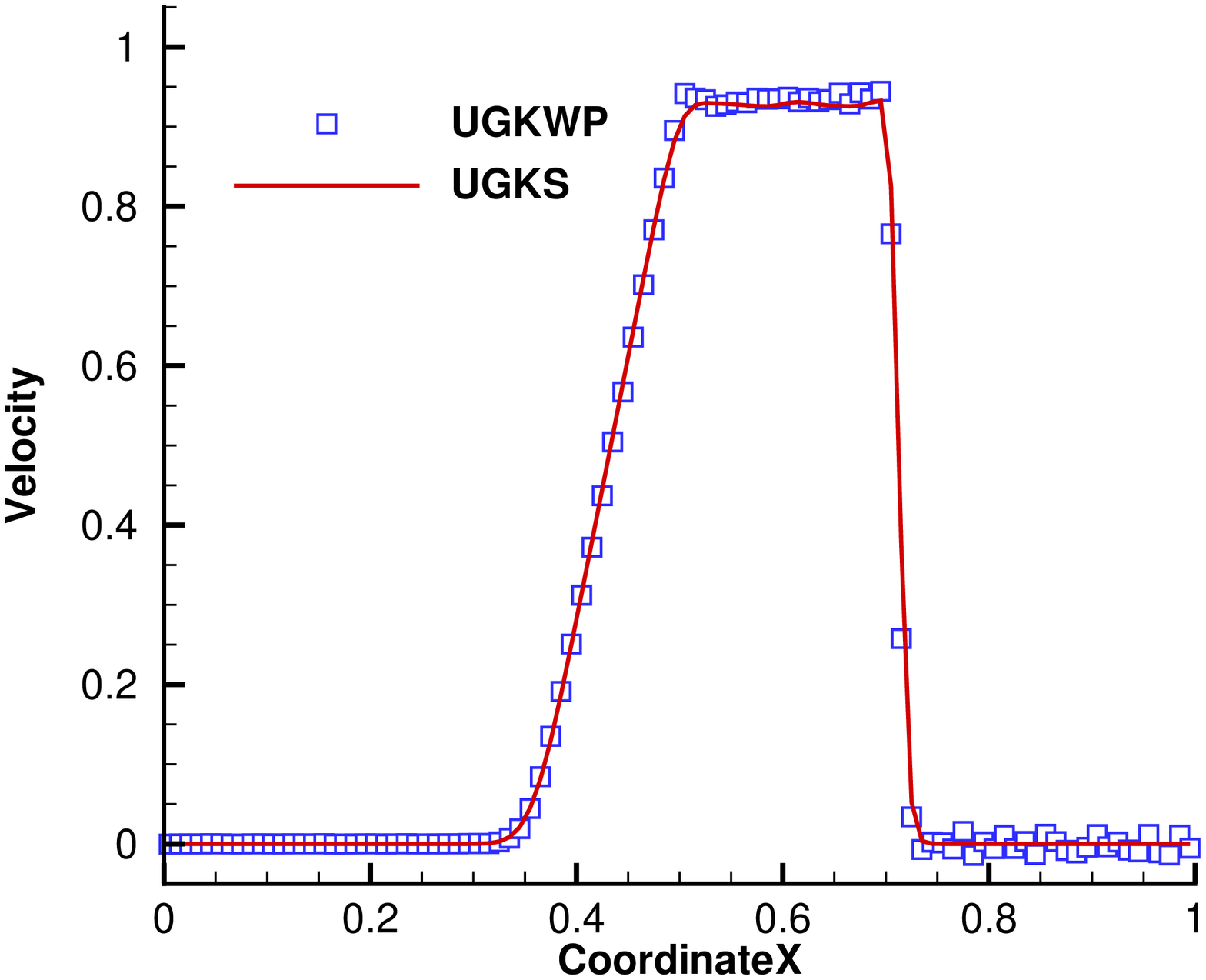}}
\subfigure[Temperature]{\includegraphics[width=0.32\textwidth]{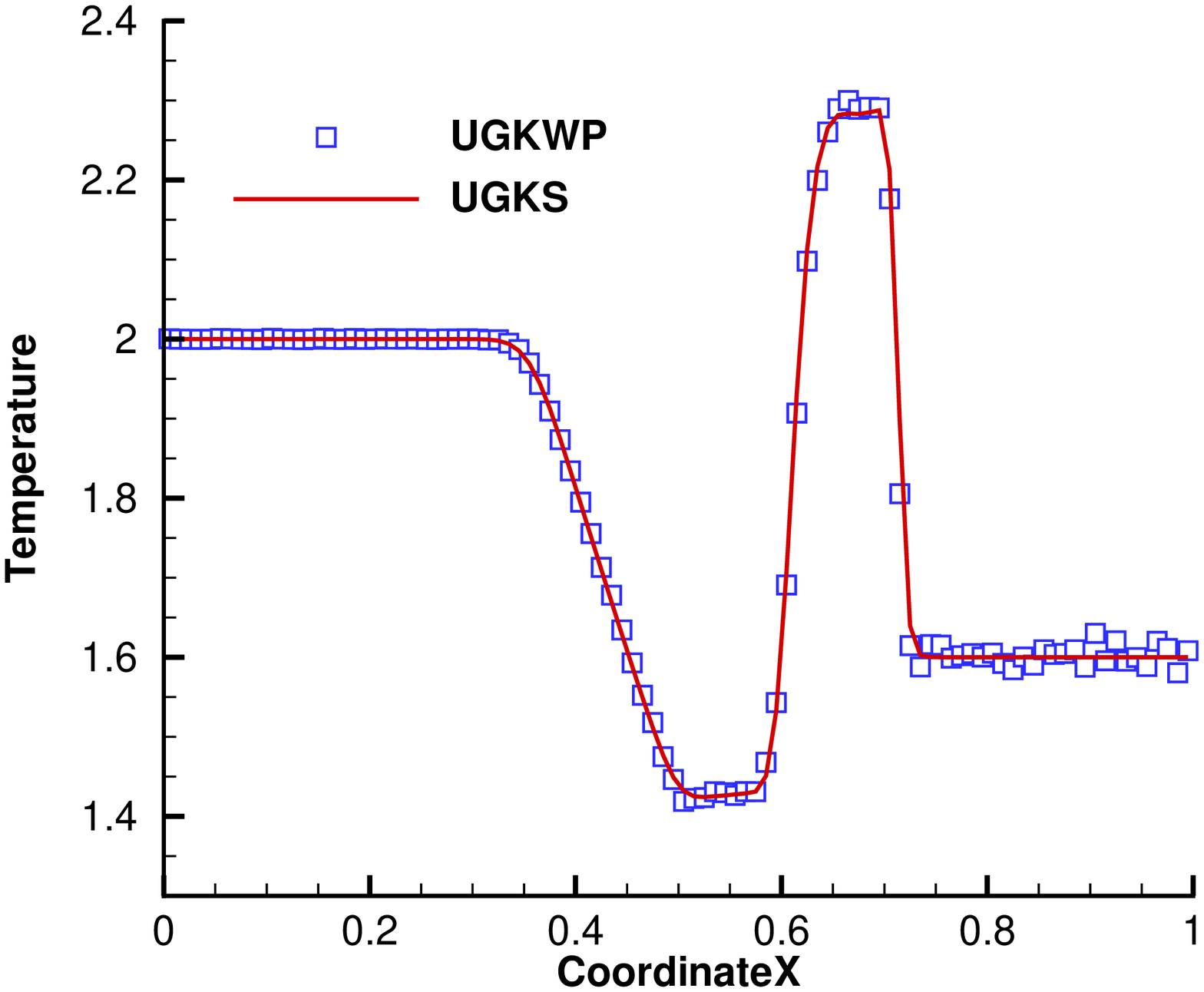}}
\caption{\label{fig:sod_kn-4}Sod test cases at ${\rm Kn} = 10^{-4}$.}
\end{figure}
\begin{figure}[H]
\subfigure[Density]{\includegraphics[width=0.32\textwidth]{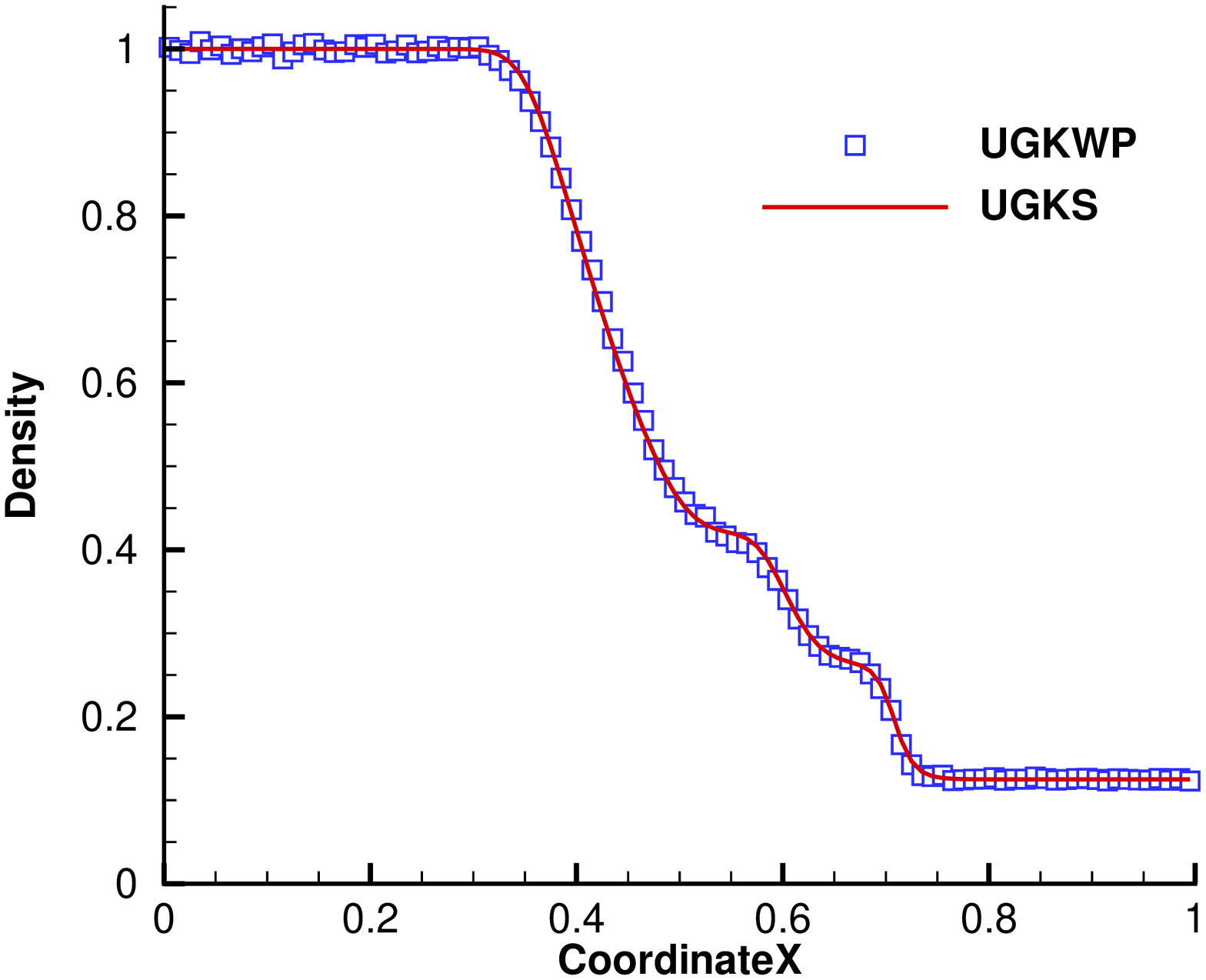}}
\subfigure[Velocity]{\includegraphics[width=0.32\textwidth]{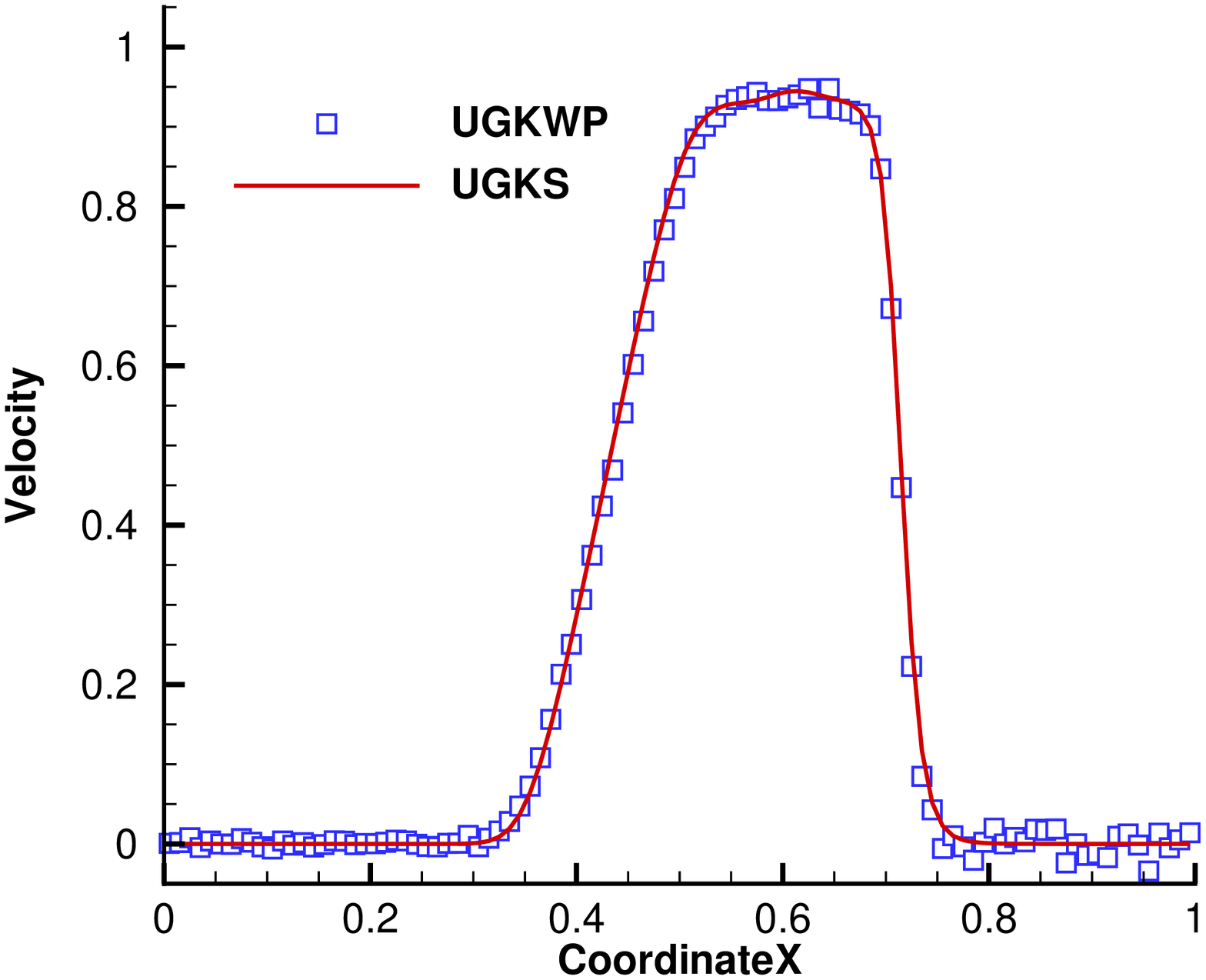}}
\subfigure[Temperature]{\includegraphics[width=0.32\textwidth]{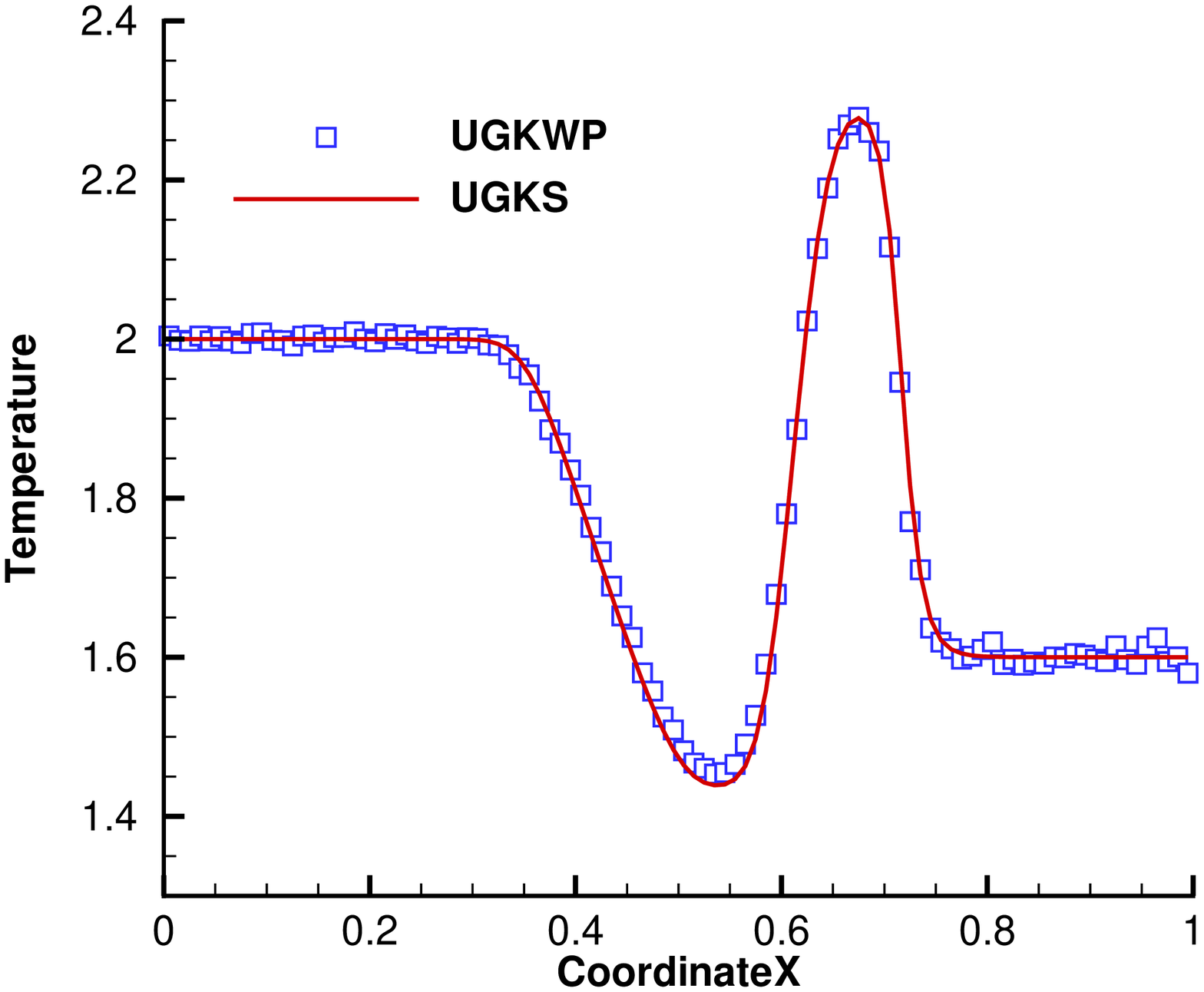}}
\caption{\label{fig:sod_kn-3}Sod test cases at ${\rm Kn} = 10^{-3}$.}
\end{figure}
\begin{figure}[H]
\subfigure[Density]{\includegraphics[width=0.32\textwidth]{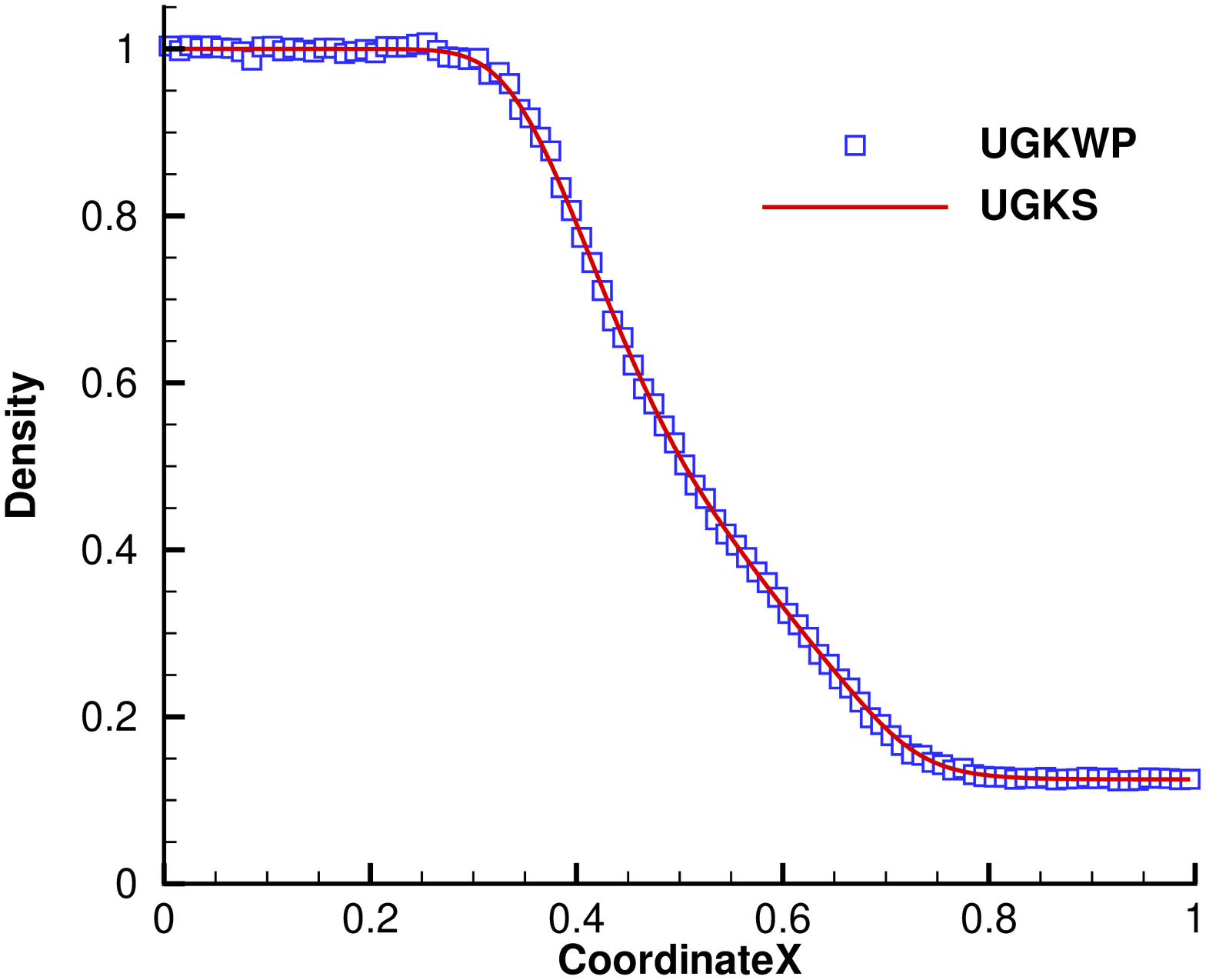}}
\subfigure[Velocity]{\includegraphics[width=0.32\textwidth]{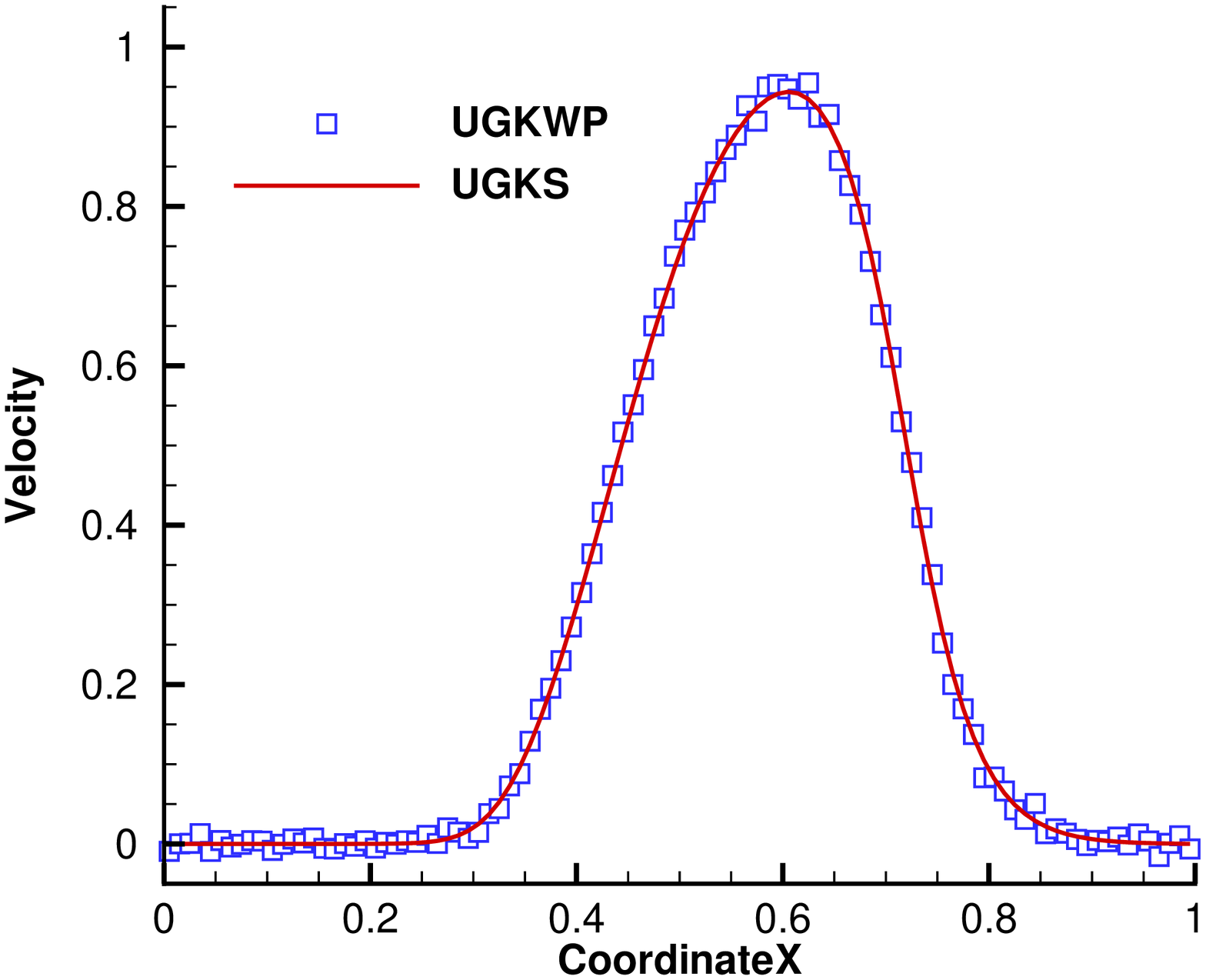}}
\subfigure[Temperature]{\includegraphics[width=0.32\textwidth]{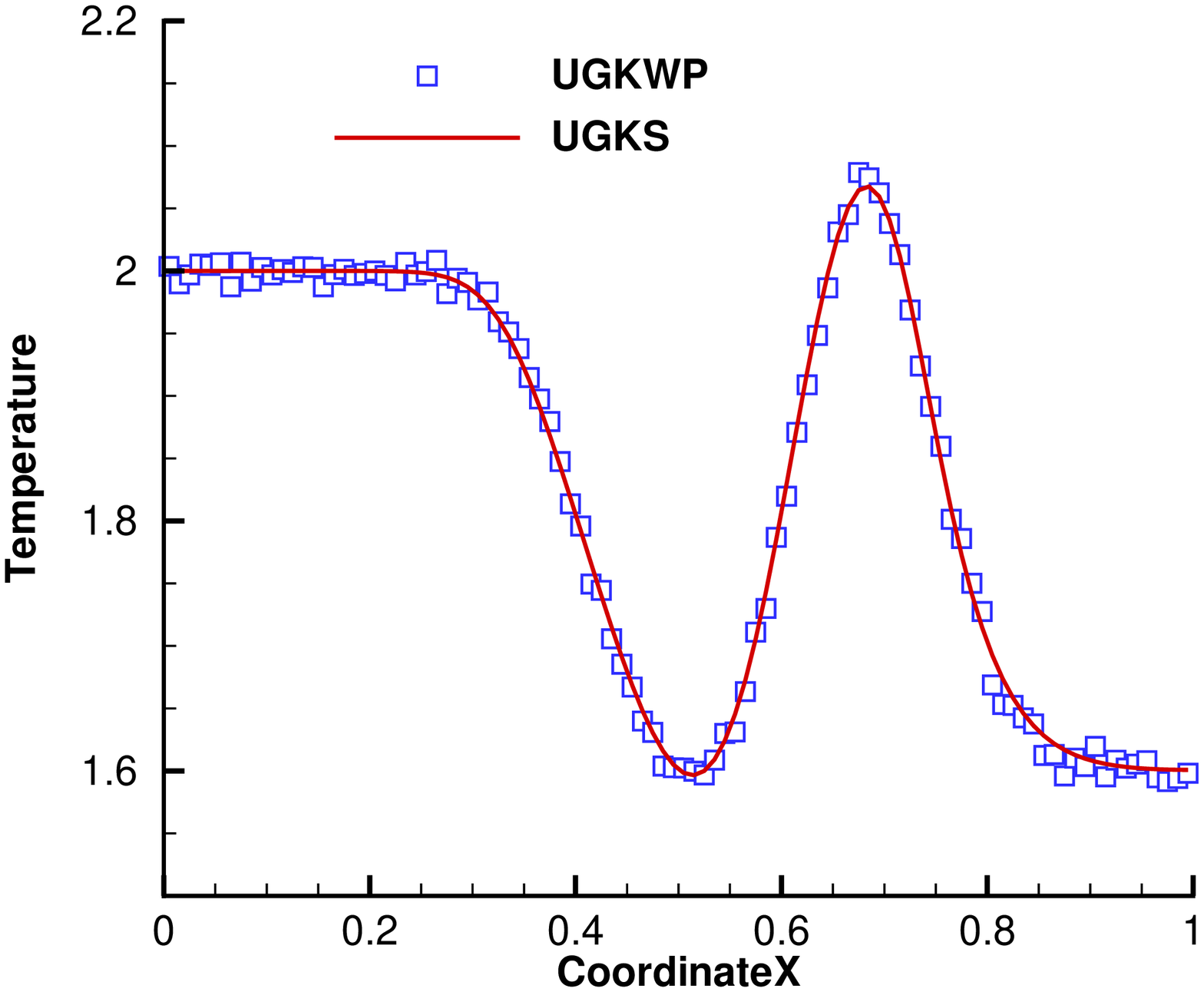}}
\caption{\label{fig:sod_kn-2}Sod test cases at ${\rm Kn} = 10^{-2}$.}
\end{figure}
\begin{figure}[H]
\subfigure[Density]{\includegraphics[width=0.32\textwidth]{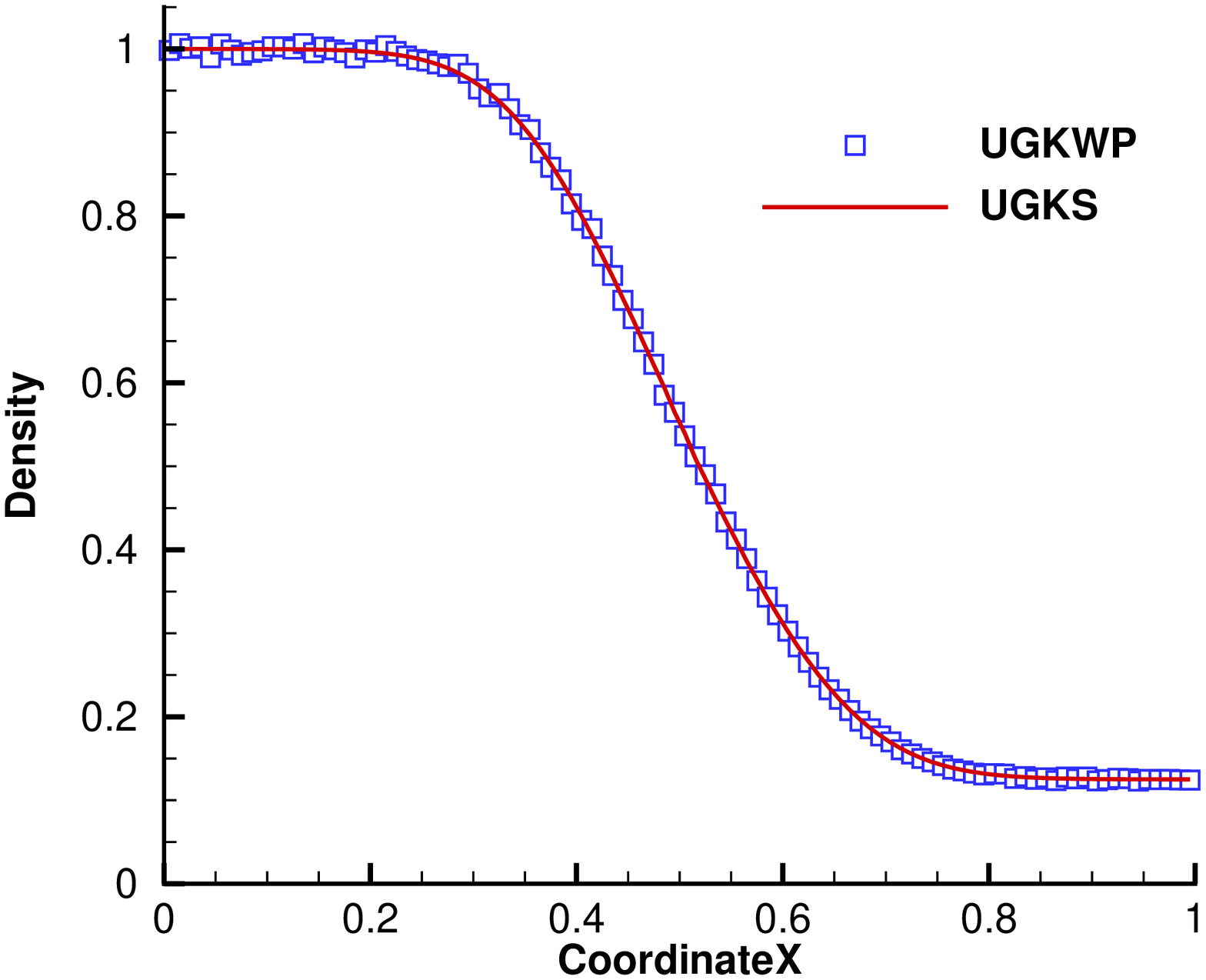}}
\subfigure[Velocity]{\includegraphics[width=0.32\textwidth]{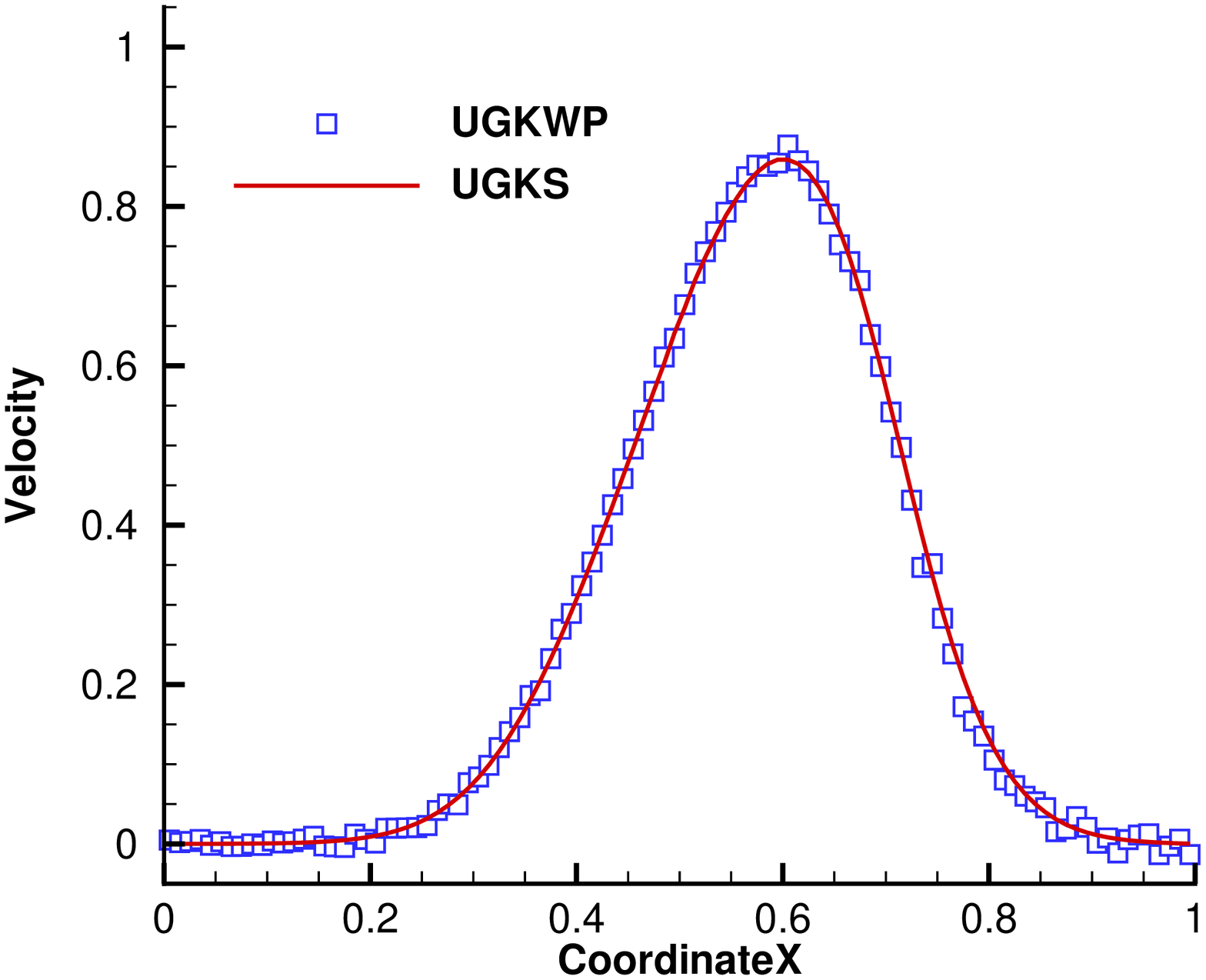}}
\subfigure[Temperature]{\includegraphics[width=0.32\textwidth]{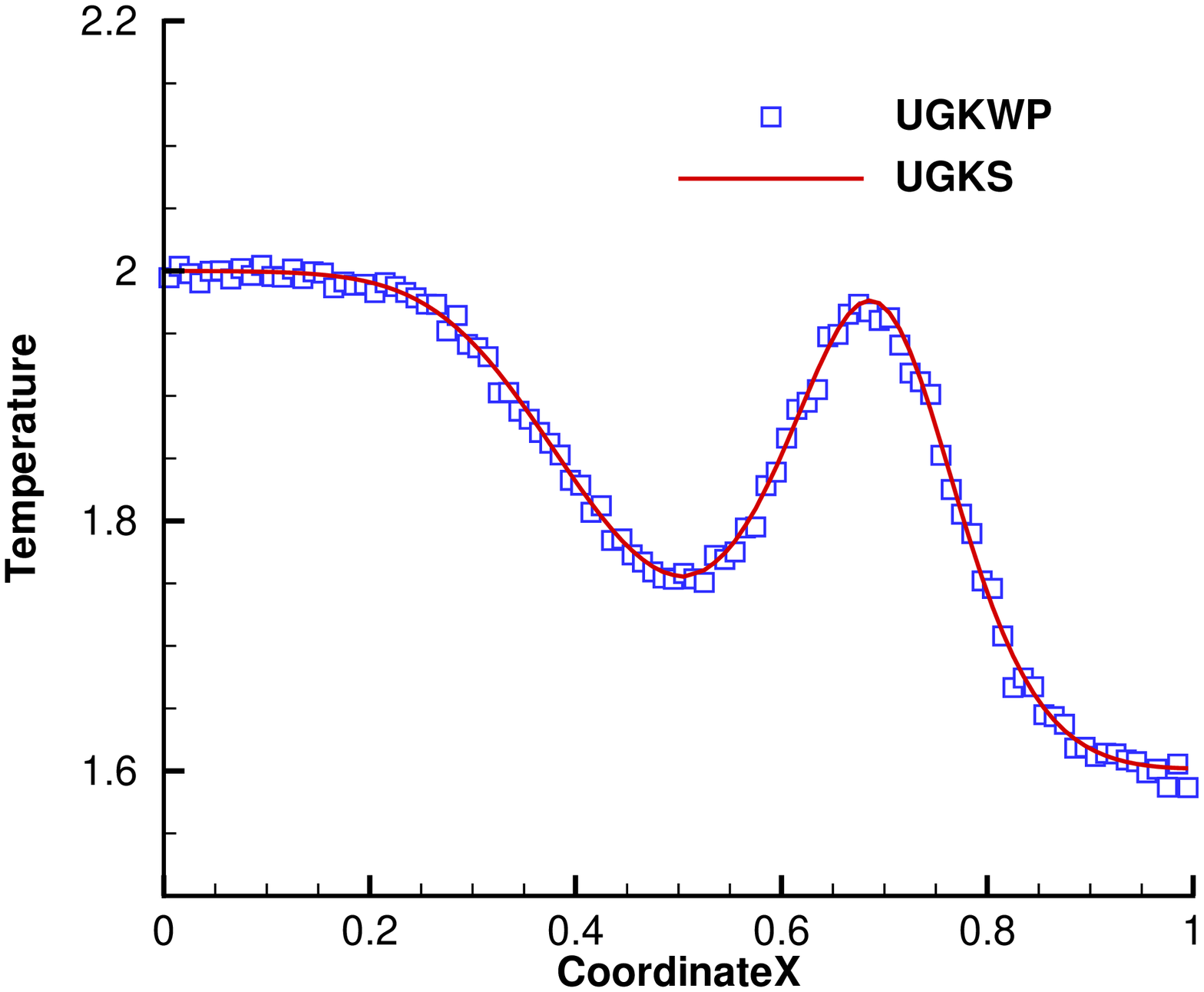}}
\caption{\label{fig:sod_kn-1}Sod test cases at ${\rm Kn} = 0.1$.}
\end{figure}
\begin{figure}[H]
\subfigure[Density]{\includegraphics[width=0.32\textwidth]{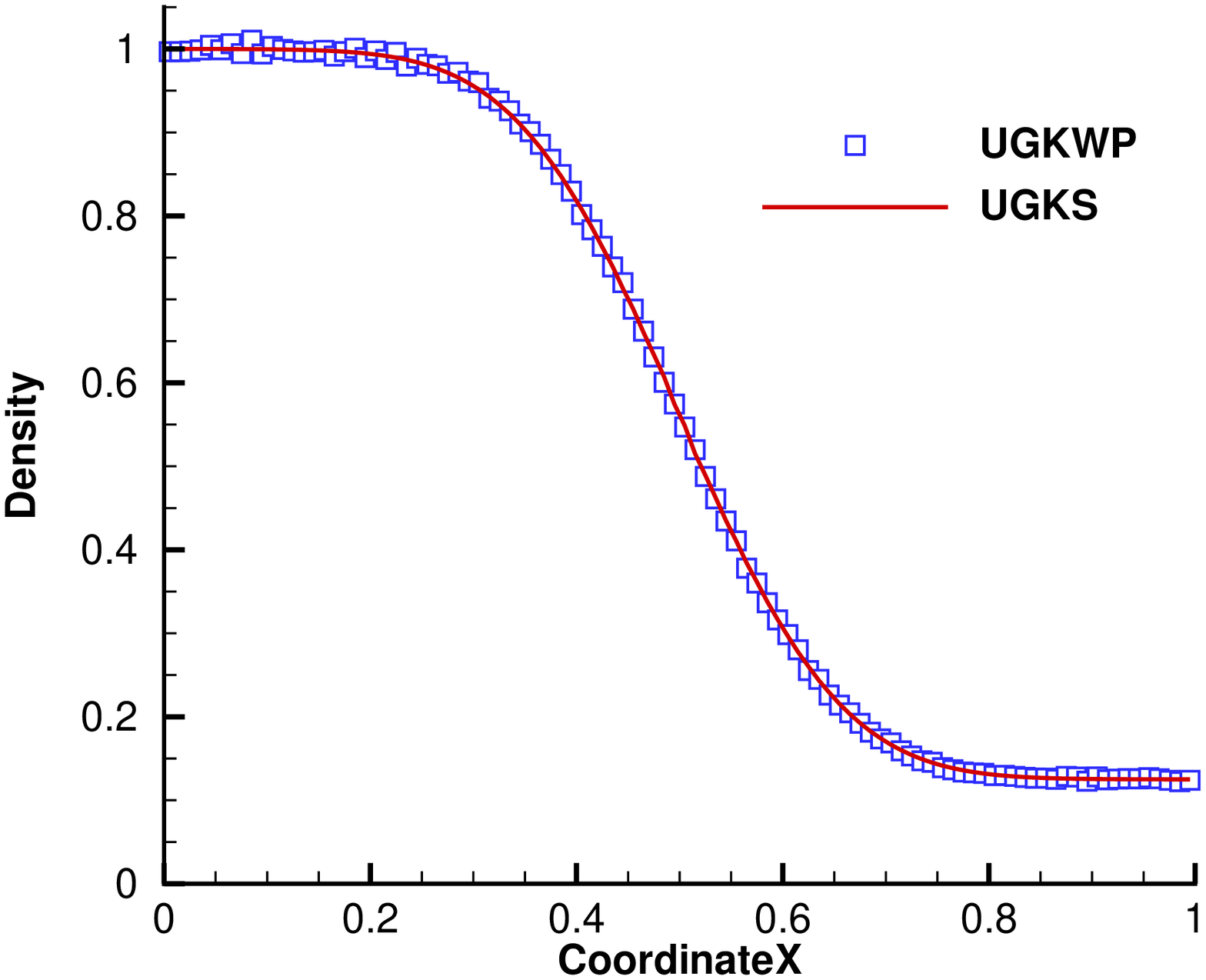}}
\subfigure[Velocity]{\includegraphics[width=0.32\textwidth]{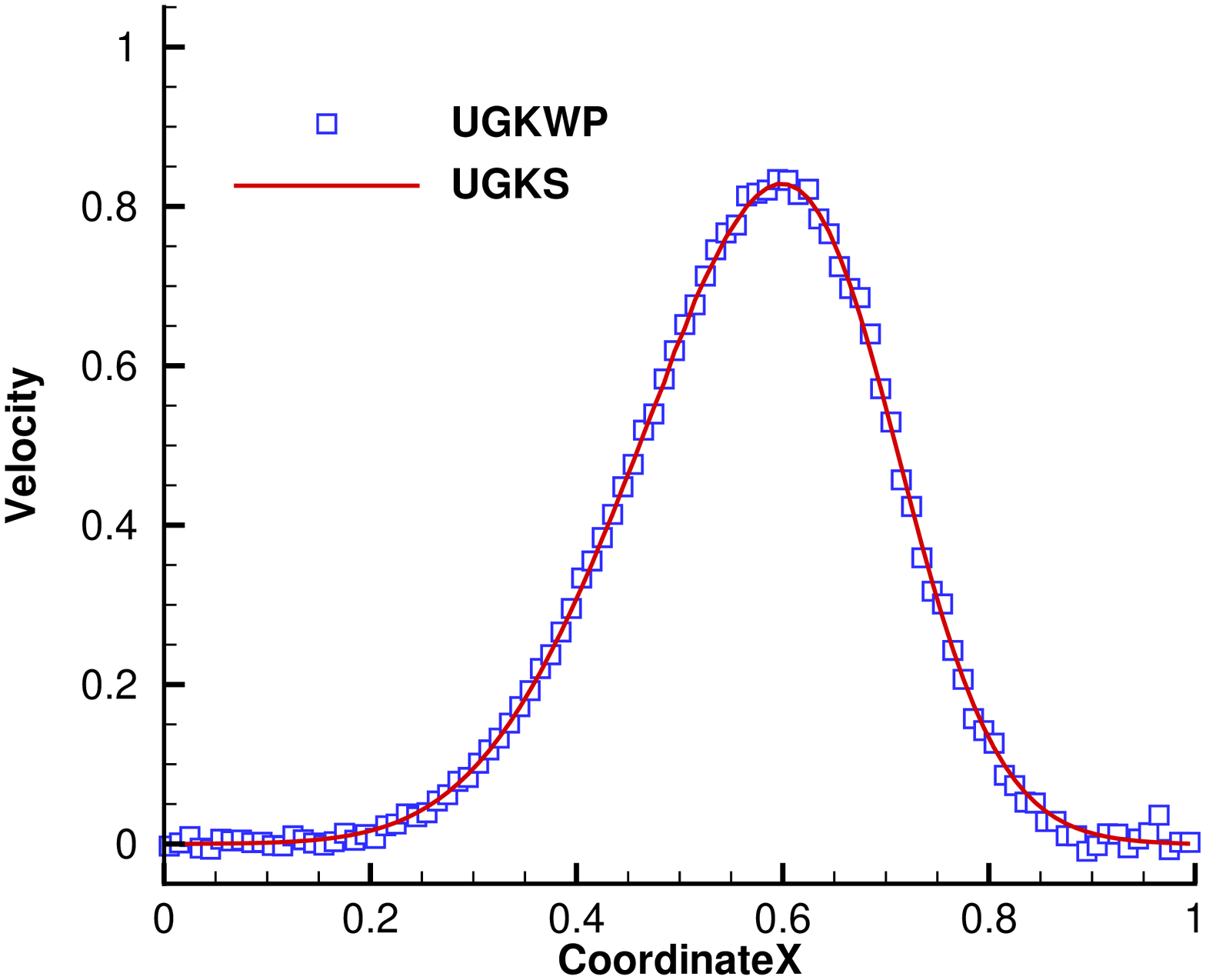}}
\subfigure[Temperature]{\includegraphics[width=0.32\textwidth]{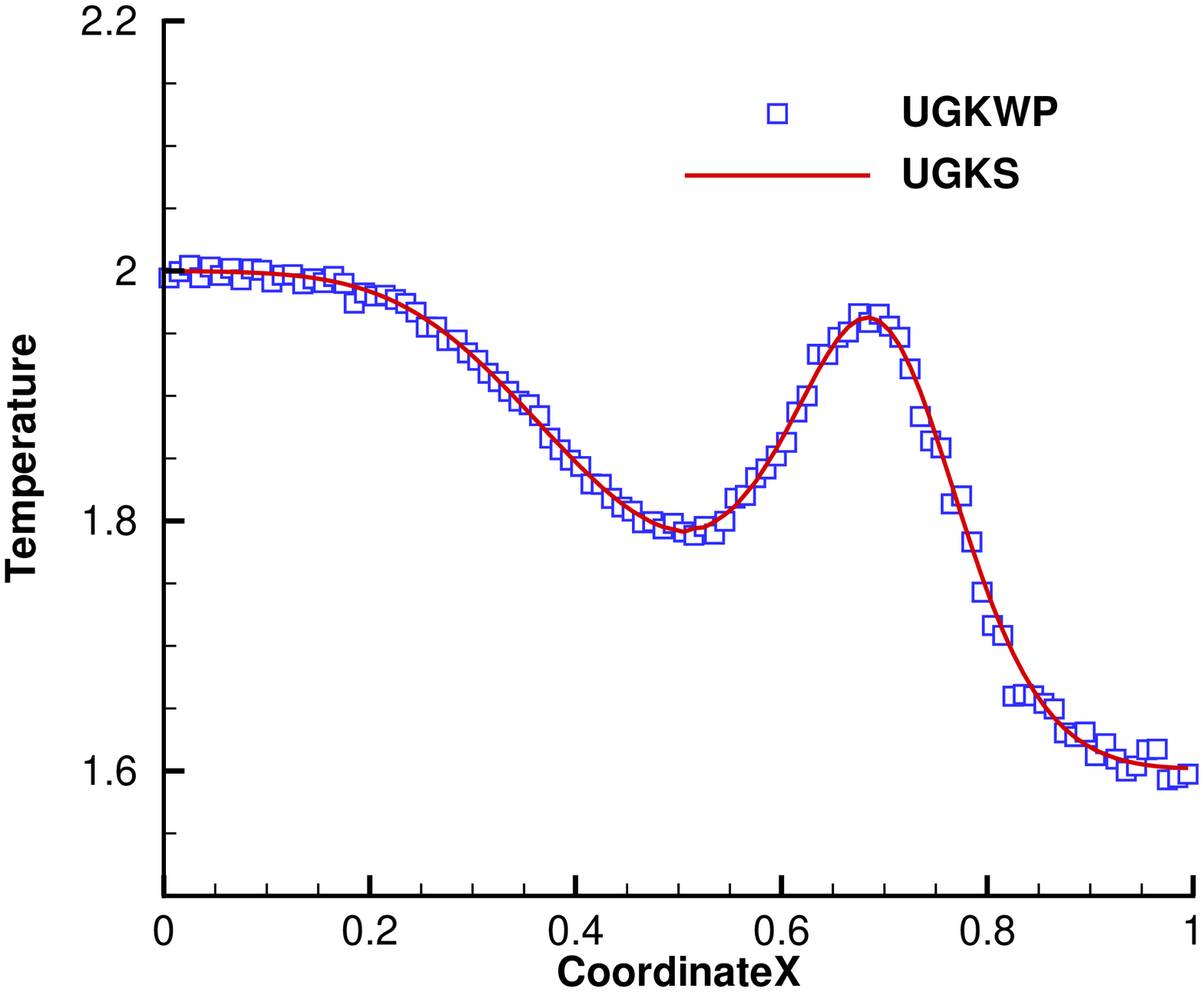}}
\caption{\label{fig:sod_kn-0}Sod test cases at ${\rm Kn} = 1$. }
\end{figure}
\begin{figure}[H]
\subfigure[Density]{\includegraphics[width=0.32\textwidth]{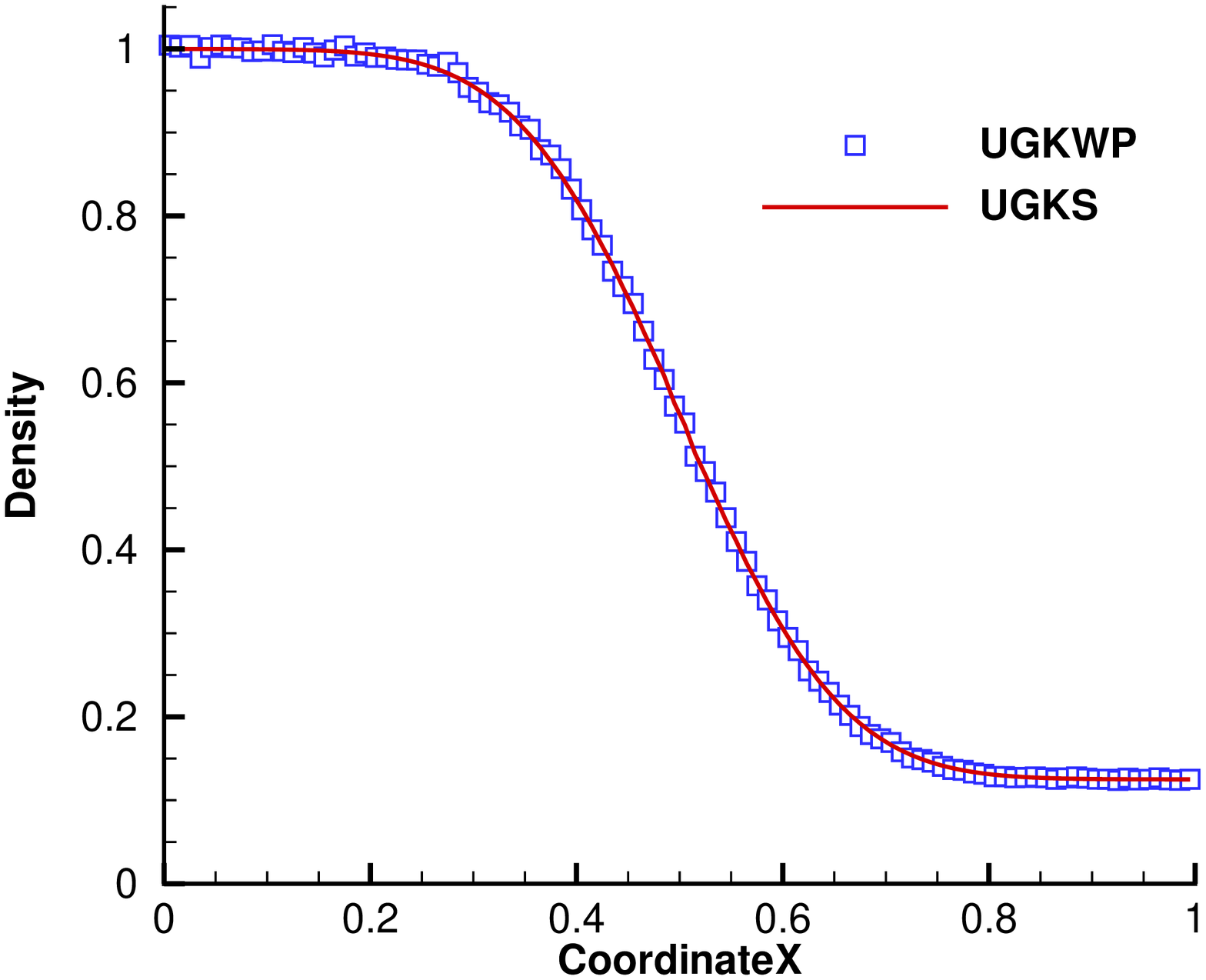}}
\subfigure[Velocity]{\includegraphics[width=0.32\textwidth]{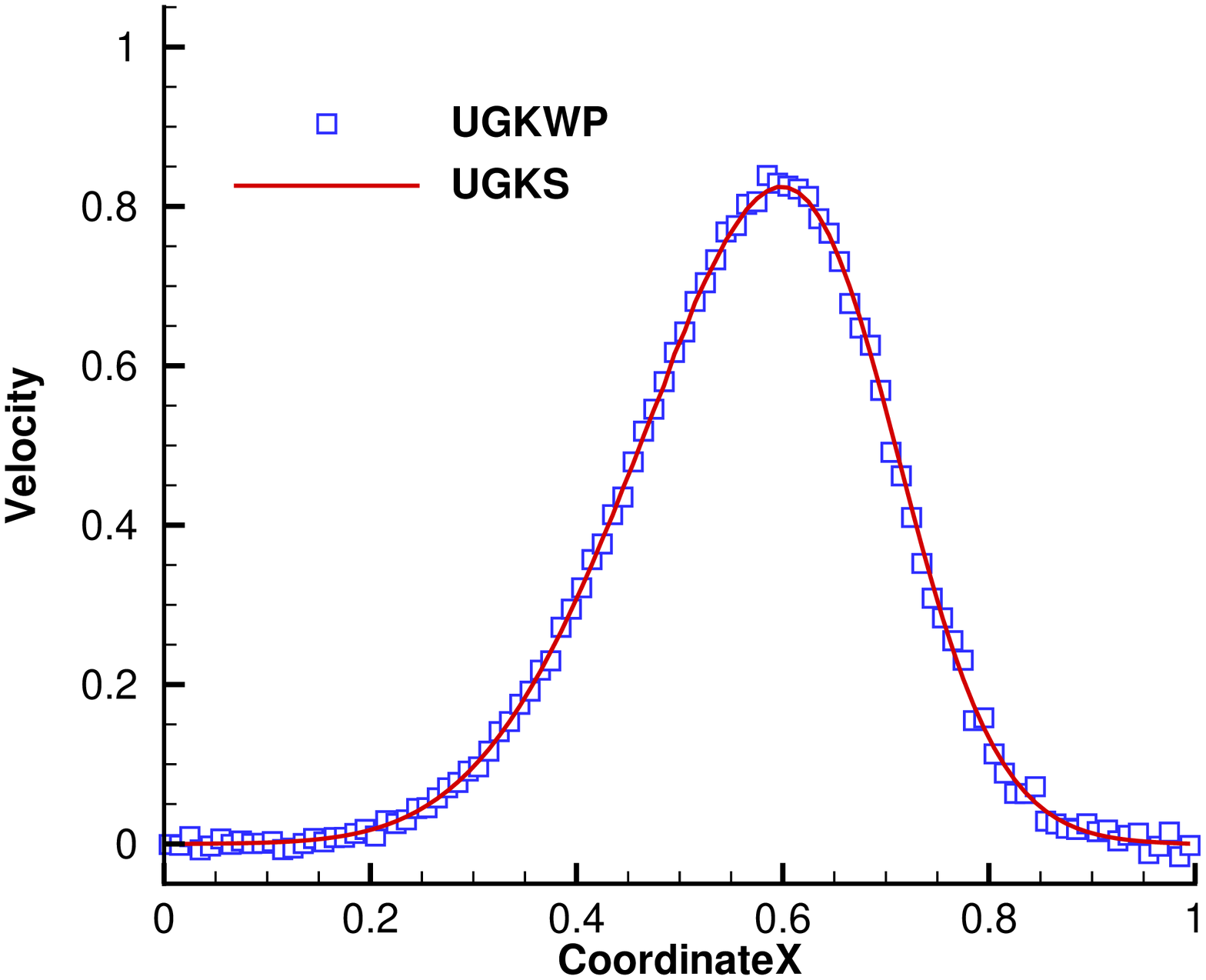}}
\subfigure[Temperature]{\includegraphics[width=0.32\textwidth]{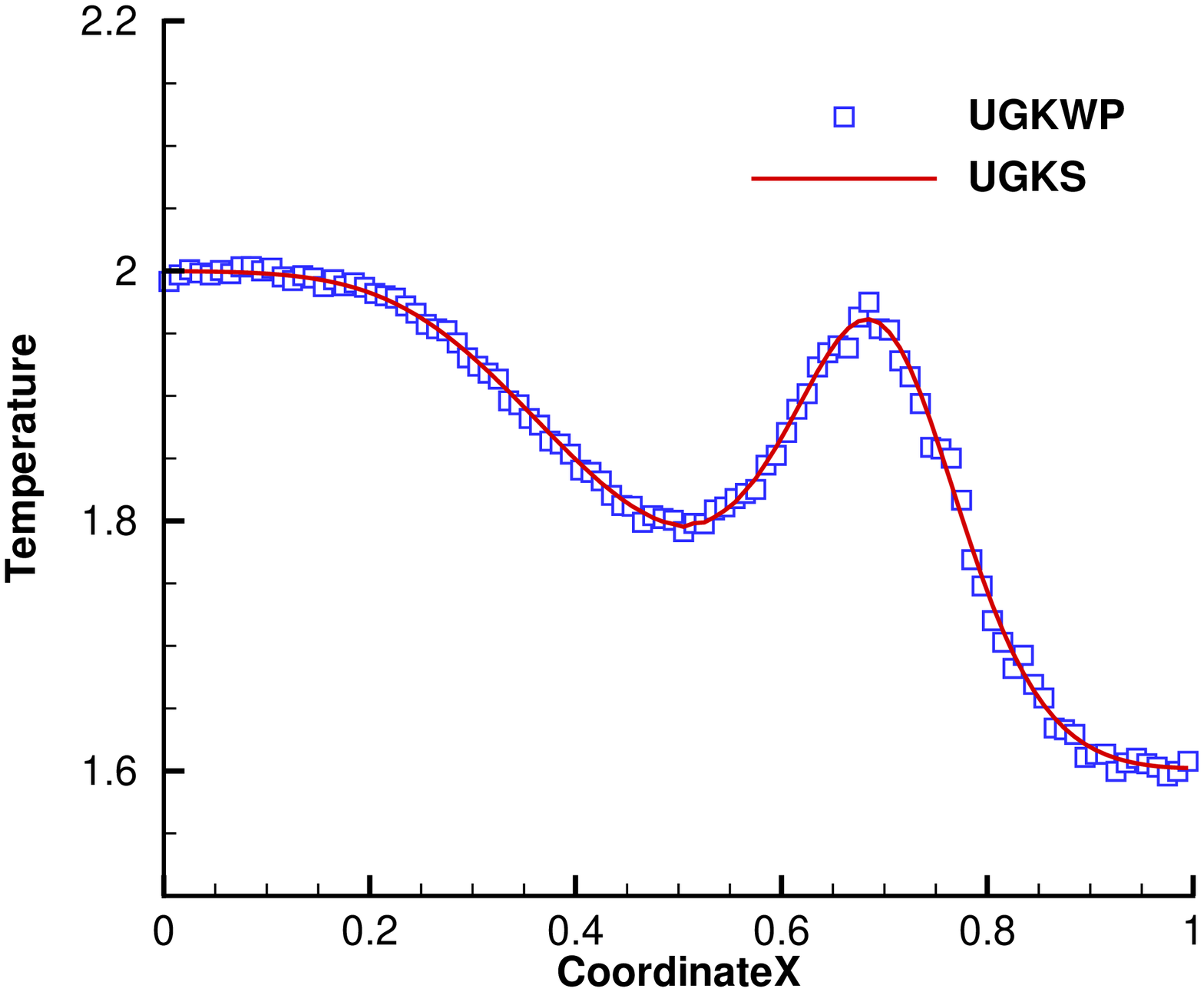}}
\caption{\label{fig:sod_kn10}Sod test cases at ${\rm Kn} = 10$.}
\end{figure}

The density, velocity and temperature obtained by the UGKS and the UGKWP method for different Knudsen number cases are plotted in Figs.~\ref{fig:sod_kn-4}--\ref{fig:sod_kn10}, where the two dimensional flow field is projected to one dimensional data in the $x$ direction by taking average over the ten triangular cells along $y$ direction.
In addition, in order to reduce the statistical noises, the unsteady flow solutions of the UGKWP method are averaged over $10$ times of computations.
It can be seen that for all the cases in different flow regimes the UGKWP solutions agree well with the UGKS data.
The capability of the UGKWP method for numerical simulations in continuum and rarefied flows are validated.

For the UGKS, once the discretization for the physical space and the velocity space is given, the computational costs for all Knudsen number cases will be same due to its unified treatment.
The memory requirement and computational time in the UGKS simulations are $1.1$ GB and $15$ minutes.
While for the UGKWP method, the overall CPU time of $10$ times of computations is about $65$ seconds for the cases with larger Knudsen numbers, and the memory cost is around $55$ MB.
Moreover, for the case at ${\rm Kn} = 10^{-4}$ in continuum flow, since the portion of hydrodynamic waves increases and much fewer discrete particles are needed to be sampled and tracked, the computational cost of the UGKWP method gets lower to $12$ seconds with $11$ MB.
Generally speaking, in comparison with the UGKS, order-of-magnitude in efficiency increment and memory reduction can be achieved by the UGKWP method for the two dimensional Sod shock tube problem in the continuum and rarefied flows.

\subsection{Cavity flow}
The low-speed micro flow in a lid-driven cavity is computed at Knudsen numbers $0.1$, $1$ and $10$.
The Knudsen number is defined as the ratio of molecular mean free path to the length of side wall.
The argon gas with molecular mass $m_0 = 6.63\times 10^{-26} {\rm kg}$ is studied and the variable hard sphere model (VHS) is used for all three cases.
The lid velocity is set to $50 {\rm m/s}$.
Isothermal boundary condition is applied with a fixed temperature $T_w = 273 {\rm K}$.
The dynamic viscosity is computed by $\mu = \mu_0 (T/T_0)^{0.81}$.

\begin{figure}[H]
\centering
\subfigure[Velocity X]{\includegraphics[width=0.32\textwidth]{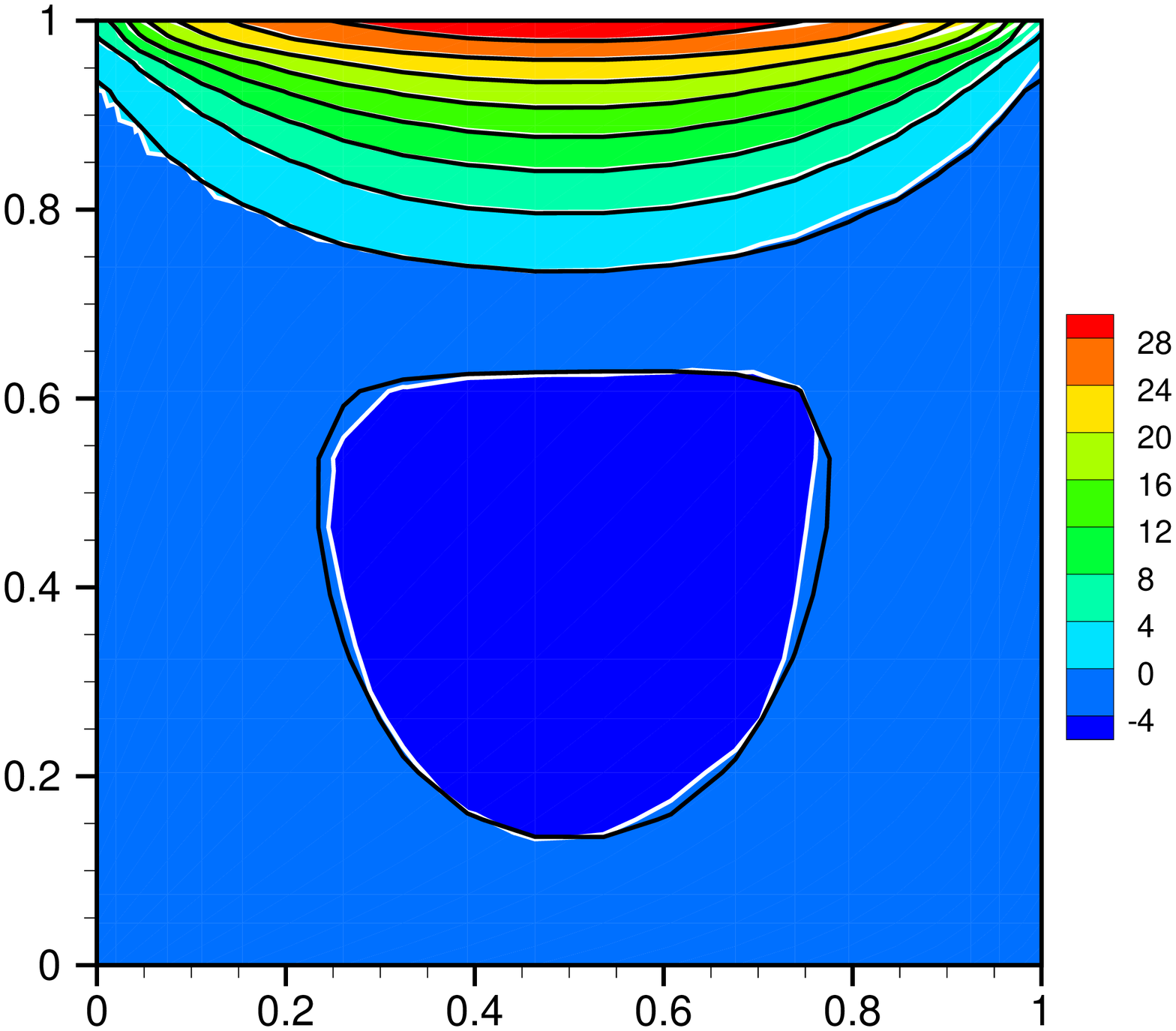}}
\subfigure[Velocity Y]{\includegraphics[width=0.32\textwidth]{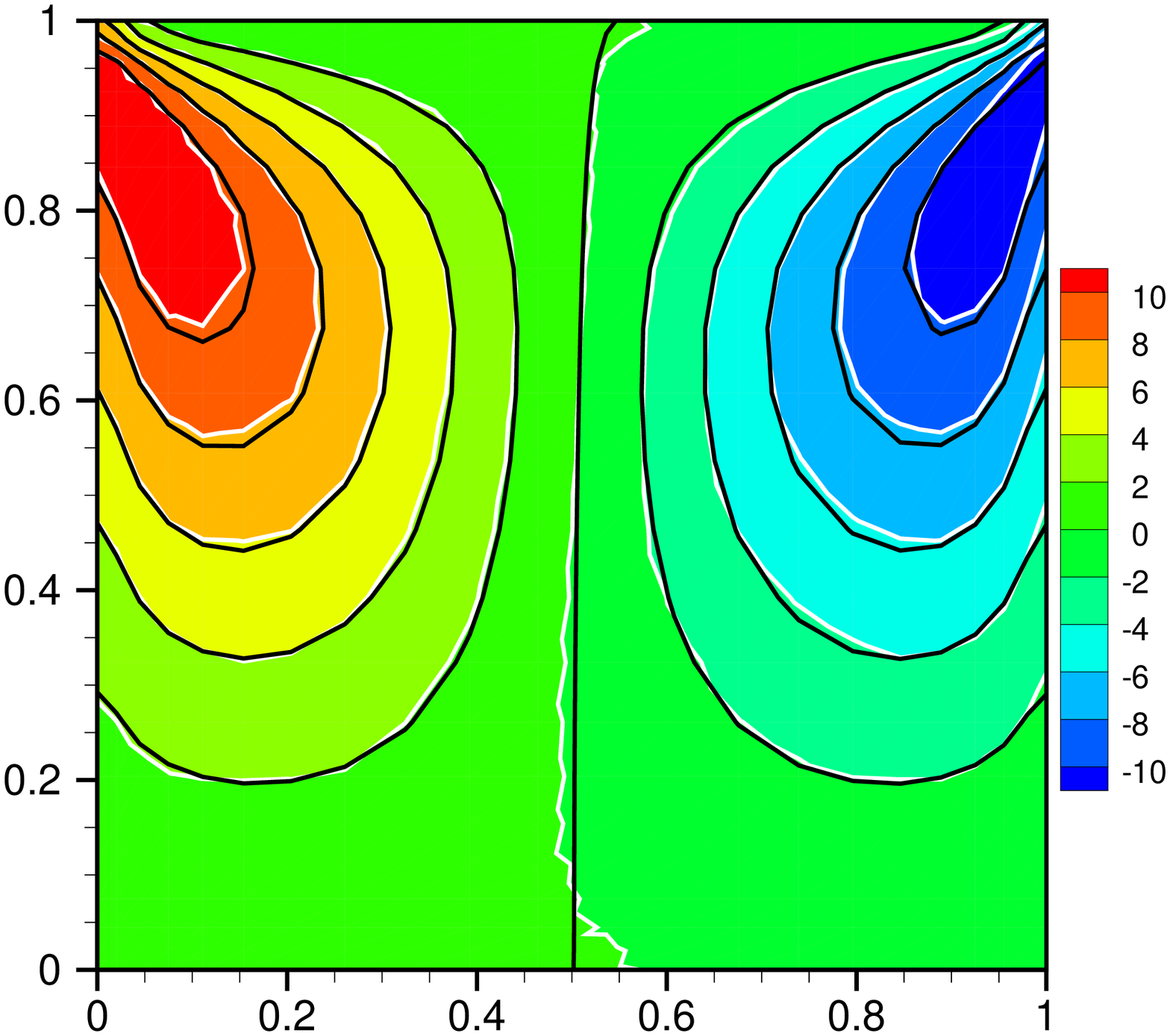}}
\subfigure[Temperature]{\includegraphics[width=0.32\textwidth]{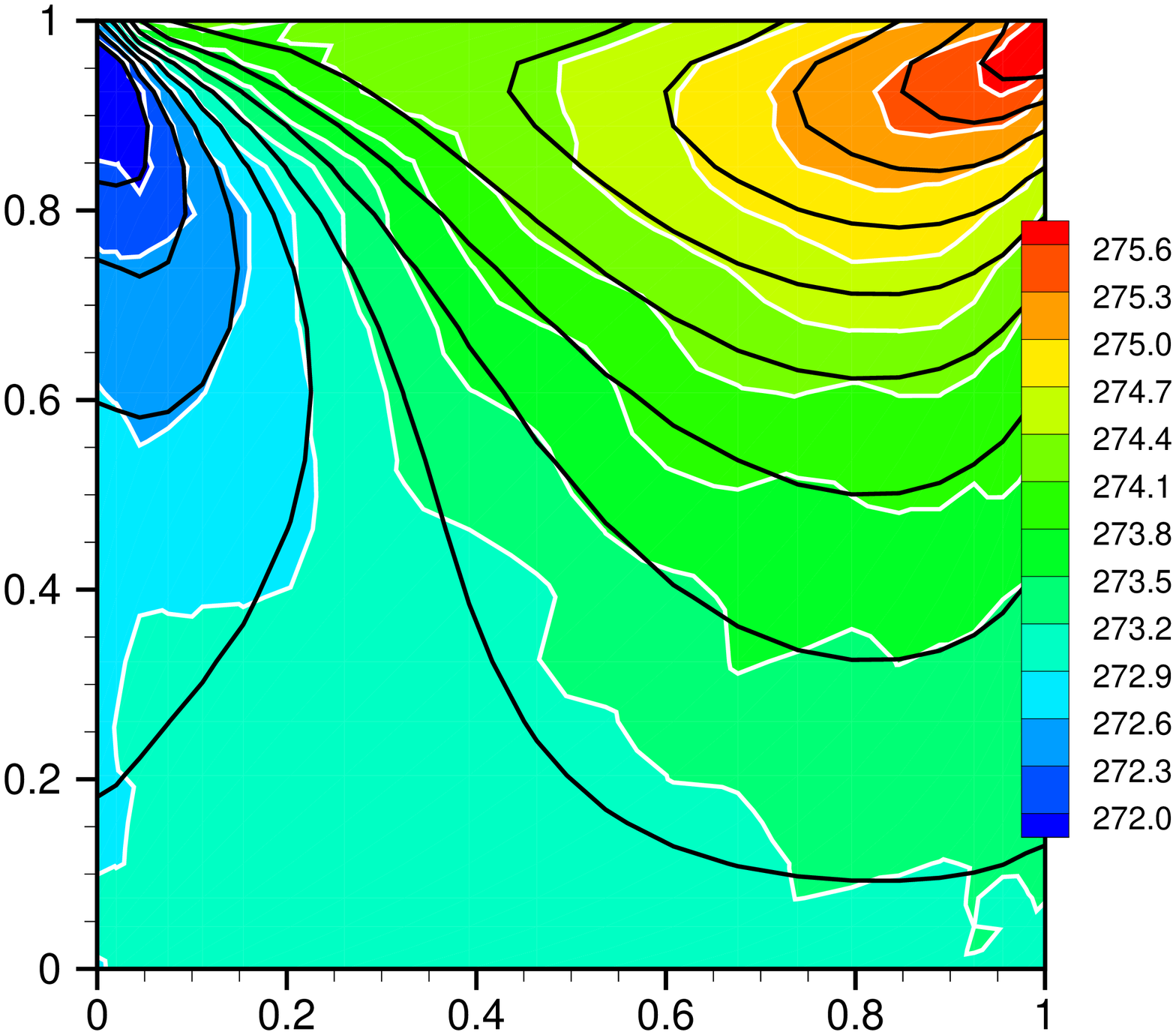}}
\caption{\label{fig:cavity_kn01}Cavity flow at ${\rm Kn} = 0.1$. The background with white lines denotes the UGKWP results and the solid lines are UGKS solutions.}
\end{figure}
\begin{figure}[H]
	\centering
	\subfigure[Velocity X]{\includegraphics[width=0.32\textwidth]{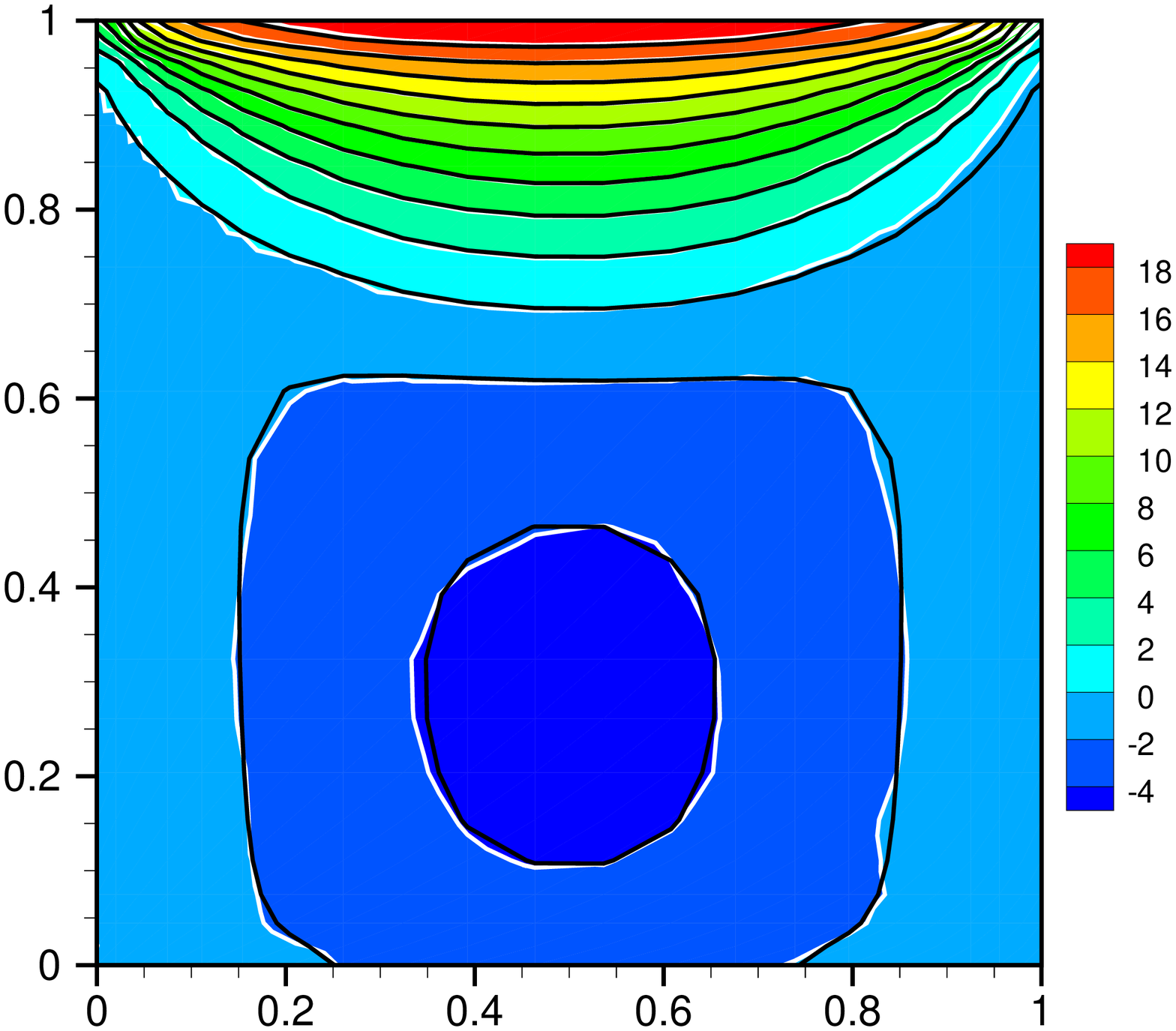}}
	\subfigure[Velocity Y]{\includegraphics[width=0.32\textwidth]{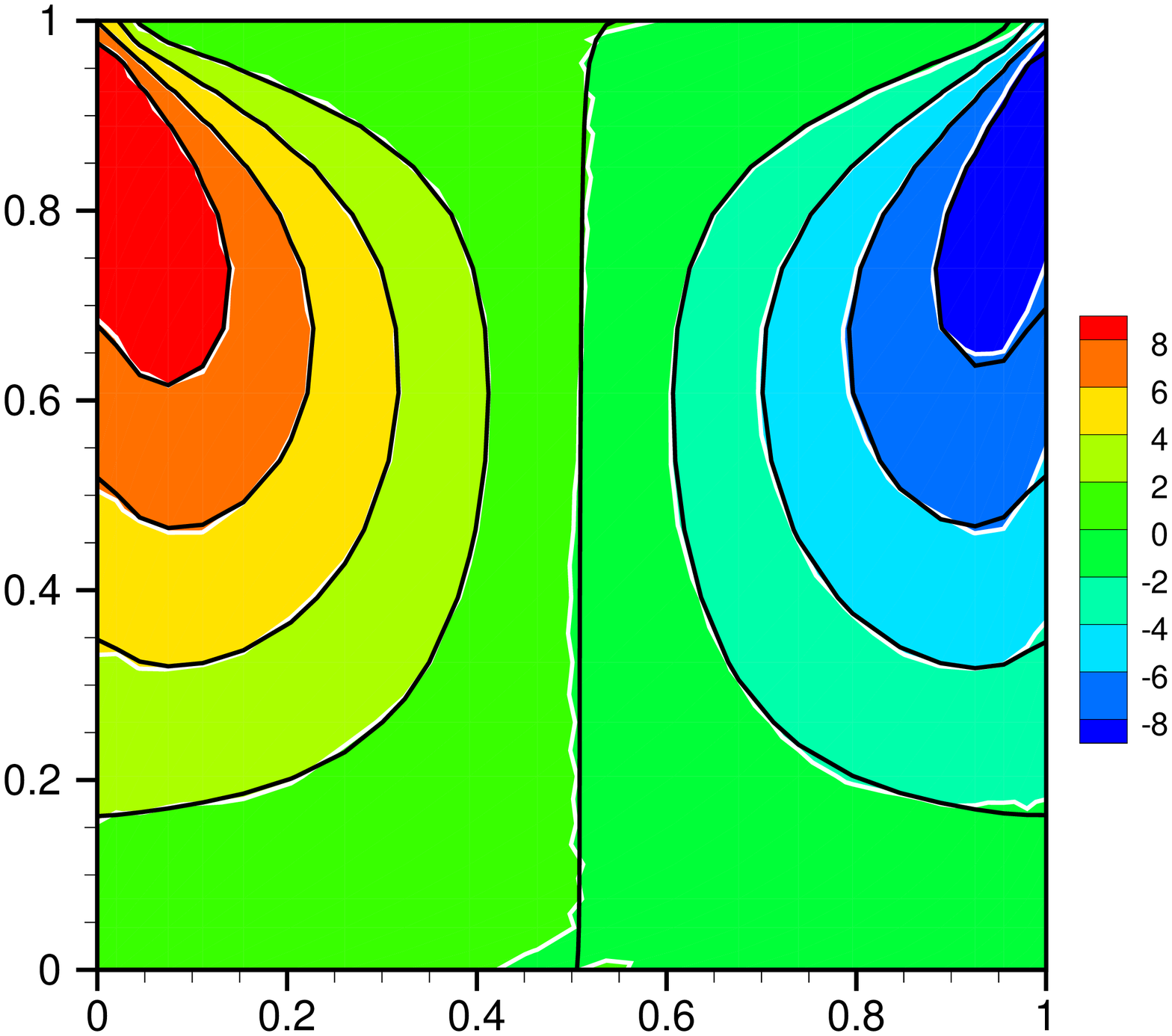}}
	\subfigure[Temperature]{\includegraphics[width=0.32\textwidth]{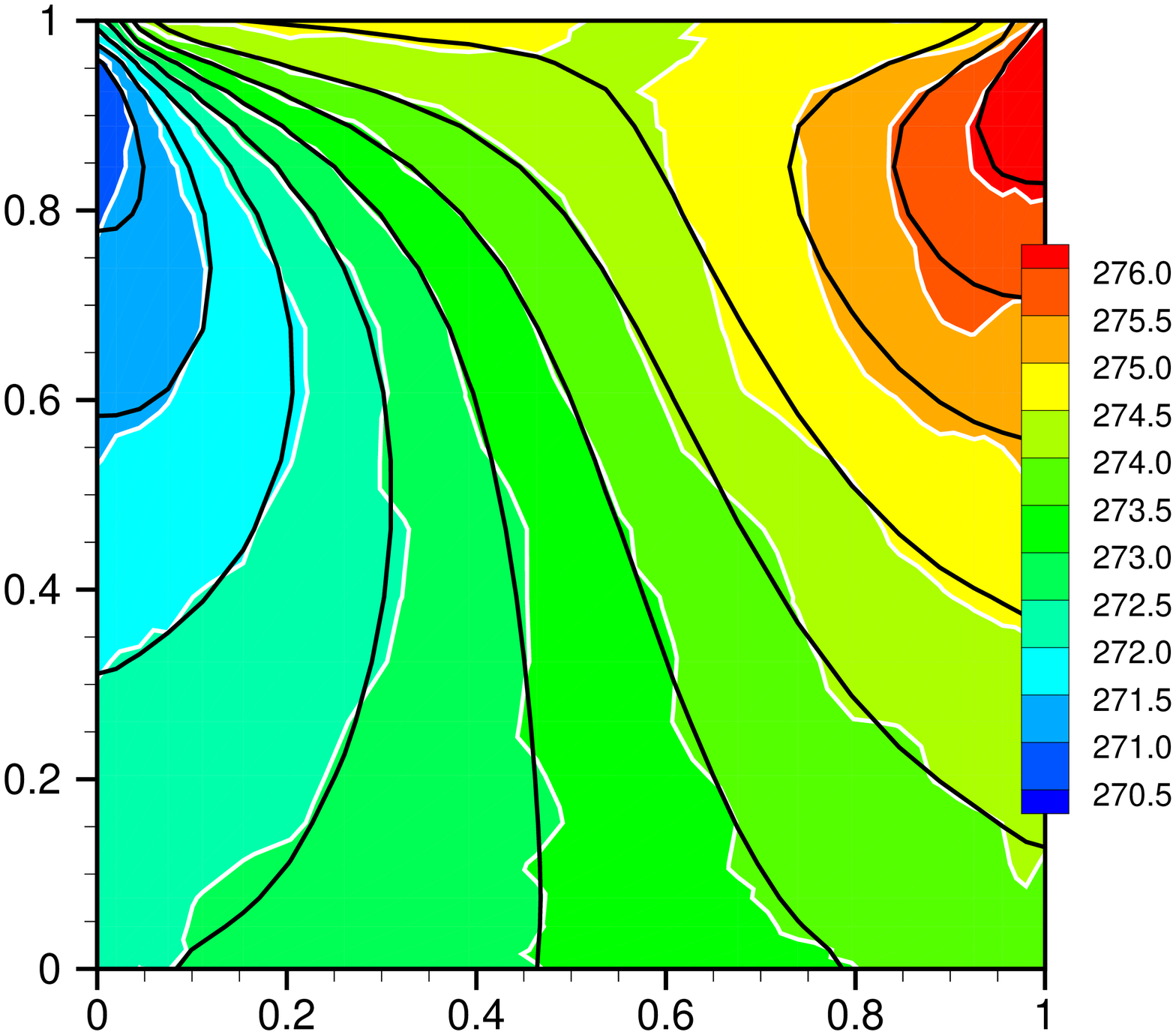}}
	\caption{\label{fig:cavity_kn1}Cavity flow at ${\rm Kn} = 1$. The background with white lines denotes the UGKWP results and the solid lines are UGKS solutions.}
\end{figure}
\begin{figure}[H]
	\centering
	\subfigure[Velocity X]{\includegraphics[width=0.32\textwidth]{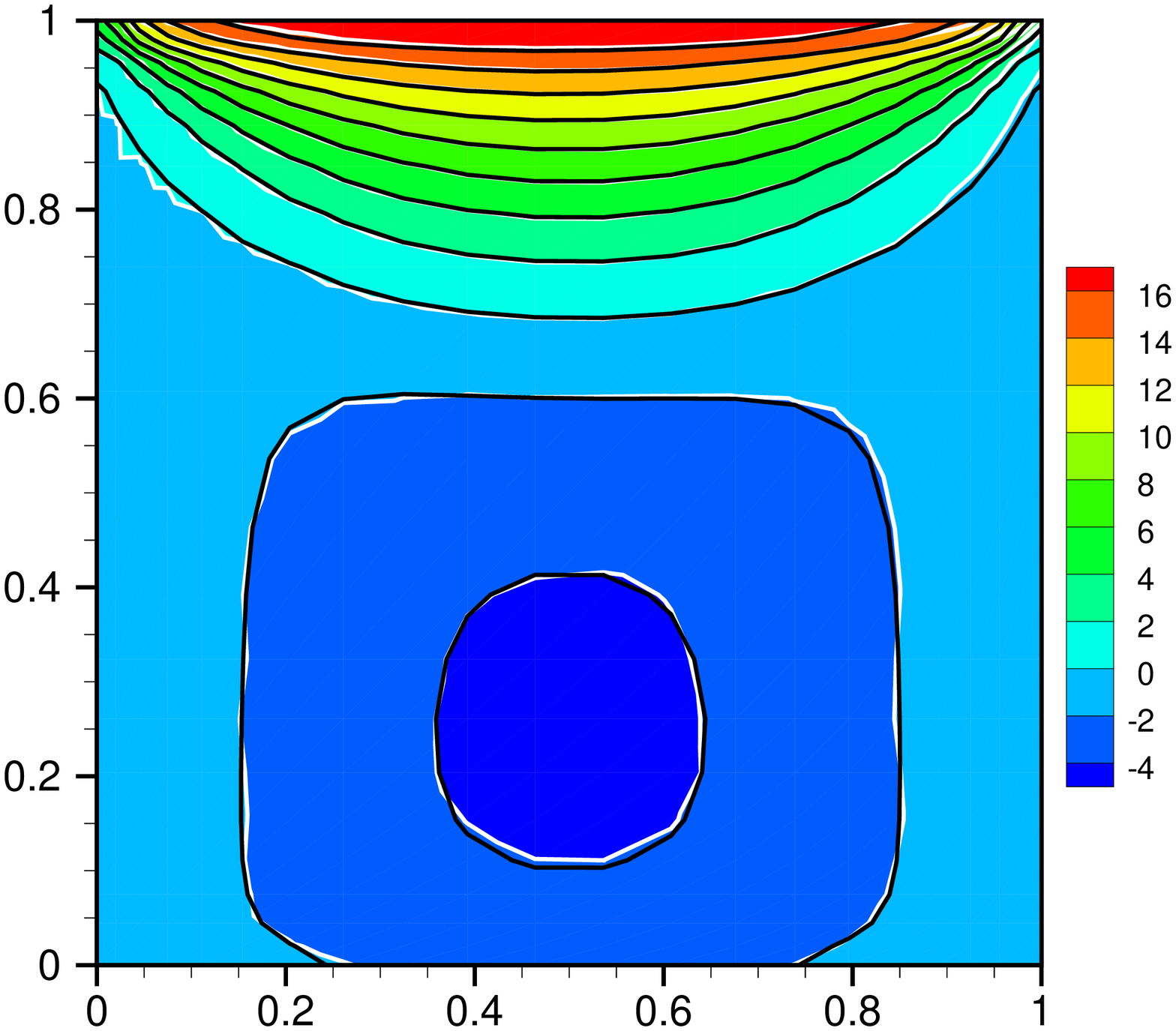}}
	\subfigure[Velocity Y]{\includegraphics[width=0.32\textwidth]{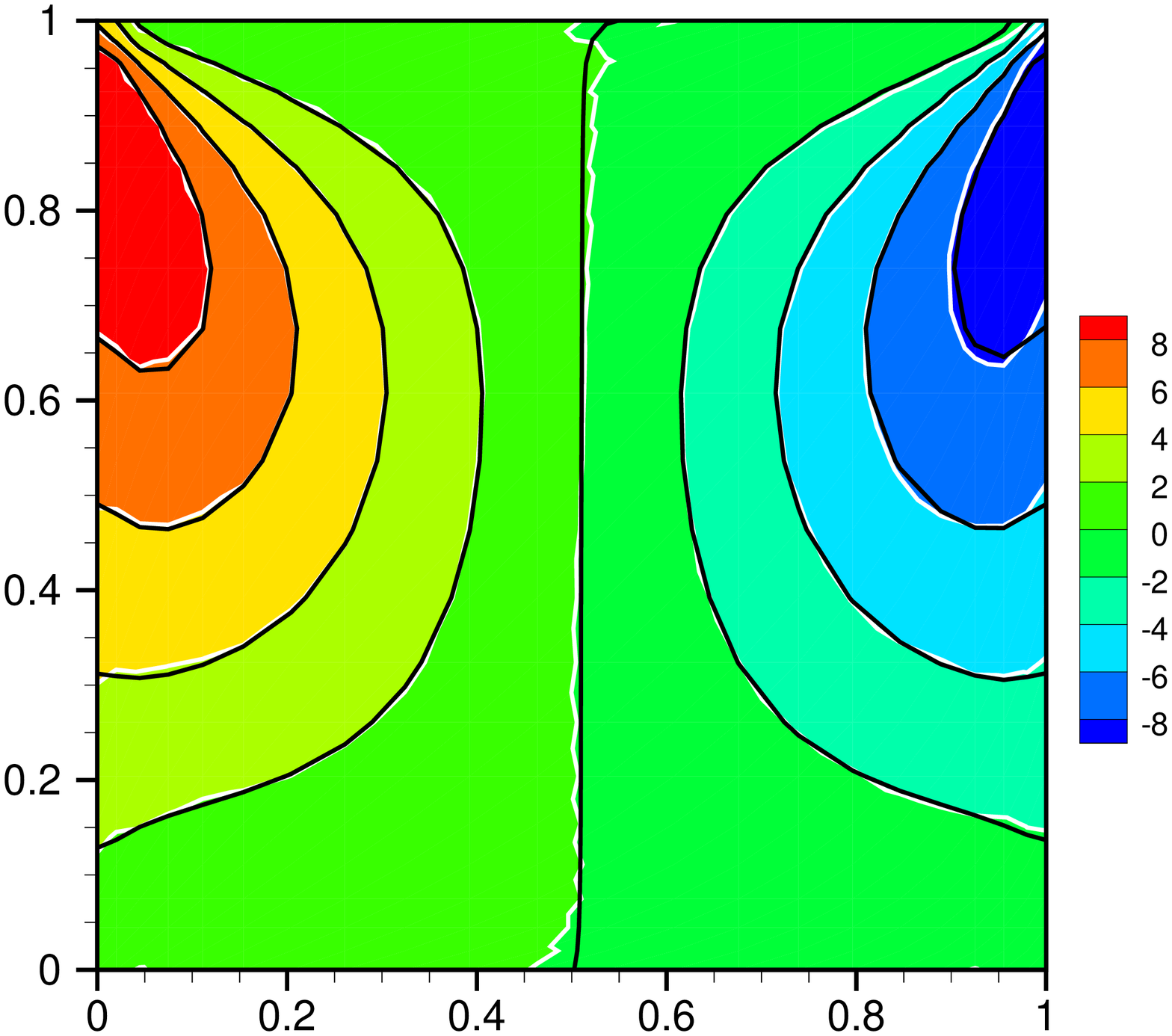}}
	\subfigure[Temperature]{\includegraphics[width=0.32\textwidth]{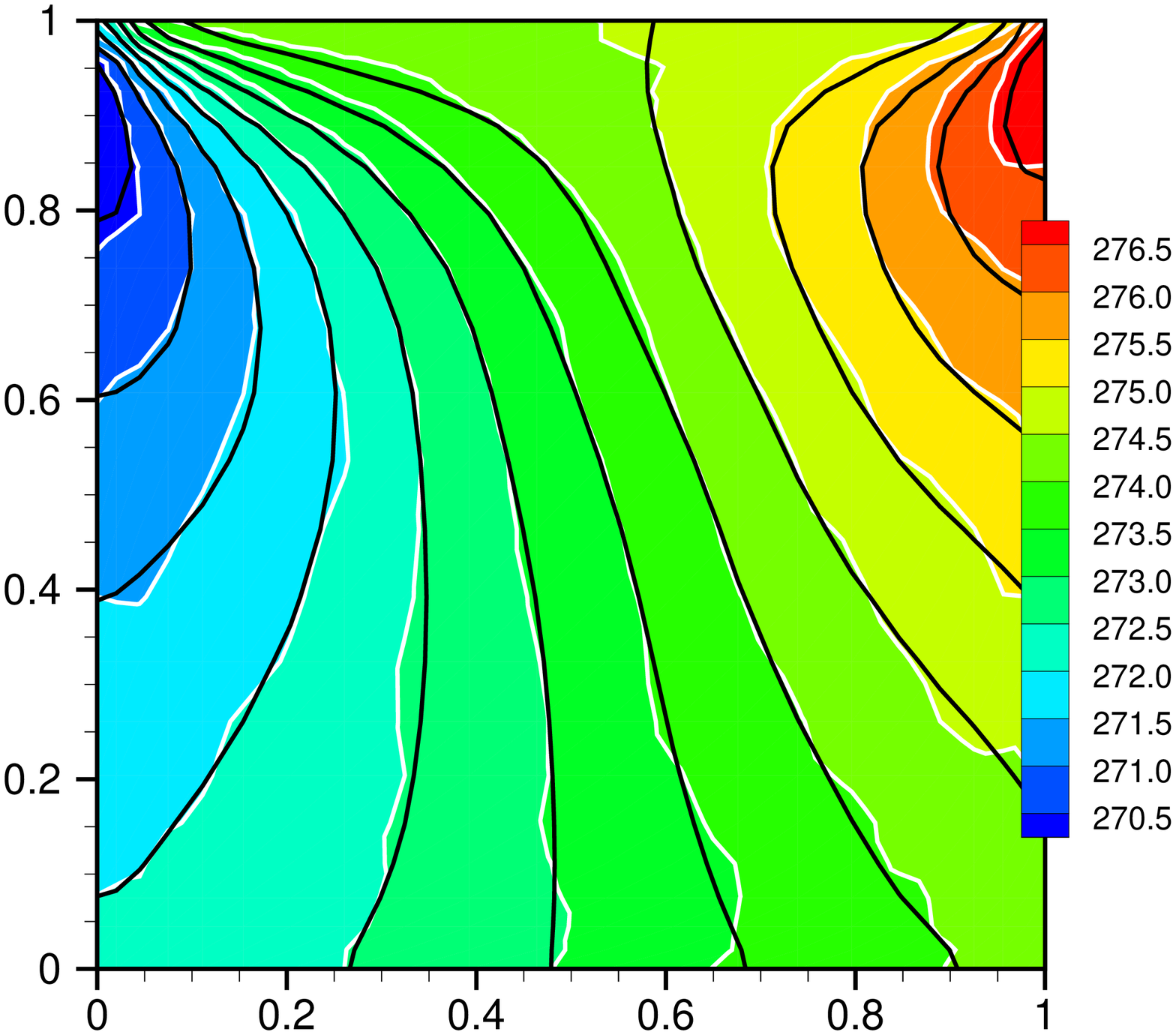}}
	\caption{\label{fig:cavity_kn10}Cavity flow at ${\rm Kn} = 10$. The background with white lines denotes the UGKWP results and the solid lines are UGKS solutions.}
\end{figure}

\begin{figure}[H]
	\centering
	\subfigure[\label{fig:cavity_mesh}Mesh]{\includegraphics[width=0.48\textwidth]{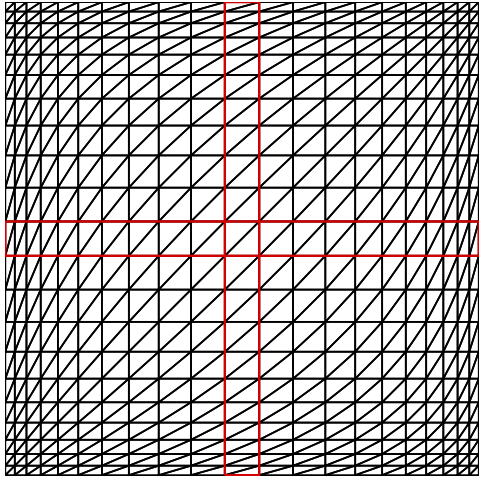}}
	\subfigure[${\rm Kn}=0.1$]{\includegraphics[width=0.48\textwidth]{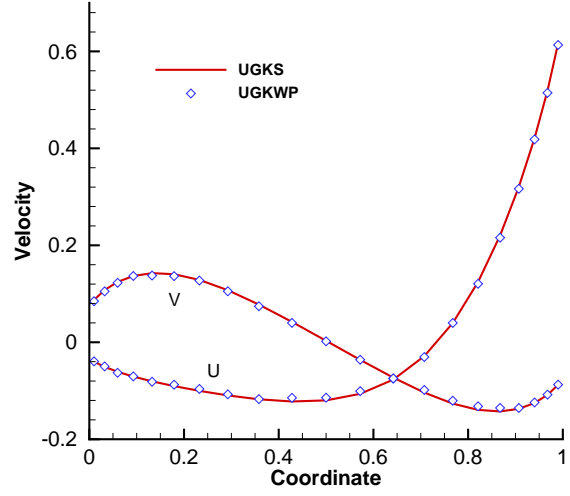}}\\
	\subfigure[${\rm Kn}=1$]{\includegraphics[width=0.48\textwidth]{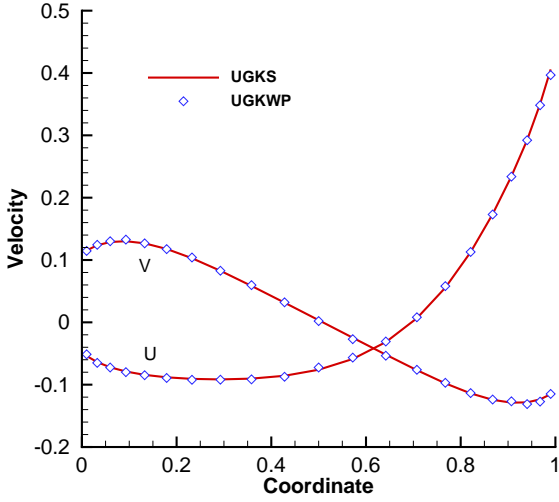}}
	\subfigure[${\rm Kn}=10$]{\includegraphics[width=0.48\textwidth]{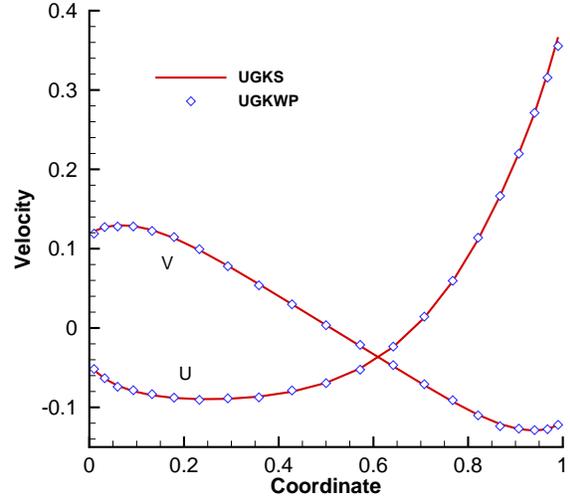}}
	\caption{\label{fig:cavity_velocity}Velocity profiles along the central lines for cavity flows.}
\end{figure}

The computational domain is discretized into $21\times21\times2$ triangular cells as shown in Fig.~\ref{fig:cavity_mesh}.
For the UGKS computations, $100\times100$ discrete velocity points are employed in the velocity space; and for the UGKWP method, we initially set the reference number of particles $N_r$ for each cell as $5000$.
The numerical results are plotted in the Fig.~\ref{fig:cavity_kn01} -- Fig.~\ref{fig:cavity_kn10}, where the distributions of the velocity and temperature are compared between the UGKWP and UGKS solutions.
Moreover, the velocity profiles along the central lines of the cavity are extracted by taking average of two neighboring triangular cells.
From Fig.~\ref{fig:cavity_velocity}, it can be seen that satisfactory results are obtained for these three cases.
For the low speed rarefied flow, we employ a large number of simulation particles and do many averaging  process to reduce the statistical noises so that the high-order quantity, such as the temperature distribution, can be obtained.
It takes about $5$ hours for the UGKWP method to obtain the current results.
For the UGKS, with the acceleration techniques, such as implicit algorithm and multigrid method \cite{zhu2016implicit,zhu2017unified}, the convergent solution with no statistical noises can be obtained within $5$ minutes.
Therefore, the deterministic method with acceleration techniques would still be a better choice for low-speed rarefied flow studies, which has much higher efficiency than the stochastic related method.

\subsection{Laminar boundary layer}
\begin{figure}[H]
	\centering
	\includegraphics[width=0.8\textwidth]{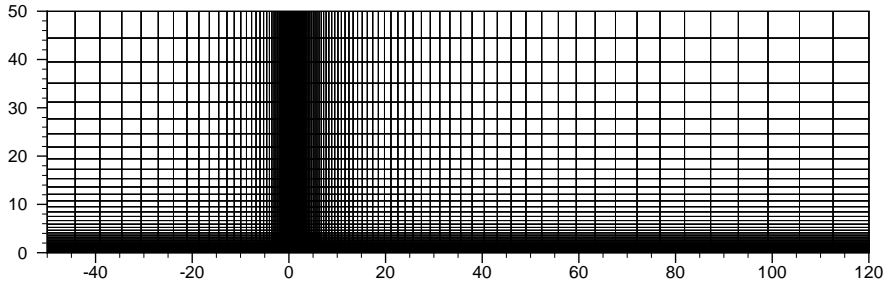}
	\caption{\label{fig:bl_mesh}Computational mesh for laminar boundary layer simulation.}
\end{figure}
\begin{figure}[H]
	\centering
	\subfigure[Density]{\includegraphics[width=0.32\textwidth]{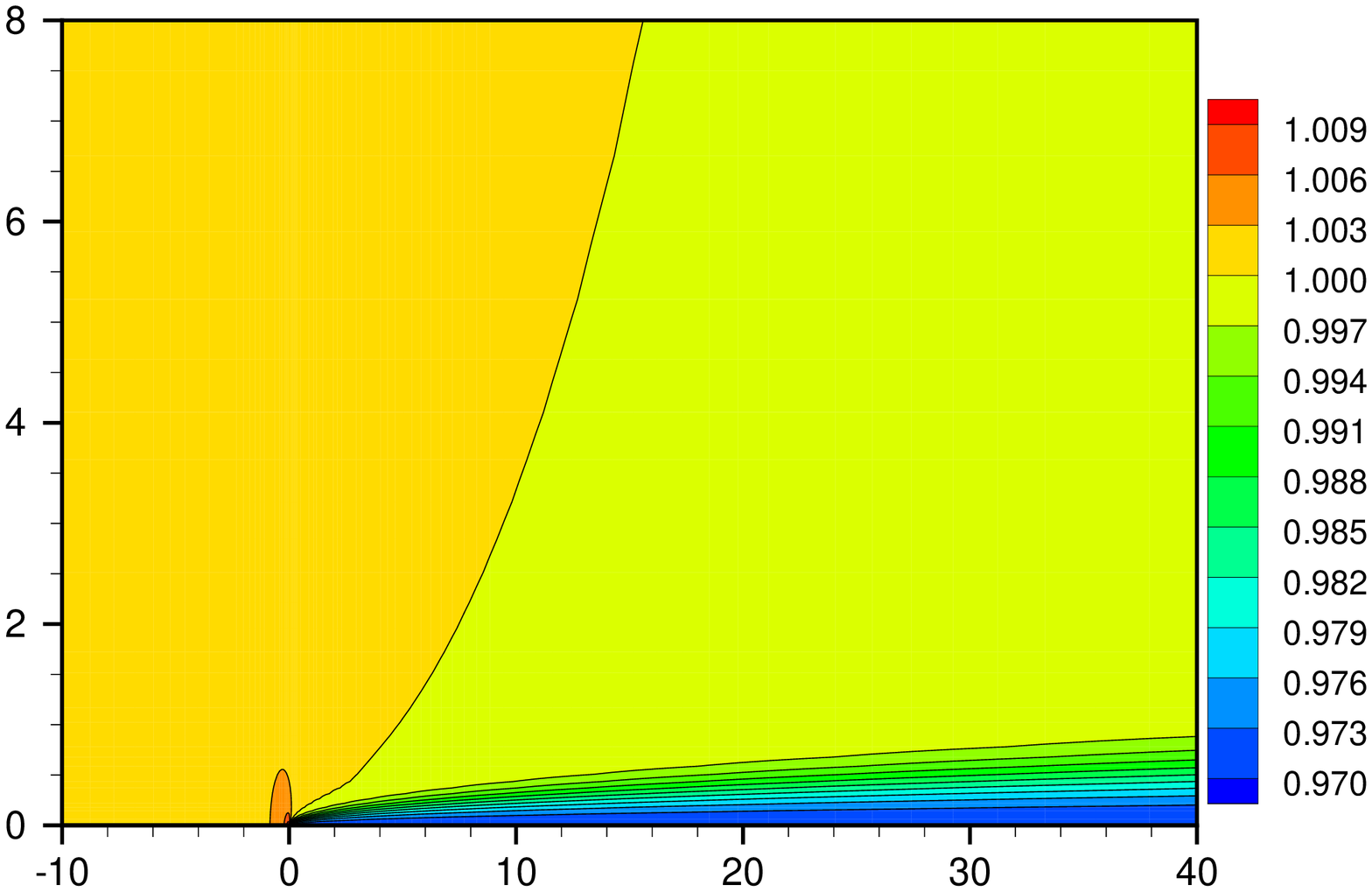}}
	\subfigure[Velocity X]{\includegraphics[width=0.32\textwidth]{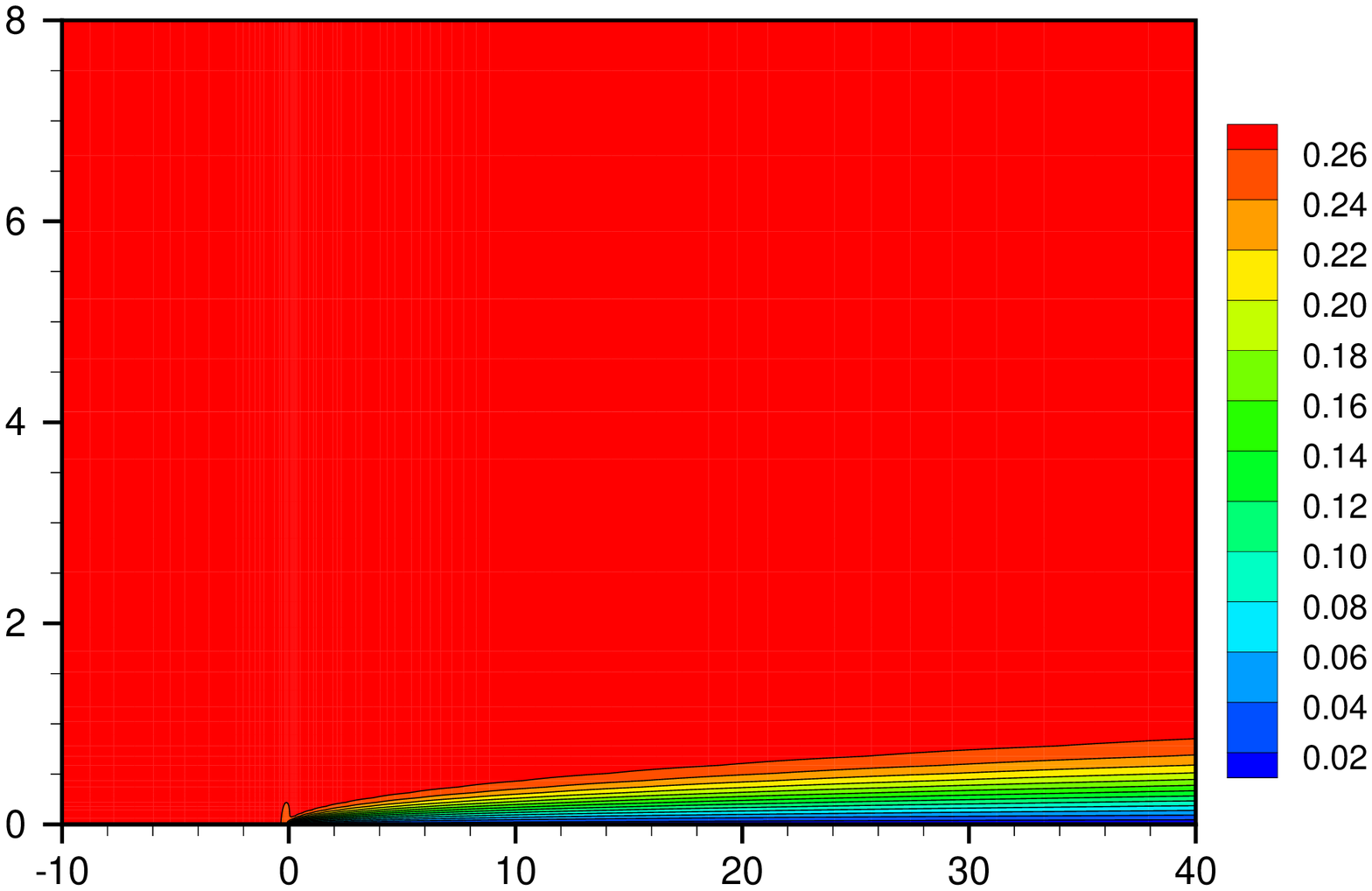}}
	\subfigure[Velocity Y]{\includegraphics[width=0.32\textwidth]{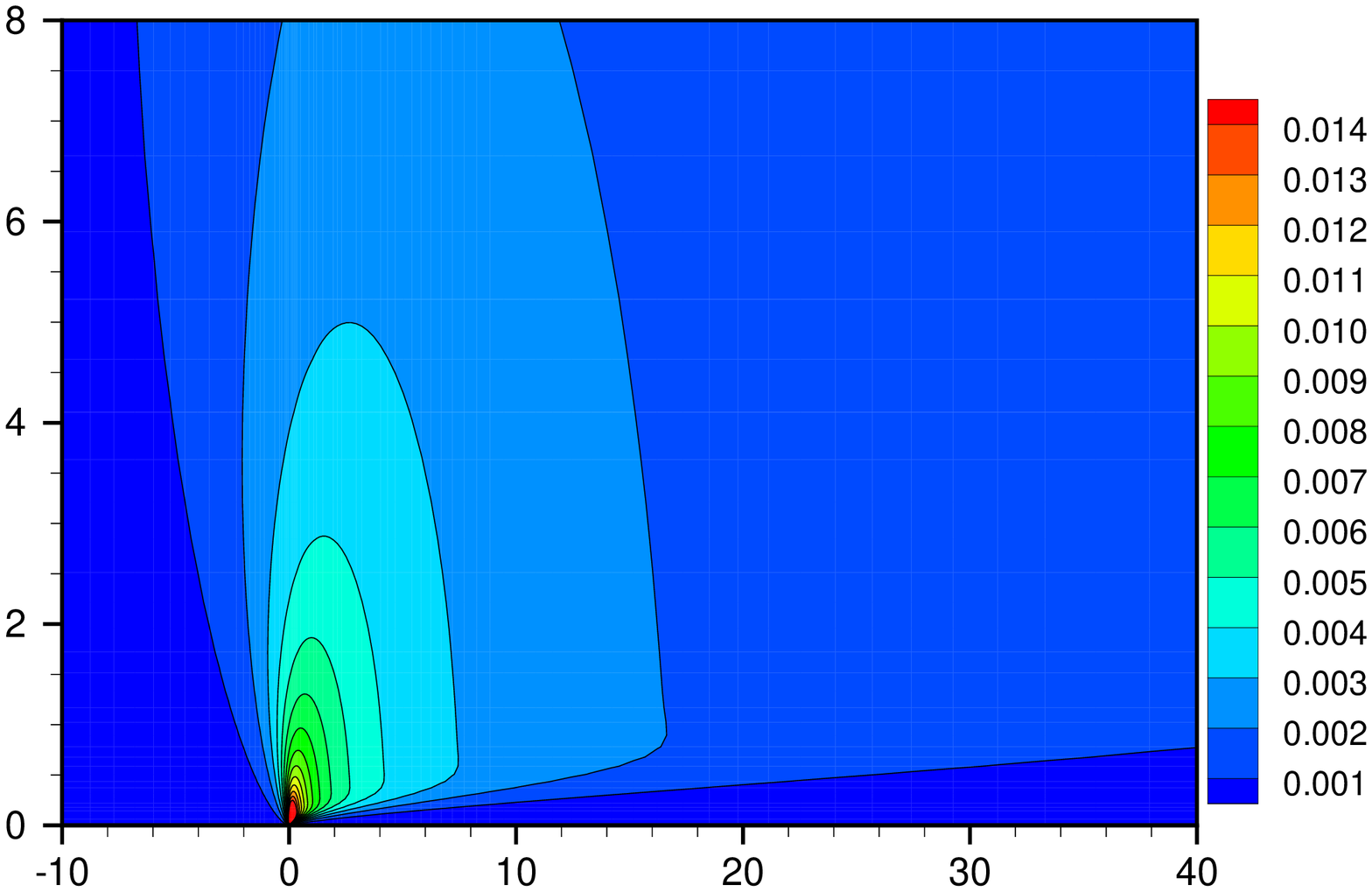}}
	\caption{\label{fig:bl_flow}Flow around the leading edge of flat plate.}	
\end{figure}
\begin{figure}[H]
	\centering
	\subfigure[Velocity X]{\includegraphics[width=0.48\textwidth]{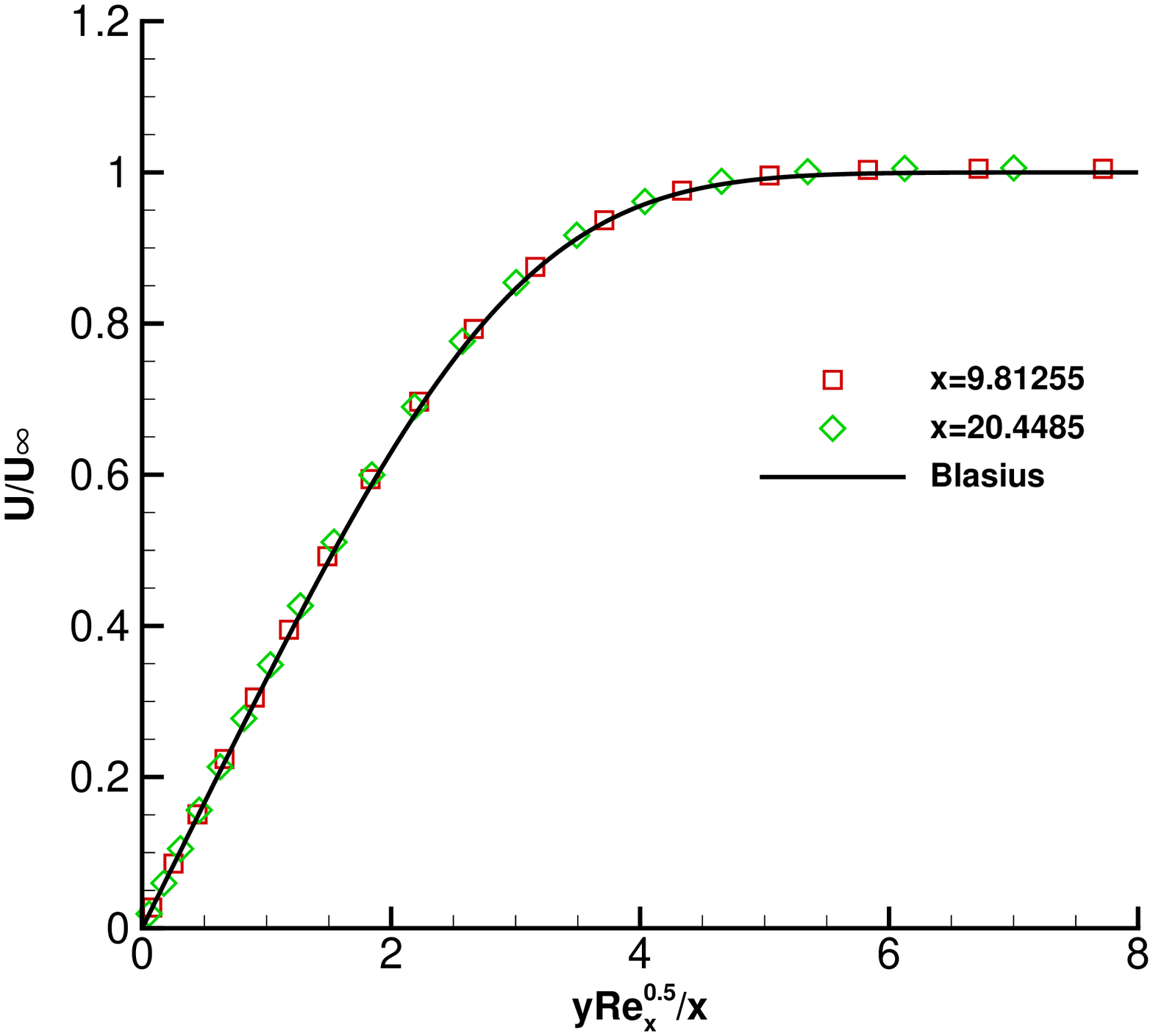}}
	\subfigure[Velocity Y]{\includegraphics[width=0.48\textwidth]{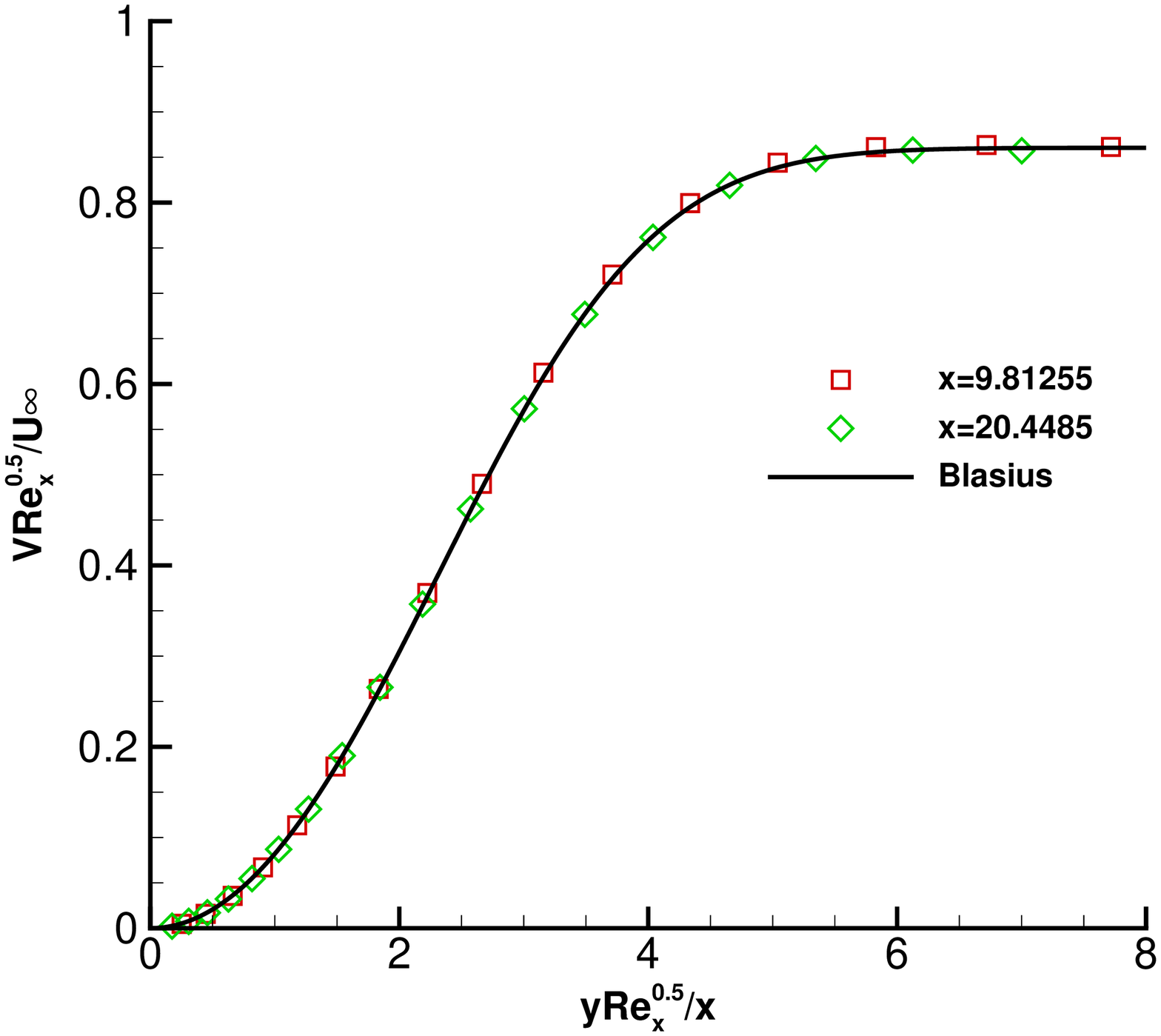}}
	\caption{\label{fig:bl_line}Velocity distribution in the laminar boundary layer obtained by the UGKWP method.}	
\end{figure}

The laminar boundary layer over a flat plate is computed to validate the current multiscale method for viscous NS solutions in the continuum limit.
The computational domain is $[-50, 120]\times[0, 50]$ as shown in Fig.~\ref{fig:bl_mesh}.
A non-uniform mesh with $120\times50$ cells is employed.
The free stream is monatomic gas flow at Reynolds number ${\rm Re} = 10^{5}$ and Mach number ${\rm Ma}=0.3$ with constant viscosity.
The Reynolds number and Mach number is defined with respect to the length of the flat plate $L = 120 L_0$.
The reference variables $U_0$ and $t_0$ to non-dimensionalize the velocity and time are given by $U_0 = \sqrt{2 k_B T_0 / m_0}$ and $t_0 = {L_0}/{U_0}$, where $T_0$ is the temperature in the free stream.
The flow field at time $t=1000 t_0$ is given as the convergent steady state solution in Fig.~\ref{fig:bl_flow}, where the distribution of the density and velocity around the leading edge is enlarged in the $y$ direction.
Comparison between the UGKWP results and the Blasius solutions is given in the Fig.~\ref{fig:bl_line}.


In the computation, the time step $\Delta t$ and particle collision time $\tau$ are $0.02t_0$ and $6.57\times10^{-4}t_0$, respectively.
Since the ratio $e^{-\Delta t/\tau}$ has a very small value of $6\times10^{-14}$, the hydrodynamic wave is dominant and the particle contribution can be neglected.
The computational time for $50000$ step simulations is $15$ minutes and the memory cost is $24$ MB.
Under such condition, the present UGKWP method automatically becomes a hydrodynamic fluid solver, such as the GKS \cite{xu2001gks,xu1994numerical}.
Due to the multiscale transport, the UGKWP method can recover NS solutions without the requirement of the mesh size and the time step being less than the mean free path and the particle collision time.
Moreover, the computational cost is comparable to the hydrodynamic fluid solver in the continuum regime.

\subsection{Cylinder flow}
Hypersonic flow past a circle cylinder at ${\rm Ma} = 5$ and $30$ are simulated to show the capability of the current method for high-speed rarefied flow simulations.
The free stream is initialized with the monatomic gas flow of argon with an initial temperature $T_{\infty} = 273 {\rm K}$.
The diameter $D$ of the cylinder is $1{\rm m}$ long.
The solid boundaries are isothermal walls with a constant temperature $T_{w} = 273 {\rm K}$.
The Knudsen number is defined with respect to the diameter of the circle cylinder.

\begin{figure}[H]
	\centering
	\subfigure[Computational domain]
	{\includegraphics[width=0.4\textwidth]{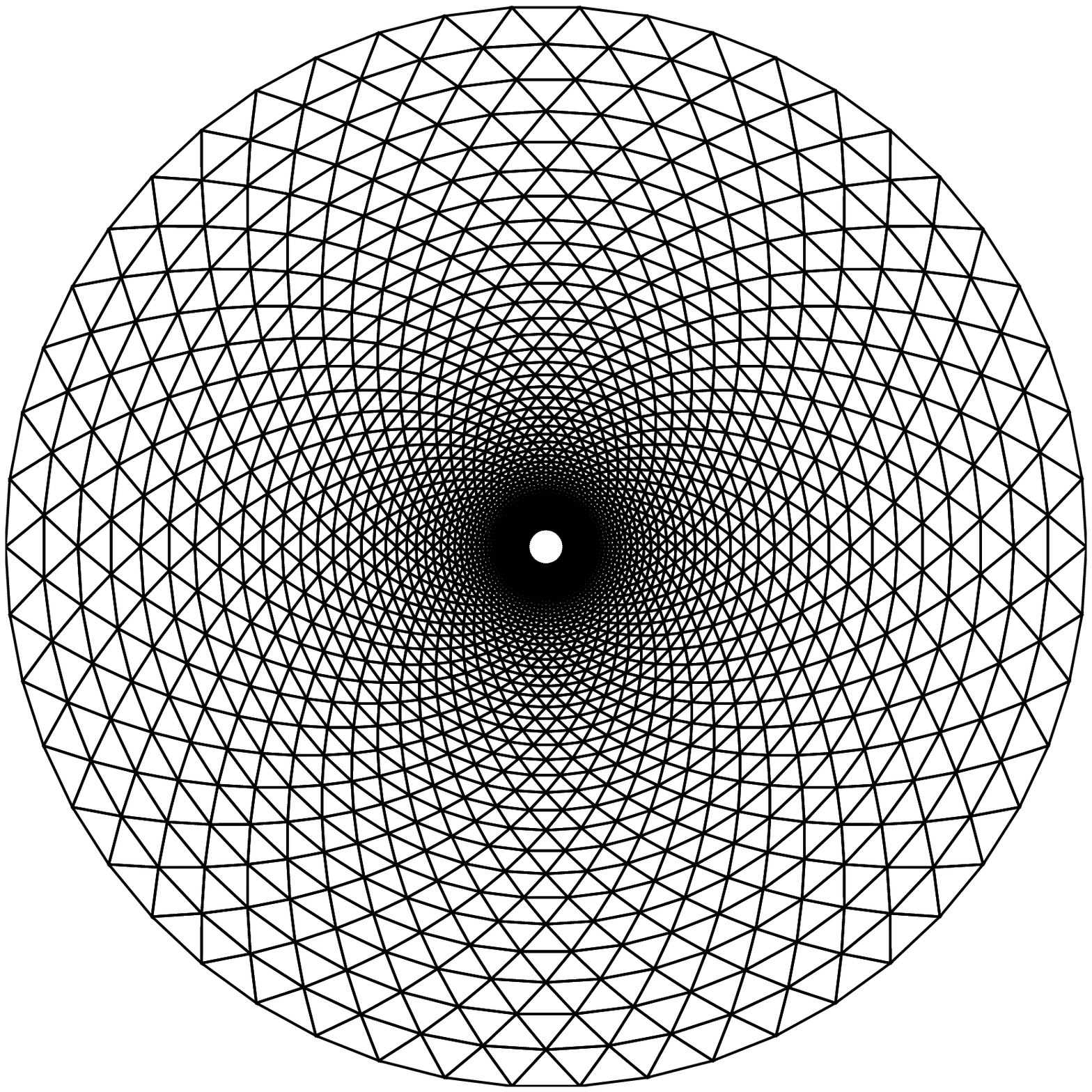}}\hspace{0.1\textwidth}
	\subfigure[Details near the boundaries]
	{\includegraphics[width=0.4\textwidth]{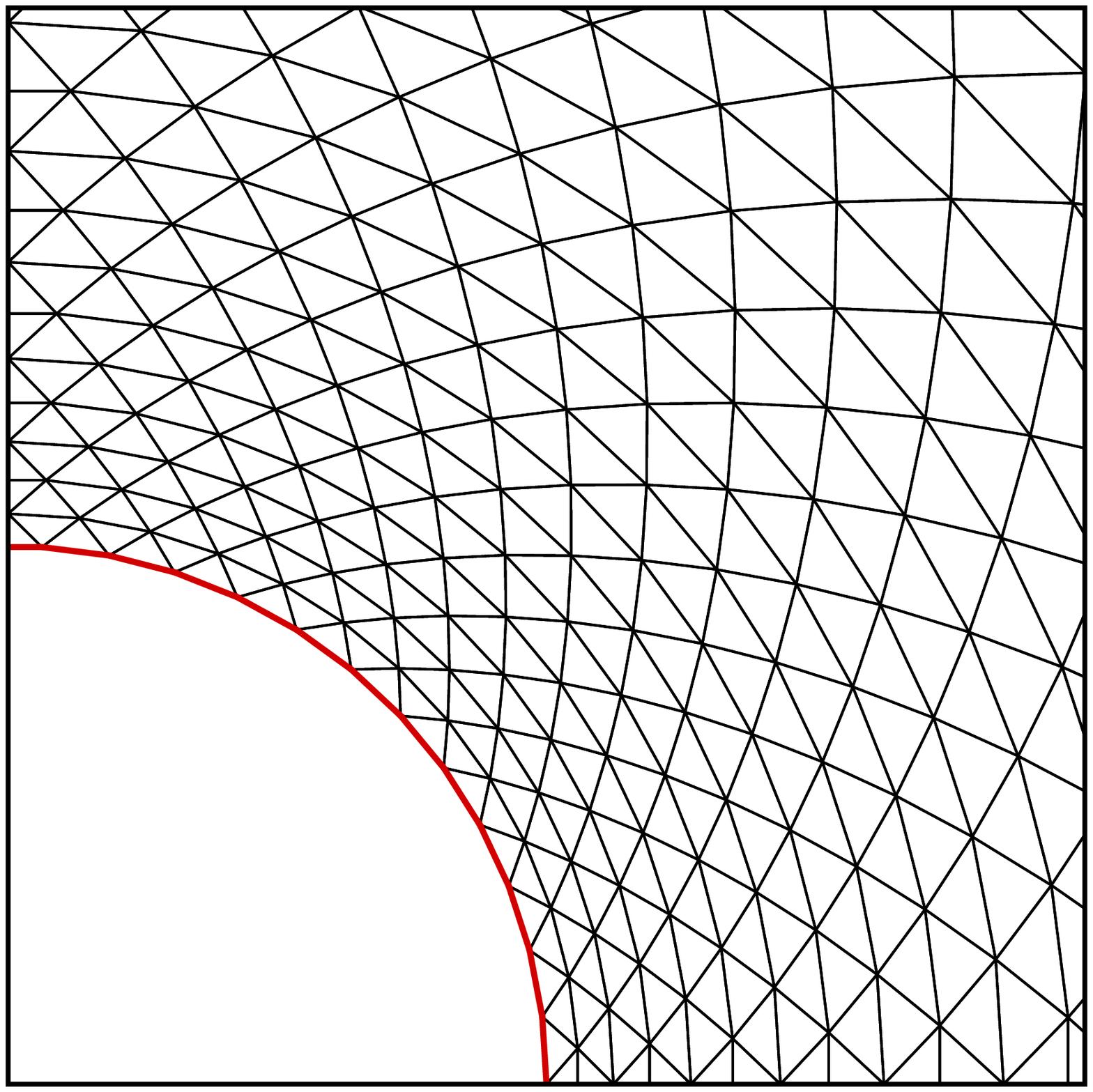}}
	\caption{\label{fig:cylinder_mesh} The computational mesh for the cylinder flows at ${\rm Kn}=1$.}
\end{figure}

\begin{figure}[H]
	\centering
	\subfigure[Velocity X]{\includegraphics[width=0.32\textwidth]{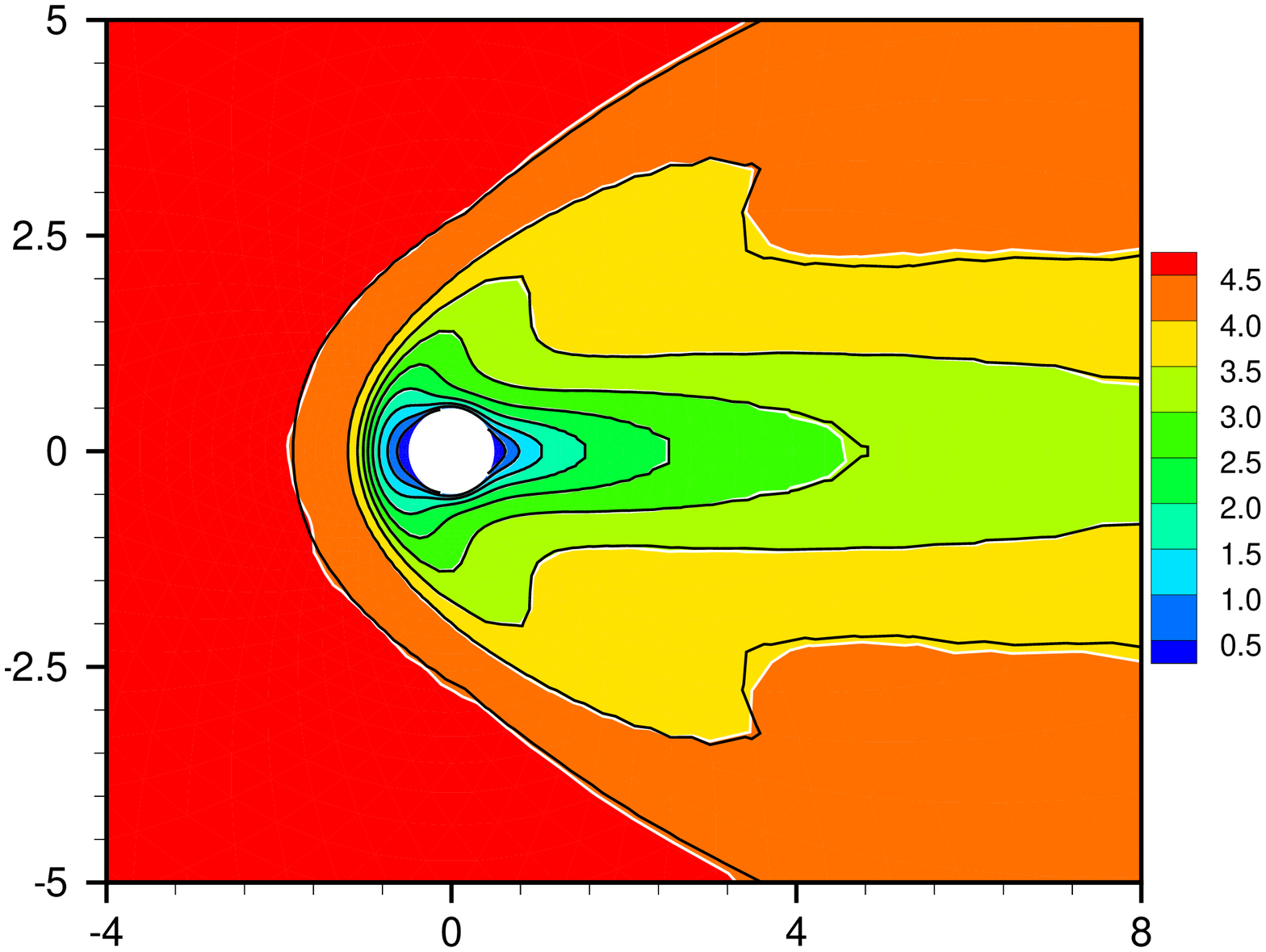}}
	\subfigure[Velocity Y]{\includegraphics[width=0.32\textwidth]{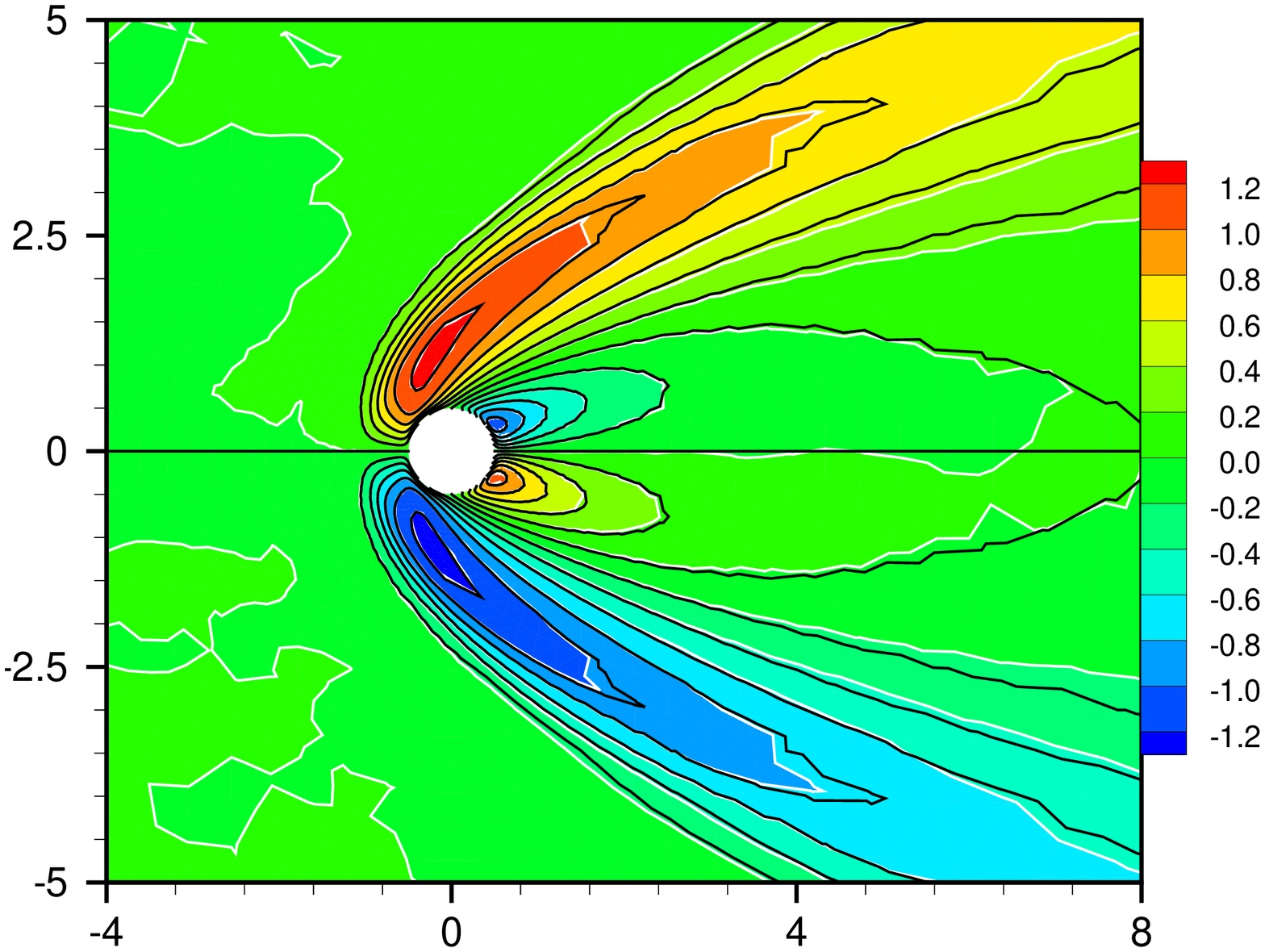}}
	\subfigure[Temperature]{\includegraphics[width=0.32\textwidth]{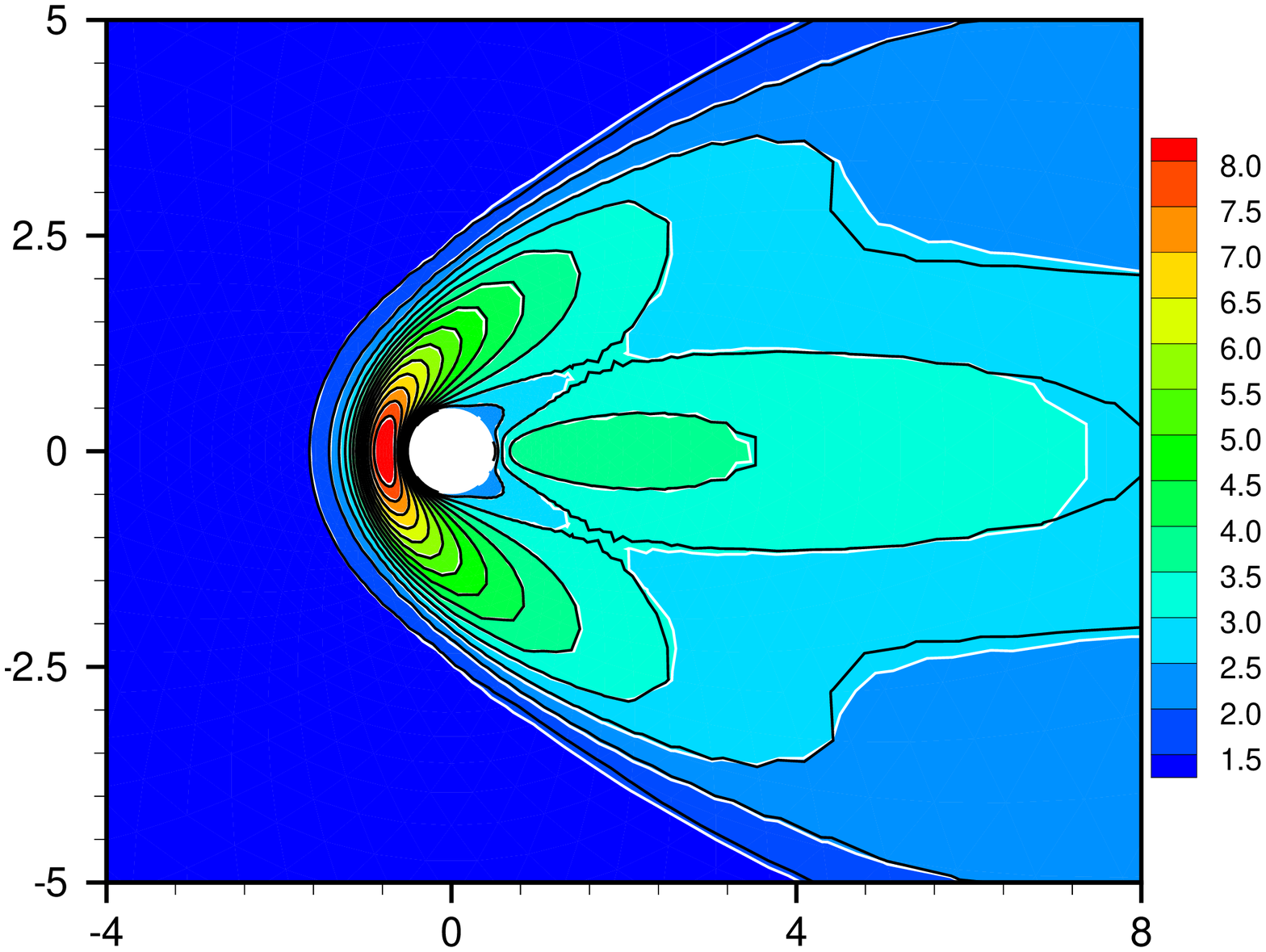}}
	\caption{\label{fig:cylinder_kn01}Hypersonic flow at ${\rm Ma}=5$ around a circle cylinder for ${\rm Kn} = 0.1$. The background is the UGKWP results and the black solid lines denote the UGKS solutions. The velocities are normalized by the most probable speed $C_{\infty} = \sqrt{2 k_B T_{\infty} / m_0} = 337 {\rm m/s}$ and the temperature is normalized by the free stream temperature $T_{\infty} = 273{\rm K}$.}
\end{figure}

\begin{figure}[H]
	\centering
	\subfigure[Pressure]{\includegraphics[width=0.32\textwidth]{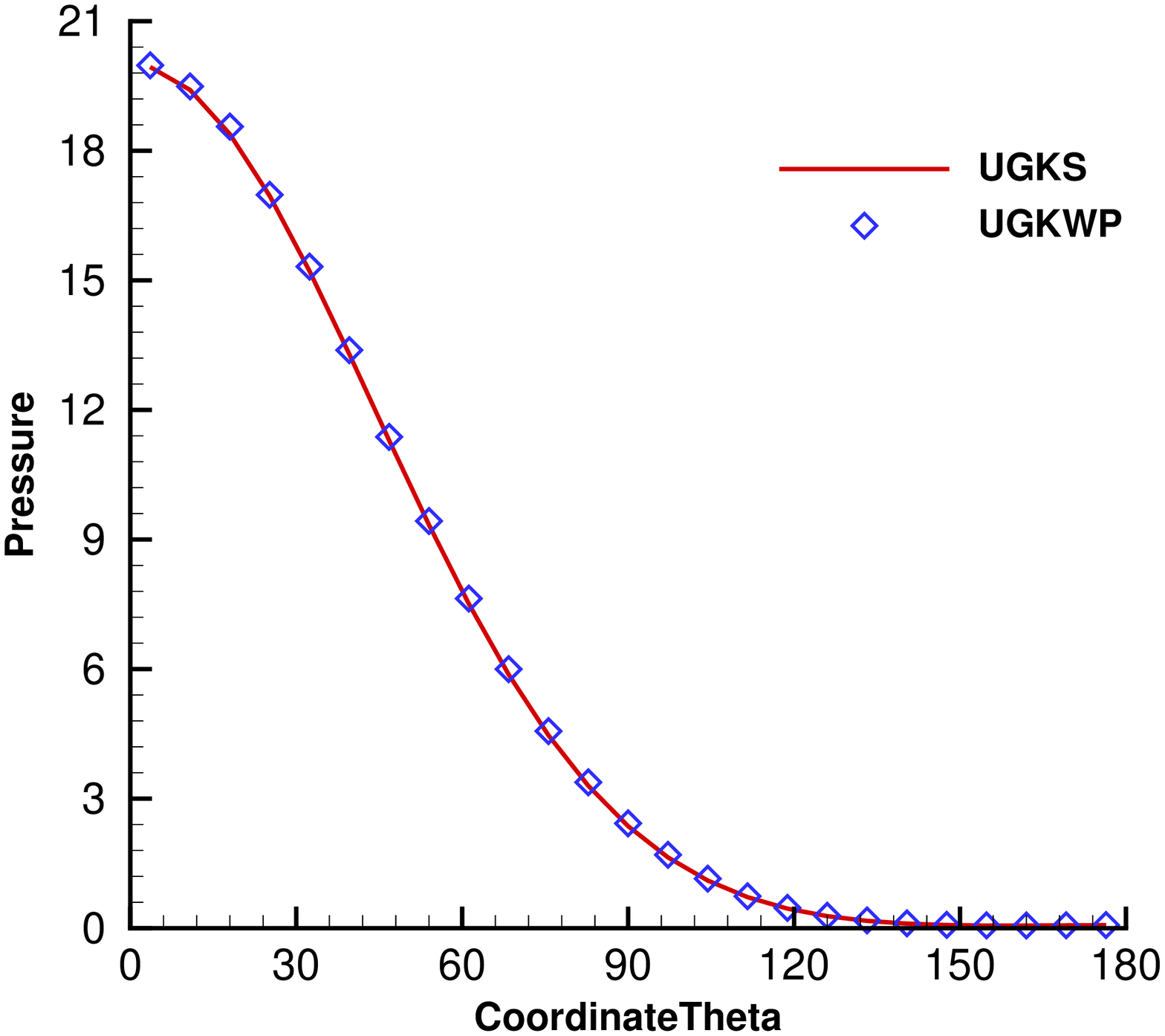}}
	\subfigure[Shear stress]{\includegraphics[width=0.32\textwidth]{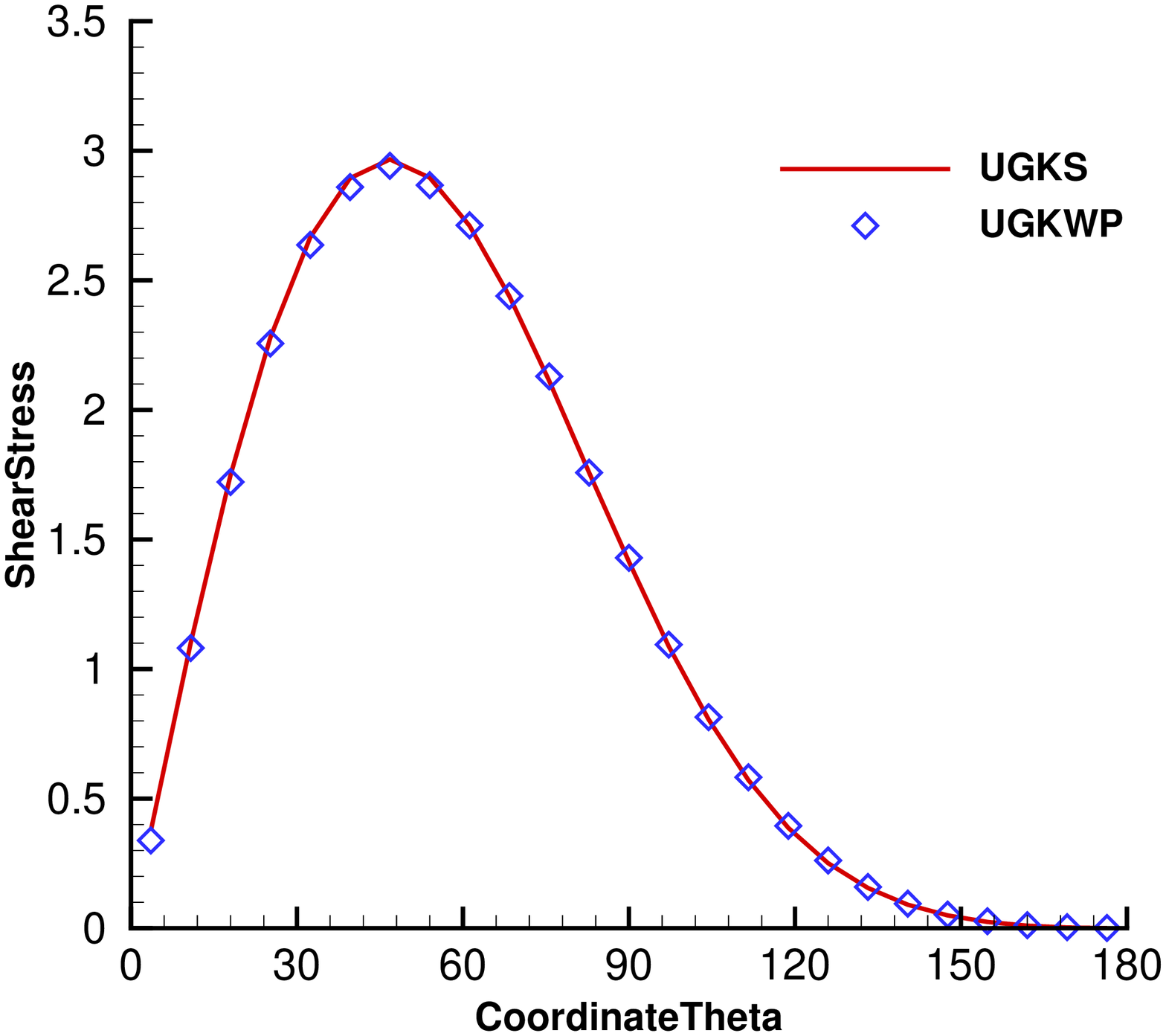}}
	\subfigure[Heat flux]{\includegraphics[width=0.32\textwidth]{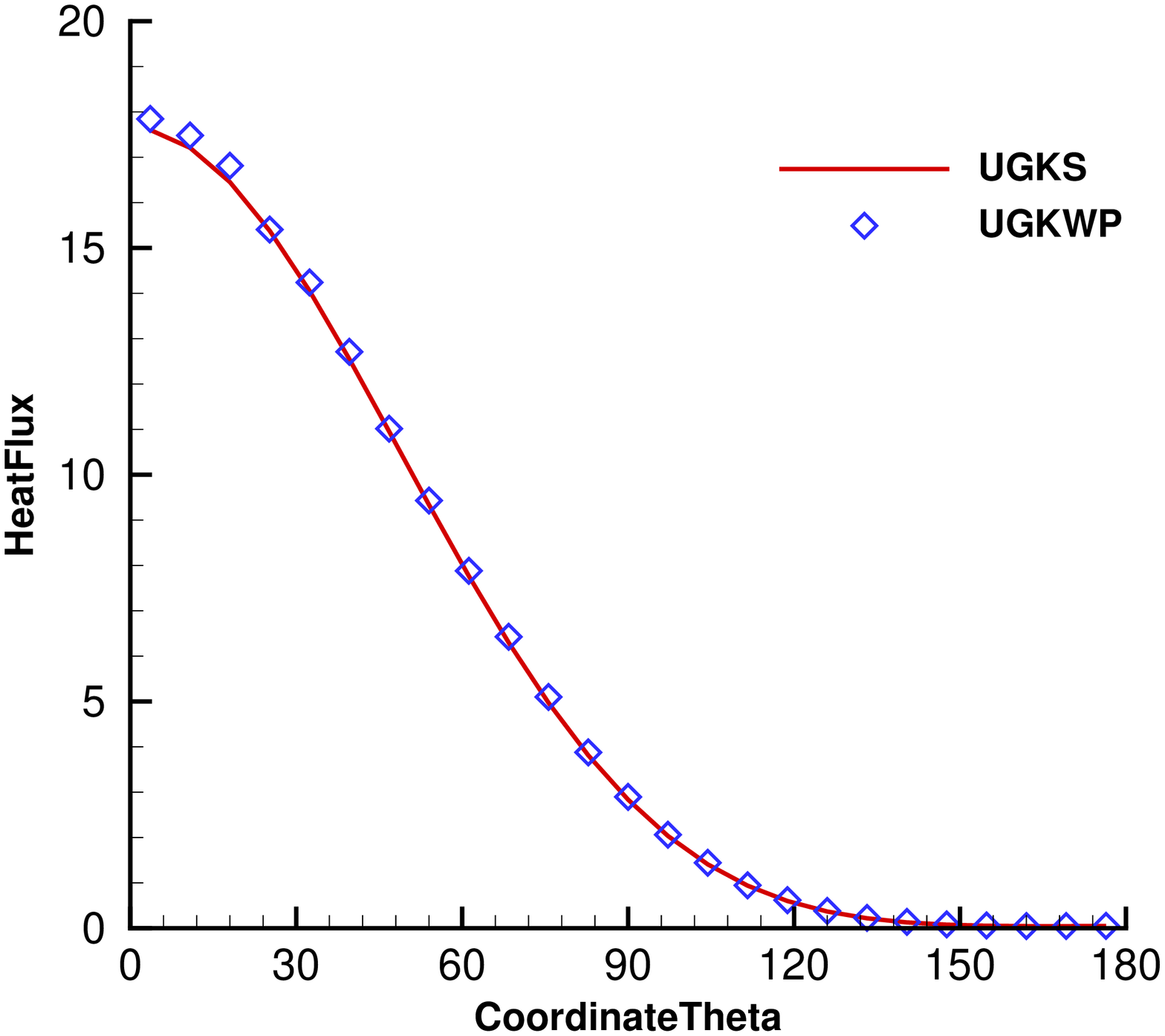}}
	\caption{\label{fig:cylinder_surface_kn01}Surface quantities around the circle cylinder at ${\rm Ma}=5$ and ${\rm Kn}=0.1$. The pressure and shear stress are normalized by $\rho_{\infty} C_{\infty}^2$, and the heat flux is normalized by $\rho_{\infty} C_{\infty}^3$. $C_{\infty} = \sqrt{2 k_B T_{\infty} / m_0} = 337 {\rm m/s}$ is the most probable speed of the free stream.}
\end{figure}

\begin{figure}[H]
	\centering
	\subfigure[VelocityX]{\includegraphics[width=0.32\textwidth]{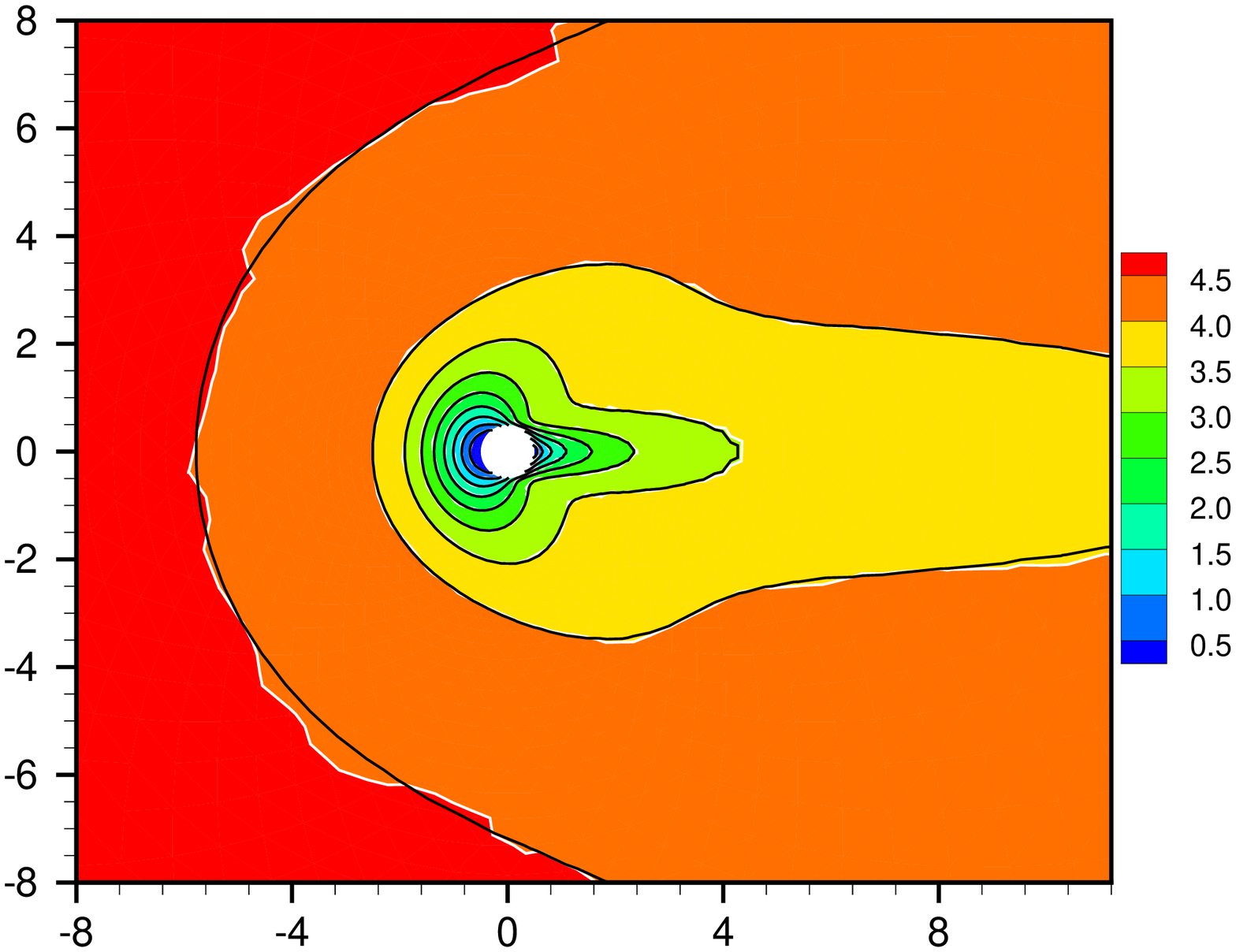}}
	\subfigure[VelocityY]{\includegraphics[width=0.32\textwidth]{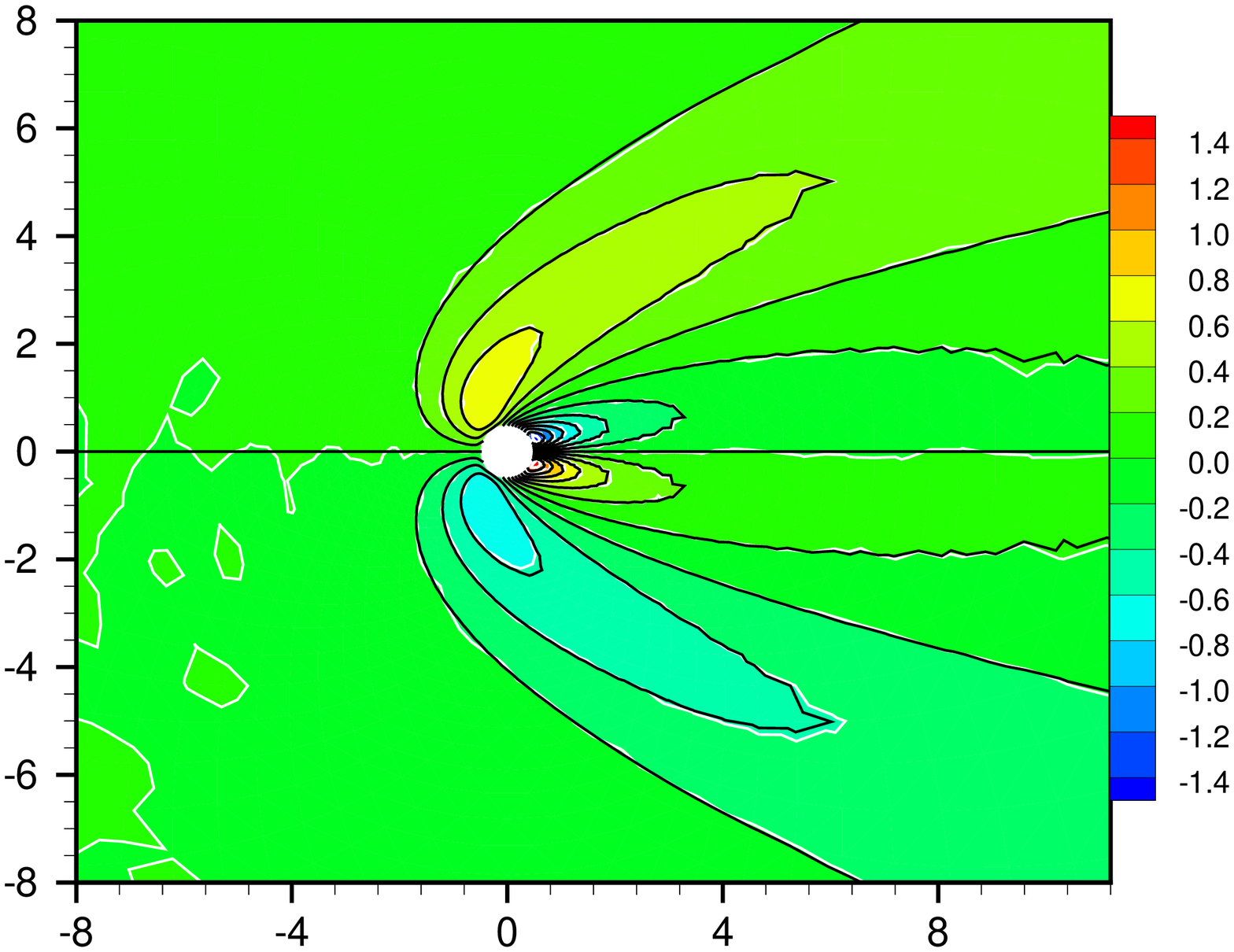}}
	\subfigure[Temperature]{\includegraphics[width=0.32\textwidth]{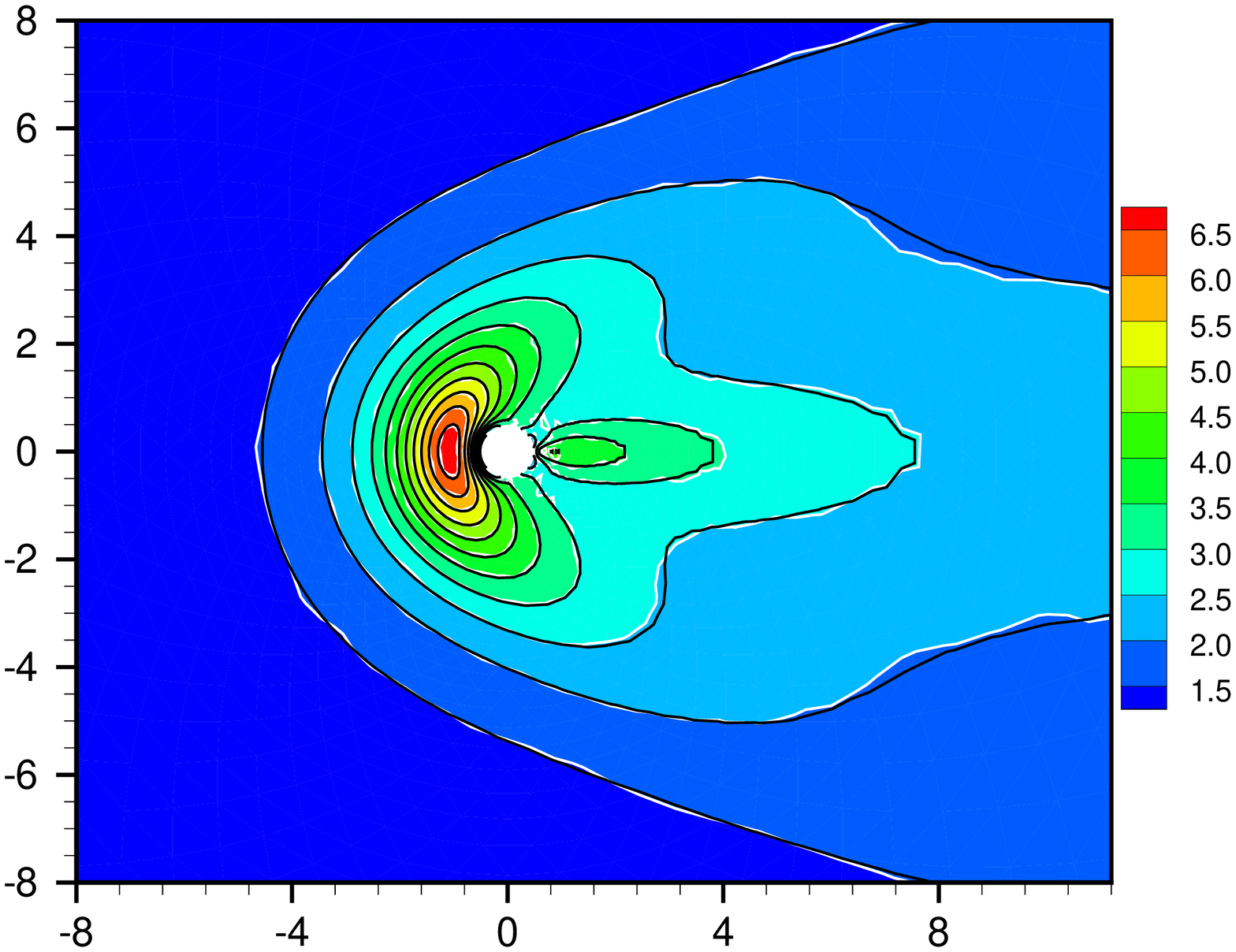}}
	\caption{\label{fig:cylinder_kn1}Hypersonic flow at ${\rm Ma}=5$  around a circle cylinder for ${\rm Kn} = 1$. The background is the UGKWP results and the black solid lines denote the UGKS solutions. The velocities are normalized by the most probable speed $C_{\infty} = \sqrt{2 k_B T_{\infty} / m_0} = 337 {\rm m/s}$ and the temperature is normalized by the free stream temperature $T_{\infty} = 273{\rm K}$.}
\end{figure}

\begin{figure}[H]
	\centering
	\subfigure[Pressure]{\includegraphics[width=0.32\textwidth]{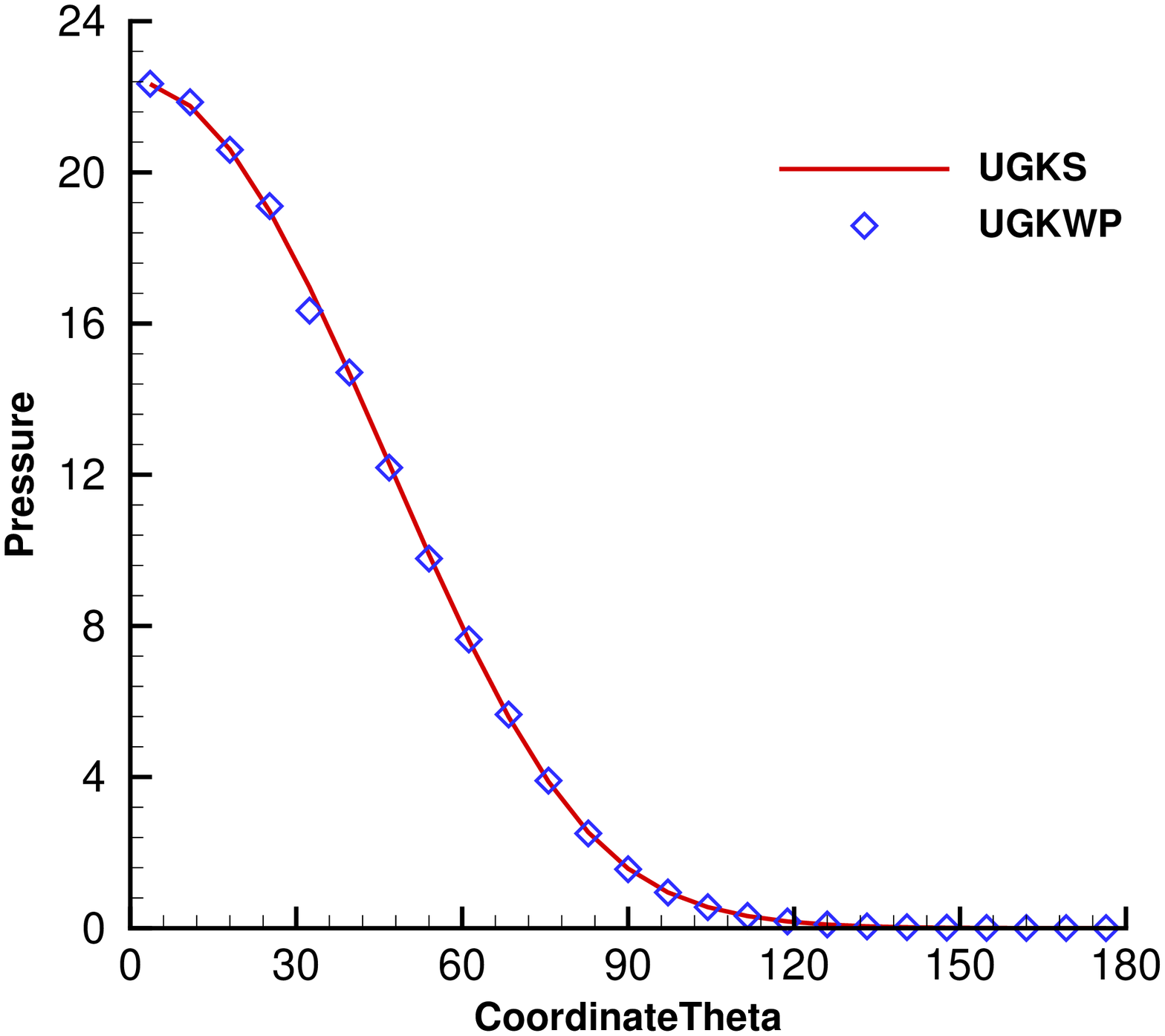}}
	\subfigure[Shear stress]{\includegraphics[width=0.32\textwidth]{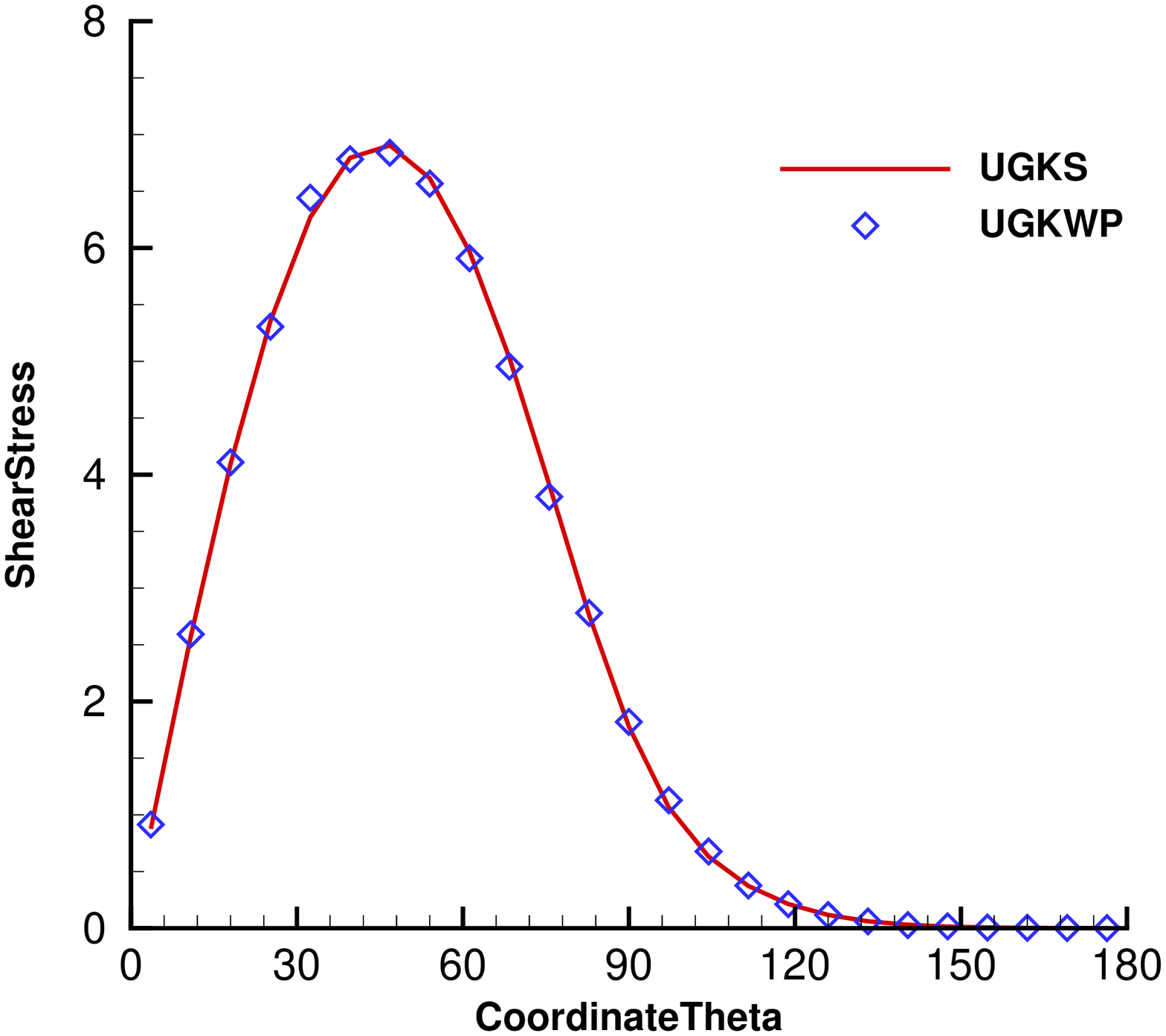}}
	\subfigure[Heat flux]{\includegraphics[width=0.32\textwidth]{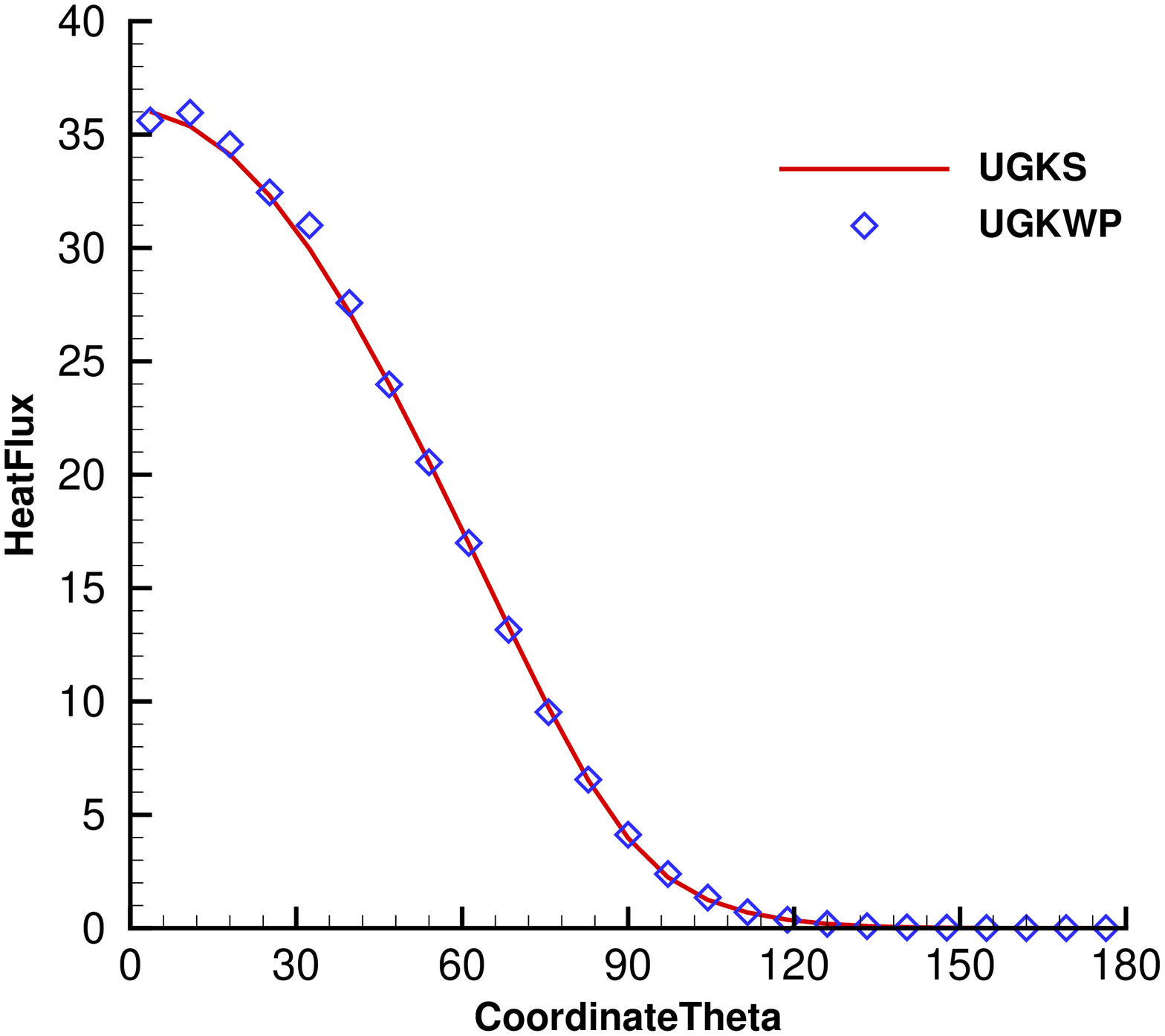}}
	\caption{\label{fig:cylinder_surface_kn1}Surface quantities around the circle cylinder at ${\rm Ma}=5$ and ${\rm Kn}=1$. The pressure and shear stress are normalized by $\rho_{\infty} C_{\infty}^2$, and the heat flux is normalized by $\rho_{\infty} C_{\infty}^3$. $C_{\infty} = \sqrt{2 k_B T_{\infty} / m_0} = 337 {\rm m/s}$ is the most probable speed of the free stream.}
\end{figure}

\begin{table}[h]
	\caption{\label{tab:cylinder_kn01}Computational cost for hypersonic flow at ${\rm Ma} = 5$ and ${\rm Kn}=0.1$ around a circle cylinder.}
	\begin{ruledtabular}	
		\begin{tabular}{cccc}
			~& UGKS & UGKWP & ratio \\
			\hline
			CPU time & $45{\rm h}$ $16{\rm min}$ & $2{\rm h}$ $10{\rm min}$ & $20.9$\\
			Memory   & $4.9 {\rm GB}$            & $277{\rm MB}$            & $18.1$\\
			Steps    & $5000$\footnotemark[1] $+$ $35000$ & $30000$ $+$ $10000$\footnotemark[2]& \\
		\end{tabular}
	\end{ruledtabular}
	\footnotetext[1]{$5000$ steps of first-order calculation for a better initial state in UGKS computation.}
	\footnotetext[2]{$10000$ steps of averaging process in the UGKWP simulation.}
\end{table}
\begin{table}[h]
	\caption{\label{tab:cylinder_kn1}Computational cost for hypersonic flow at ${\rm Ma} = 5$ and ${\rm Kn}=1$ around a circle cylinder.}
	\begin{ruledtabular}	
		\begin{tabular}{cccc}
			~& UGKS & UGKWP & ratio \\
			\hline
			CPU time & $45{\rm h}$ $16{\rm min}$ & $2{\rm h}$ $42{\rm min}$ & $16.8$\\
			Memory   & $4.9 {\rm GB}$            & $310{\rm MB}$            & $16.2$\\
			Steps    & $5000$\footnotemark[1] $+$ $35000$ & $30000$ $+$ $10000$\footnotemark[2]& \\
		\end{tabular}
	\end{ruledtabular}
		\footnotetext[1]{$5000$ steps of first-order calculation for a better initial state in UGKS computation.}
		\footnotetext[2]{$10000$ steps of averaging process in the UGKWP simulation.}		
\end{table}

For the free stream with a relatively low Mach number ${\rm Ma}=5$, the cases at the Knudsen numbers $0.1$ and $1$ are computed.
The computational domain is discretized by $50 \times 50 \times2$ triangular cells as shown in Fig.~\ref{fig:cylinder_mesh}, which covers a region of $\pi (15D)^2$.
Along the radial direction, the minimum heights of the triangles near the boundaries are $0.01{\rm m}$ and $0.03{\rm m}$ for ${\rm Kn} = 0.1$ and $1$, respectively.
The UGKS employs $100\times100$ velocity points in the velocity space, and the initial reference number of particles $N_r$ for the UGKWP method is set as $400$.
In comparison with the UGKS solutions, the flow fields computed by the UGKWP method are shown in Fig.~\ref{fig:cylinder_kn01} and Fig.~\ref{fig:cylinder_kn1}.
It can be seen that the UGKWP results agree well with those obtained from the UGKS computations.
Detailed comparisons of the surface quantities, such as the pressure, shear stress, and heat flux, are given in Fig.~\ref{fig:cylinder_surface_kn01} and Fig.~\ref{fig:cylinder_surface_kn1}.
The computational cost is listed in Tables ~\ref{tab:cylinder_kn01} and \ref{tab:cylinder_kn1}.
The UGKS solutions are fully recovered by the UGKWP method on the unstructured meshes, but with one-order-of-magnitude lower in computational cost and memory consumption from UGKWP.

\begin{figure}[H]
	\centering
	\subfigure[VelocityX]{\includegraphics[width=0.32\textwidth]{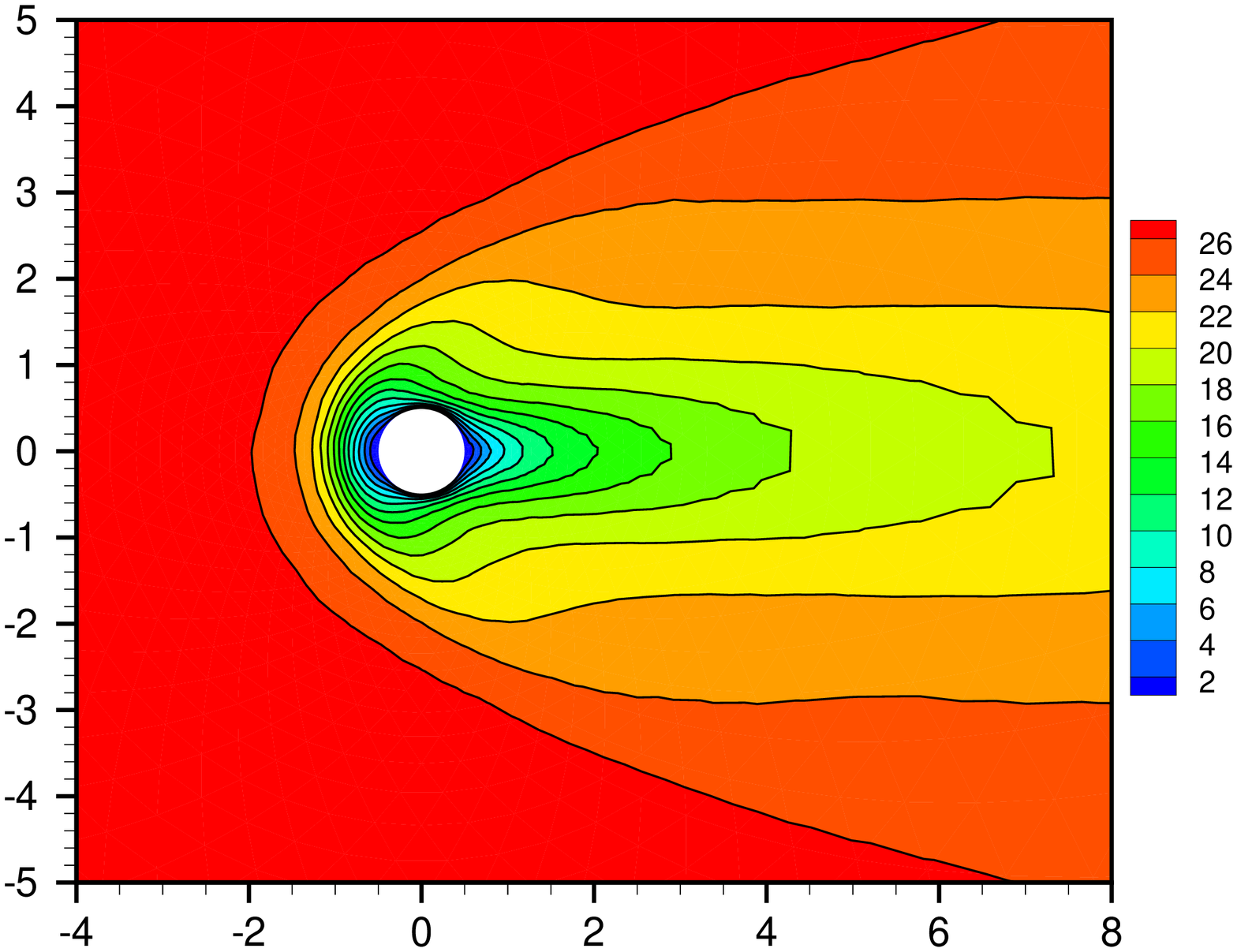}}
	\subfigure[VelocityY]{\includegraphics[width=0.32\textwidth]{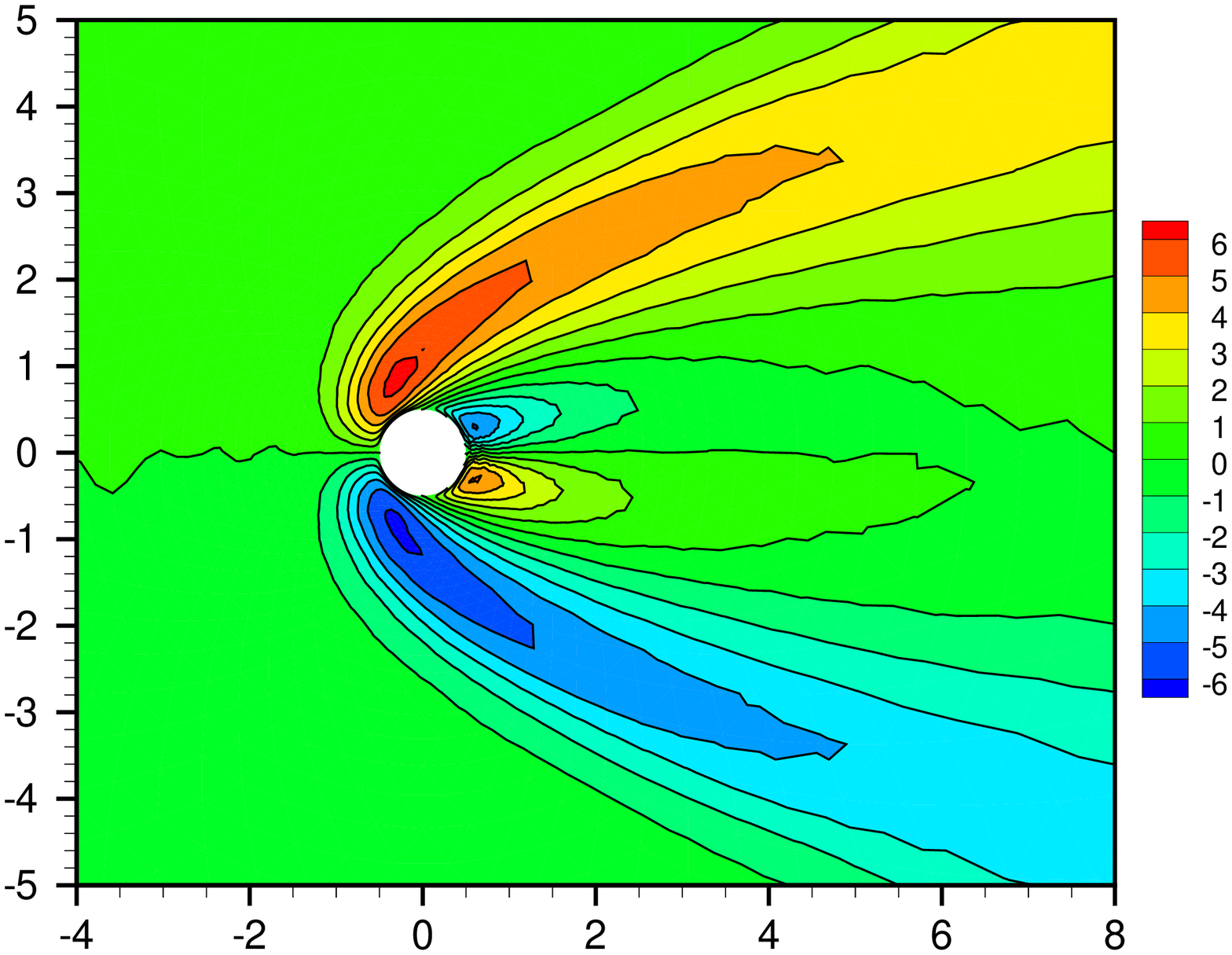}}
	\subfigure[Temperature]{\includegraphics[width=0.32\textwidth]{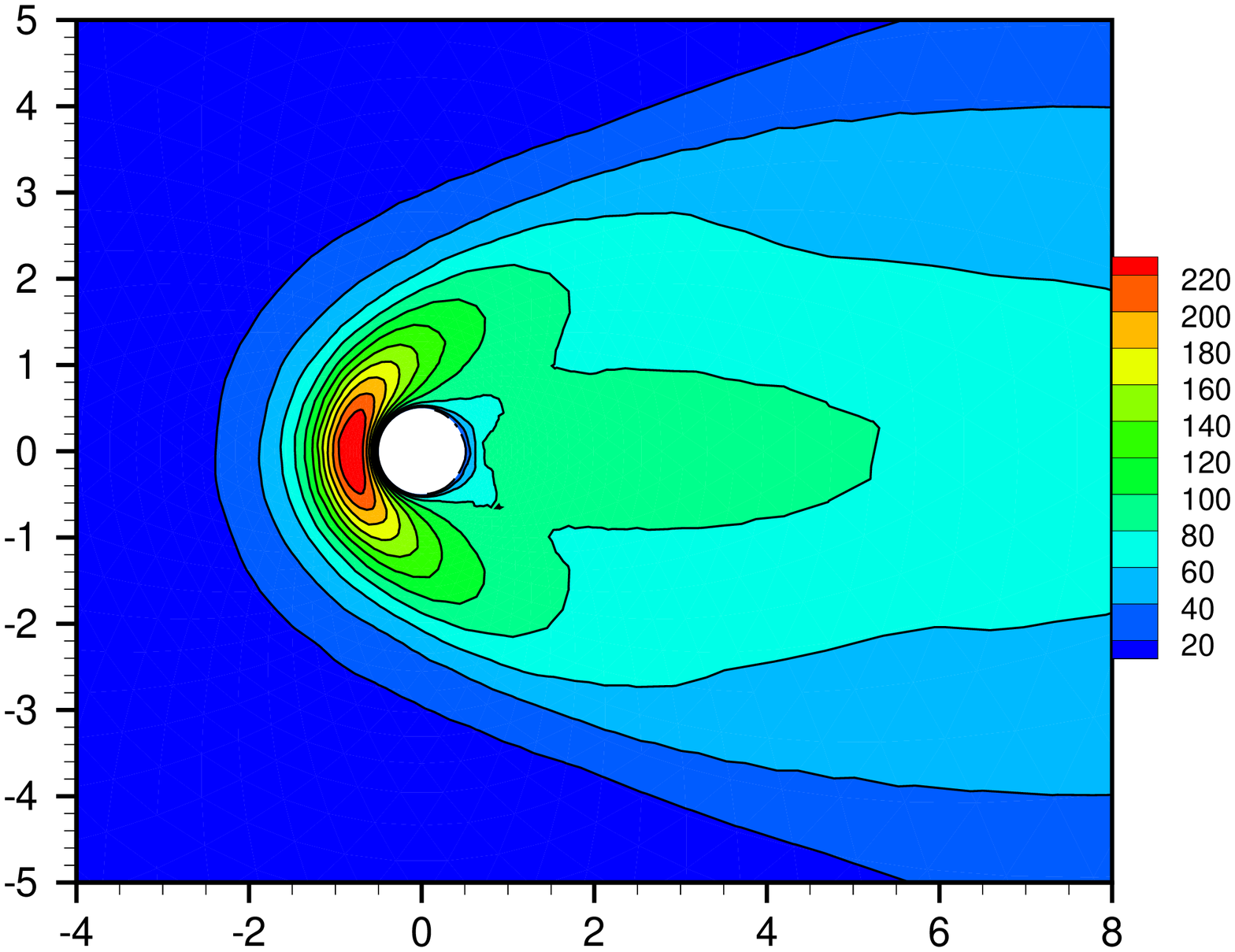}}
	\caption{\label{fig:cylinder_ma30}Hypersonic flow at ${\rm Ma}=30$ around a circle cylinder for ${\rm Kn} = 0.1$ obtained by the UGKWP method. The velocities are normalized by the most probable speed $C_{\infty} = \sqrt{2 k_B T_{\infty} / m_0} = 337 {\rm m/s}$ and the temperature is normalized by the free stream temperature $T_{\infty} = 273{\rm K}$.}
\end{figure}

Furthermore, a very high speed flow at ${\rm Ma}=30$ is computed for the case ${\rm Kn} = 0.1$ on the same unstructured mesh.
Since the memory requirement of the discrete velocity points for the UGKS is unaffordable for such high Mach number computation, we only show the results of the UGKWP method in Fig.~\ref{fig:cylinder_ma30}.
In the computation, the memory cost of the UGKWP method is only $375$ MB.
The advantage of the particle method with a nature adaptivity in the phase space through particles is well inherited by the UGKWP method for high speed rarefied flow computations.

\section{Discussion and Conclusion}
In this paper, we introduce the unified gas-kinetic wave-particle (UGKWP) method on unstructured mesh for flow simulation in all Knudsen regimes.
Similar to the UGKS methodology,
the direct modeling of the flow physics on numerical mesh size and time step is carried out to construct the multiscale algorithm.
The early discrete velocity-based UGKS is further developed to the purely particle-based UGKP and wave-particle-based UGKWP methods.
In the UGKP method, based on the integral solution of the kinetic model equation,
the free transport and collision processes are well described for the evolution of simulation particles in a statistical point of view.
Different from the DSMC method where simulation particles stream for a whole time step $\Delta t$ and then get possible collision,
the free transport time of the simulation particles in both UGKP and UGKWP methods is obtained from the integral solution, and the free streaming convection is constrained due to particles' interaction in different flow regimes.
The collision process is handled by re-sampling simulation particles from a Maxwellian distribution according to the conservation laws.
 Due to the multiscale transport modeling in the UGKS methodology, the UGKWP method has no kinetic scale related 
 time step and cell size limitations which are imposed in many other kinetic equation solvers and particle methods.

A novel wave-particle adaptive formulation is introduced to describe the microscopic gas distribution function.
Specifically, the flow state in each cell contains the deterministic hydrodynamic waves and the stochastic simulation particles, and the proportion between the waves and particles evolves adaptively according to the local flow physics.
Unified treatment can be carried out for all finite volume cells in the computational domain.
In the continuum regimes, the hydrodynamic waves are dominant and the UGKWP method goes to a hydrodynamic fluid solver, such as GKS;
while in the highly rarefied flow, the UGKWP method performs as same as the stochastic particle method.
The wave-particle adaptivity makes the UGKWP method very efficient in different flow regimes by inheriting the advantages of the deterministic method and the stochastic method, and the advantages of kinetic particle transport and hydrodynamic continuum wave evolution.

Numerical test cases, including Sod problem at different Knudser numbers, low-speed micro cavity flow, laminar boundary layer for viscous NS solutions, and high-speed cylinder flow,
 are computed to validate the current method.
It shows that the UGKWP method can recover the UGKS solutions in all flow regimes.
For low-speed micro flow with a small temperature variation, 
the UGKS with acceleration techniques shows obvious advantages over the stochastic UGKWP method on the aspects of efficiency and accuracy because the deterministic UGKS does not suffer from the statistical noises.
For the continuum flows at small Knudsen numbers and the rarefied gas flow at high Mach numbers, the UGKWP method can achieve much higher efficiency and requires lower memory cost.
The unified treatment, multiscale property, high efficiency of the UGKWP method make it a very promising tool to study the multiscale problems in real engineering applications, such as the reentry of space vehicles and the high-speed near-space flights.

In the current study, we only consider the BGK model equation with a unit Prandtl number.
It would not be difficult to apply other kinetic models in the current method to obtain more accurate results or to develop a more realistic method for real gas simulations.
Moreover, the concepts of the wave-particle adaptive formulation and the direct modeling on the mesh size and time step scales could be used in other multiscale transport processes, such as plasma, granular flow, and radiation, in the construction of multiscale multi-efficiency methods.

\section{Acknowledgement}

The work of Zhu and Zhong is supported by the National Natural Science Foundation of China (Grant No. 11472219), the 111 Project of China (B17037) as well as the ATCFD Project (2015-F-016).
The research of Liu and Xu is supported Hong Kong research grant council (16206617) and National Natural Science Foundation of China (11772281, 91852114).

\bibliography{ugkwp}

\begin{thebibliography}{10}

\bibitem{bird1994book}
Graeme~Austin Bird.
\newblock {\em Molecular gas dynamics and the direct simulation of gas flows}.
\newblock Oxford University Press, USA, 1994.

\bibitem{oran1998direct}
ES~Oran, CK~Oh, and BZ~Cybyk.
\newblock Direct simulation {Monte} {Carlo}: recent advances and applications.
\newblock {\em Annual Review of Fluid Mechanics}, 30(1):403--441, 1998.

\bibitem{fan2001statistical}
Jing Fan and Ching Shen.
\newblock Statistical simulation of low-speed rarefied gas flows.
\newblock {\em Journal of Computational Physics}, 167(2):393--412, 2001.

\bibitem{shen2006rarefied}
Ching Shen.
\newblock {\em Rarefied gas dynamics: fundamentals, simulations and micro
  flows}.
\newblock Springer Science \& Business Media, 2006.

\bibitem{sun2002direct}
Quanhua Sun and Iain~D Boyd.
\newblock A direct simulation method for subsonic, microscale gas flows.
\newblock {\em Journal of Computational Physics}, 179(2):400--425, 2002.

\bibitem{baker2005variance}
Lowell~L Baker and Nicolas~G Hadjiconstantinou.
\newblock Variance reduction for {Monte} {Carlo} solutions of the {Boltzmann}
  equation.
\newblock {\em Physics of Fluids}, 17(5):051703, 2005.

\bibitem{homolle2007low}
Thomas~MM Homolle and Nicolas~G Hadjiconstantinou.
\newblock A low-variance deviational simulation {Monte} {Carlo} for the
  {Boltzmann} equation.
\newblock {\em Journal of Computational Physics}, 226(2):2341--2358, 2007.

\bibitem{degond2011moment}
Pierre Degond, Giacomo Dimarco, and Lorenzo Pareschi.
\newblock The moment-guided {Monte} {Carlo} method.
\newblock {\em International Journal for Numerical Methods in Fluids},
  67(2):189--213, 2011.

\bibitem{pareschi2001time}
Lorenzo Pareschi and Giovanni Russo.
\newblock Time relaxed {Monte} {Carlo} methods for the {Boltzmann} equation.
\newblock {\em SIAM Journal on Scientific Computing}, 23(4):1253--1273, 2001.

\bibitem{pareschi2000asymptotic}
Lorenzo Pareschi and Giovanni Russo.
\newblock Asymptotic preserving {Monte} {Carlo} methods for the {Boltzmann}
  equation.
\newblock {\em Transport Theory and Statistical Physics}, 29(3-5):415--430,
  2000.

\bibitem{ren2014asymptotic}
Wei Ren, Hong Liu, and Shi Jin.
\newblock An asymptotic-preserving {Monte} {Carlo} method for the {Boltzmann}
  equation.
\newblock {\em Journal of Computational Physics}, 276:380--404, 2014.

\bibitem{dimarco2011exponential}
Giacomo Dimarco and Lorenzo Pareschi.
\newblock Exponential {Runge}--{Kutta} methods for stiff kinetic equations.
\newblock {\em SIAM Journal on Numerical Analysis}, 49(5):2057--2077, 2011.

\bibitem{burt2008low}
Jonathan~M Burt and Iain~D Boyd.
\newblock A low diffusion particle method for simulating compressible inviscid
  flows.
\newblock {\em Journal of Computational Physics}, 227(9):4653--4670, 2008.

\bibitem{nanbu1981simulation}
Kenichi Nanbu.
\newblock On the simulation method for the {Bhatnager}--{Gross}--{Krook}
  equation.
\newblock {\em Journal of the Physical Society of Japan}, 50(9):3154--3158,
  1981.

\bibitem{gallis2000application}
Michael Gallis and John Torczynski.
\newblock The application of the {BGK} model in particle simulations.
\newblock In {\em 34th Thermophysics Conference}, page 2360, 2000.

\bibitem{macrossan2001particle}
Michael~N Macrossan.
\newblock A particle simulation method for the {BGK} equation.
\newblock In {\em AIP Conference Proceedings}, volume 585, pages 426--433. AIP,
  2001.

\bibitem{tumuklu2016particle}
Ozgur Tumuklu, Zheng Li, and Deborah~A Levin.
\newblock Particle ellipsoidal statistical {Bhatnagar}--{Gross}--{Krook}
  approach for simulation of hypersonic shocks.
\newblock {\em AIAA Journal}, pages 3701--3716, 2016.

\bibitem{fei2018particle}
Fei Fei, Jun Zhang, Jing Li, and Zhaohui Liu.
\newblock A unified stochastic particle {Bhatnagar}--{Gross}--{Krook} method
  for multiscale gas flows.
\newblock {\em arXiv preprint arXiv:1808.03801}, 2018.

\bibitem{jenny2010solution}
Patrick Jenny, Manuel Torrilhon, and Stefan Heinz.
\newblock A solution algorithm for the fluid dynamic equations based on a
  stochastic model for molecular motion.
\newblock {\em Journal of computational physics}, 229(4):1077--1098, 2010.

\bibitem{gorji2011fokker}
Mohammad~H Gorji, Maniel Torrilhon, and Patrick Jenny.
\newblock {Fokker}--{Planck} model for computational studies of monatomic
  rarefied gas flows.
\newblock {\em Journal of fluid mechanics}, 680:574--601, 2011.

\bibitem{gorji2015fokker}
M~Hossein Gorji and Patrick Jenny.
\newblock {Fokker}--{Planck}--{DSMC} algorithm for simulations of rarefied gas
  flows.
\newblock {\em Journal of Computational Physics}, 287:110--129, 2015.

\bibitem{chu1965kinetic}
CK~Chu.
\newblock Kinetic-theoretic description of the formation of a shock wave.
\newblock {\em The Physics of Fluids}, 8(1):12--22, 1965.

\bibitem{aristov2012direct}
VV~Aristov.
\newblock {\em Direct methods for solving the {Boltzmann} equation and study of
  nonequilibrium flows}, volume~60.
\newblock Springer Science \& Business Media, 2012.

\bibitem{tcheremissine2005direct}
Felix Tcheremissine.
\newblock Direct numerical solution of the {Boltzmann} equation.
\newblock In {\em AIP Conference Proceedings}, volume 762, pages 677--685. AIP,
  2005.

\bibitem{li2004gkua}
Zhi-Hui Li and Han-Xin Zhang.
\newblock Study on gas kinetic unified algorithm for flows from rarefied
  transition to continuum.
\newblock {\em Journal of Computational Physics}, 193(2):708--738, 2004.

\bibitem{li2019}
Zhi-Hui Li, Ao-Ping Peng, Qiang Ma, Lei-Ning Dang, Xiao-Wei Tang, and Xue-Zhou
  Sun.
\newblock Gas-kinetic unified algorithm for computable modeling of {Boltzmann}
  equation and application to aerothermodynamics for falling disintegration of
  uncontrolled {Tiangong}-{No.1} spacecraft.
\newblock {\em Advances in Aerodynamics, (2019) 1:4},
  https://doi.org/10.1186/s42774-019-0009-4.

\bibitem{xu2010unified}
Kun Xu and Juan-Chen Huang.
\newblock A unified gas-kinetic scheme for continuum and rarefied flows.
\newblock {\em Journal of Computational Physics}, 229(20):7747--7764, 2010.

\bibitem{guo2013discrete}
Zhaoli Guo, Kun Xu, and Ruijie Wang.
\newblock Discrete unified gas kinetic scheme for all {Knudsen} number flows:
  Low-speed isothermal case.
\newblock {\em Physical Review E}, 88(3):033305, 2013.

\bibitem{yang1995rarefied}
JY~Yang and JC~Huang.
\newblock Rarefied flow computations using nonlinear model {Boltzmann}
  equations.
\newblock {\em Journal of Computational Physics}, 120(2):323--339, 1995.

\bibitem{mieussens2000discrete}
Luc Mieussens.
\newblock Discrete-velocity models and numerical schemes for the
  {Boltzmann}-{BGK} equation in plane and axisymmetric geometries.
\newblock {\em Journal of Computational Physics}, 162(2):429--466, 2000.

\bibitem{mieussens2000implicit}
Luc Mieussens.
\newblock Discrete velocity model and implicit scheme for the {BGK} equation of
  rarefied gas dynamics.
\newblock {\em Mathematical Models and Methods in Applied Sciences},
  10(08):1121--1149, 2000.

\bibitem{zhu2016implicit}
Yajun Zhu, Chengwen Zhong, and Kun Xu.
\newblock Implicit unified gas-kinetic scheme for steady state solutions in all
  flow regimes.
\newblock {\em Journal of Computational Physics}, 315:16--38, 2016.

\bibitem{zhu2017unified}
Yajun Zhu, Chengwen Zhong, and Kun Xu.
\newblock Unified gas-kinetic scheme with multigrid convergence for rarefied
  flow study.
\newblock {\em Physics of Fluids}, 29(9):096102, 2017.

\bibitem{zhu2019implicit}
Yajun Zhu, Chengwen Zhong, and Kun Xu.
\newblock An implicit unified gas-kinetic scheme for unsteady flow in all
  {Knudsen} regimes.
\newblock {\em Journal of Computational Physics}, 2019.

\bibitem{taitano2014moment}
William~T Taitano, Dana~A Knoll, Luis Chac{\'o}n, Jon~M Reisner, and Anil~K
  Prinja.
\newblock Moment-based acceleration for neutral gas kinetics with {BGK}
  collision operator.
\newblock {\em Journal of Computational and Theoretical Transport},
  43(1-7):83--108, 2014.

\bibitem{chacon2017multiscale}
Luis Chacon, Guangye Chen, Dana~A Knoll, C~Newman, H~Park, William Taitano,
  Jeff~A Willert, and Geoffrey Womeldorff.
\newblock Multiscale high-order/low-order ({HOLO}) algorithms and applications.
\newblock {\em Journal of Computational Physics}, 330:21--45, 2017.

\bibitem{chen2017memory}
Songze Chen, Chuang Zhang, Lianhua Zhu, and Zhaoli Guo.
\newblock A unified implicit scheme for kinetic model equations. part {I}.
  memory reduction technique.
\newblock {\em Science bulletin}, 62(2):119--129, 2017.

\bibitem{yang2018memory}
LM~Yang, C~Shu, WM~Yang, and J~Wu.
\newblock An implicit scheme with memory reduction technique for steady state
  solutions of {DVBE} in all flow regimes.
\newblock {\em Physics of Fluids}, 30(4):040901, 2018.

\bibitem{mouhot2006fast}
Cl{\'e}ment Mouhot and Lorenzo Pareschi.
\newblock Fast algorithms for computing the {Boltzmann} collision operator.
\newblock {\em Mathematics of computation}, 75(256):1833--1852, 2006.

\bibitem{wu2013deterministic}
Lei Wu, Craig White, Thomas~J Scanlon, Jason~M Reese, and Yonghao Zhang.
\newblock Deterministic numerical solutions of the {Boltzmann} equation using
  the fast spectral method.
\newblock {\em Journal of Computational Physics}, 250:27--52, 2013.

\bibitem{chen2012amr}
Songze Chen, Kun Xu, Cunbiao Lee, and Qingdong Cai.
\newblock A unified gas kinetic scheme with moving mesh and velocity space
  adaptation.
\newblock {\em Journal of Computational Physics}, 231(20):6643--6664, 2012.

\bibitem{filbet2010class}
Francis Filbet and Shi Jin.
\newblock A class of asymptotic-preserving schemes for kinetic equations and
  related problems with stiff sources.
\newblock {\em Journal of Computational Physics}, 229(20):7625--7648, 2010.

\bibitem{dimarco2013imex}
Giacomo Dimarco and Lorenzo Pareschi.
\newblock Asymptotic preserving implicit-explicit {Runge}--{Kutta} methods for
  nonlinear kinetic equations.
\newblock {\em SIAM Journal on Numerical Analysis}, 51(2):1064--1087, 2013.

\bibitem{xu2015direct}
Kun Xu.
\newblock {\em Direct modeling for computational fluid dynamics: construction
  and application of unified gas-kinetic schemes}.
\newblock World Scientific, 2015.

\bibitem{xu2001gks}
Kun Xu.
\newblock A gas-kinetic {BGK} scheme for the {Navier}--{Stokes} equations and
  its connection with artificial dissipation and {Godunov} method.
\newblock {\em Journal of Computational Physics}, 171(1):289--335, 2001.

\bibitem{jiang}
Dingwu Jiang, Meiliang Mao, Jin Li, and Xiaogang Deng.
\newblock An implicit parallel {UGKS} solver for flows covering various
  regimes.
\newblock {\em Advances in Aerodynamics, (2019) 1:8},
  https://doi.org/10.1186/s42774-019-0008-5.

\bibitem{sun2015gray}
Wenjun Sun, Song Jiang, and Kun Xu.
\newblock An asymptotic preserving unified gas kinetic scheme for gray
  radiative transfer equations.
\newblock {\em Journal of Computational Physics}, 285:265--279, 2015.

\bibitem{sun2015asymptotic}
Wenjun Sun, Song Jiang, Kun Xu, and Shu Li.
\newblock An asymptotic preserving unified gas kinetic scheme for
  frequency-dependent radiative transfer equations.
\newblock {\em Journal of Computational Physics}, 302:222--238, 2015.

\bibitem{guo2016phonon}
Zhaoli Guo and Kun Xu.
\newblock Discrete unified gas kinetic scheme for multiscale heat transfer
  based on the phonon {Boltzmann} transport equation.
\newblock {\em International Journal of Heat and Mass Transfer}, 102:944--958,
  2016.

\bibitem{luo2017discrete}
Xiao-Ping Luo and Hong-Liang Yi.
\newblock A discrete unified gas kinetic scheme for phonon {Boltzmann}
  transport equation accounting for phonon dispersion and polarization.
\newblock {\em International Journal of Heat and Mass Transfer}, 114:970--980,
  2017.

\bibitem{liu2017plasma}
Chang Liu and Kun Xu.
\newblock A unified gas kinetic scheme for continuum and rarefied flows {V}:
  multiscale and multi-component plasma transport.
\newblock {\em Communications in Computational Physics}, 22(5):1175--1223,
  2017.

\bibitem{liu2019granular}
Chang Liu, Zhao Wang, and Kun Xu.
\newblock A unified gas-kinetic scheme for continuum and rarefied flows {VI}:
  Dilute disperse gas-particle multiphase system.
\newblock {\em Journal of Computational Physics}, 2019.

\bibitem{li2018ugkp}
Weiming Li, Chang Liu, Yajun Zhu, Jiwei Zhang, and Kun Xu.
\newblock A unified gas-kinetic particle method for multiscale photon
  transport.
\newblock {\em arXiv preprint arXiv:1810.05984}, 2018.

\bibitem{liu2018ugkwp}
Chang Liu, Yajun Zhu, and Kun Xu.
\newblock Unified gas-kinetic wave-particle methods {I}: Continuum and rarefied
  gas flow.
\newblock {\em arXiv preprint arXiv:1811.07141}, 2018.

\bibitem{wadsworth1990one}
Deanc Wadsworth and Daniela Erwin.
\newblock One-dimensional hybrid continuum/particle simulation approach for
  rarefied hypersonic flows.
\newblock In {\em 5th Joint Thermophysics and Heat Transfer Conference}, page
  1690, 1990.

\bibitem{sun2004hybrid}
Quanhua Sun, Iain~D Boyd, and Graham~V Candler.
\newblock A hybrid continuum/particle approach for modeling subsonic, rarefied
  gas flows.
\newblock {\em Journal of Computational Physics}, 194(1):256--277, 2004.

\bibitem{degond2010multiscale}
Pierre Degond, Giacomo Dimarco, and Luc Mieussens.
\newblock A multiscale kinetic--fluid solver with dynamic localization of
  kinetic effects.
\newblock {\em Journal of Computational Physics}, 229(13):4907--4933, 2010.

\bibitem{bhatnagar1954model}
Prabhu~Lal Bhatnagar, Eugene~P Gross, and Max Krook.
\newblock A model for collision processes in gases. {I}. small amplitude
  processes in charged and neutral one-component systems.
\newblock {\em Physical review}, 94(3):511, 1954.

\bibitem{huang2012unified}
Juan-Chen Huang, Kun Xu, and Pubing Yu.
\newblock A unified gas-kinetic scheme for continuum and rarefied flows {II}:
  multi-dimensional cases.
\newblock {\em Communications in Computational Physics}, 12(3):662--690, 2012.

\bibitem{haselbacher2007efficient}
Andreas Haselbacher, Fady~M Najjar, and James~P Ferry.
\newblock An efficient and robust particle-localization algorithm for
  unstructured grids.
\newblock {\em Journal of Computational Physics}, 225(2):2198--2213, 2007.

\bibitem{xu1994numerical}
Kun Xu and Kevin~H Prendergast.
\newblock Numerical {Navier}--{Stokes} solutions from gas kinetic theory.
\newblock {\em Journal of Computational Physics}, 114(1):9--17, 1994.

\end{thebibliography}
\bibliographystyle{unsrt}
\end{document}